%% file: N4_BCFT.tex
\RequirePackage{pdf14}

\documentclass[letterpaper]{article}
\pdfoutput=1
\usepackage[usenames,dvipsnames,table]{xcolor}
\usepackage{graphicx}
\usepackage{caption}
\usepackage{subcaption}
\usepackage{array,setspace,mathrsfs,amsthm,amsfonts,amsmath,yfonts,dsfont,bbm,colonequals,mathtools}
\usepackage{./aux/jheppub}
\usepackage{afterpage}
\usepackage{xspace}
\usepackage[normalem]{ulem}
\usepackage{mathrsfs}
\DeclareMathAlphabet{\mathpzc}{OT1}{pzc}{m}{it}
\makeatletter
\newlength{\apb@width}
\newcommand{\autoparbox}[2][c]{\settowidth{\apb@width}{#2}\parbox[#1]{\apb@width}{#2}}

\makeatother

\newdimen\yysquaresize
\newdimen\yyrsquaresize
\newdimen\yythickness
\newdimen\yyskip
\def\yysquare#1{%
\setlength{\yyrsquaresize}{\yysquaresize}%
\addtolength{\yyrsquaresize}{-2\yythickness}%
\vrule width \yythickness%
\vbox to \yysquaresize{%
  \hrule height \yythickness\vss%
  \hbox to \yyrsquaresize{\hss#1\hss}%
  \vss\hrule height \yythickness}%
\vrule width \yythickness}

\def\yyyoung#1{\vtop{\baselineskip0pt\lineskip-\yythickness\halign{\tabskip-\yythickness&\yysquare{##}\cr #1}}}

\newcommand{\young}[1]{\hskip\yyskip\mbox{\yyyoung{#1}}\hskip\yyskip}

\input{./aux/topmatter}

\title{
Bootstrap equations for $\mathcal{N}=4$ SYM with defects
}
\author[1]{Pedro Liendo,}
\author[2]{Carlo Meneghelli}

\affiliation[1]{IMIP, Humboldt-Universit{\"a}t zu Berlin, IRIS Adlershof, Zum Gro{\ss}en Windkanal 6, 12489 Berlin, Germany}
\affiliation[2]{Simons Center for Geometry and Physics, Stony Brook University, Stony Brook, NY 11794-3636, USA}

\emailAdd{pliendo@physik.hu-berlin.de}
\emailAdd{cmeneghelli@scgp.stonybrook.edu}

\preprint{HU-EP-16/28}

\bigskip
\abstract{
This paper focuses on the analysis of $4d$ $\mathcal{N}=4$ superconformal theories in the presence of a defect from the point of view of the conformal bootstrap. 
We will concentrate  first on the case of codimension one, where the defect is a boundary that preserves half of the supersymmetry.
After studying the constraints imposed by supersymmetry, we will obtain the Ward identities associated to two-point functions of $\tfrac{1}{2}$-BPS operators and write their solution as a superconformal block expansion.
Due to a surprising connection between spacetime and R-symmetry conformal blocks, our results not only apply to $4d$ $\mathcal{N}=4$ superconformal theories with a boundary, but also to three more systems that have the same symmetry algebra: $4d$ $\mathcal{N}=4$ superconformal theories with a line defect, $3d$ $\mathcal{N}=4$ superconformal theories with no defect, and $OSP(4^*|4)$ superconformal quantum mechanics.
The superconformal algebra implies that all these systems possess a closed subsector of
operators in which the bootstrap equations become polynomial constraints on the CFT data. We derive 
these truncated equations and initiate the study of their solutions.
}


\begin{document}
\setcounter{tocdepth}{2}
\maketitle
\setcounter{page}{1}

\input{sections/1_intro}

\input{sections/2_superspace}
\input{sections/3_superblocks}

\input{sections/4_bootstrap}

\input{sections/5_conclusions}

\input{sections/acknowledgments}

\appendix
\addtocontents{toc}{\protect\setcounter{tocdepth}{1}}
\input{sections/A_appendix}

\input{sections/B_appendix}

\input{sections/C_appendix}

\bibliography{./aux/biblio}
\bibliographystyle{./aux/JHEP}

\end{document}

%% file: aux/topmatter.tex
\newcommand\be{\begin{equation}}
\newcommand\bea{\begin{eqnarray}}
\newcommand\eea{\end{eqnarray}}
\newcommand\ee{\end{equation}}

\makeatletter
\def\maketag@@@#1{\hbox{\m@th\normalfont\normalsize#1}}
\makeatother

\def\Bm{{\mathcal{B}}}
\def\Nm{{\mathcal{N}}}



%% file: sections/1_intro.tex

\section{Introduction}

The conformal bootstrap has seen a revival in recent years and given a plethora of results both numerical and analytical.
It is a non-perturbative approach to quantum field theory in which symmetry principles play a fundamental role without relying on weak-coupling expansions. Some of the highlights that the modern reincarnation of the bootstrap has given include high precision estimates of critical exponents \cite{ElShowk:2012ht,El-Showk:2014dwa,Kos:2014bka}, universal behavior of CFTs at large spin \cite{Komargodski:2012ek,Fitzpatrick:2012yx}, and the discovery of solvable subsectors in superconformal theories \cite{Beem:2013sza,Beem:2014kka,Chester:2014mea,Beem:2016cbd}.

An important class of systems and the main focus of this work are CFTs in the presence of an extended operator or ``defect''. Conformal defects have numerous applications in condensed matter systems, holography, and formal aspects of superconformal field theory.
From the bootstrap point of view most of the work has been on correlation functions of local operators in the presence of a defect.
Apart from the rich and well understood case of $2d$ CFTs \cite{Cardy:1984bb,Cardy:1989ir,Frohlich:2006ch}, 
two early papers in the higher dimensional case are \cite{McAvity:1993ue,McAvity:1995zd} where two-point functions of CFTs near a boundary were studied. More recently, the authors of \cite{Billo:2016cpy} presented a thorough analysis of two-point functions in the presence of a defect of any codimension. 

The most interesting feature of defect CFTs is the presence of additional \textit{CFT data}. In addition to the standard OPE coefficients of bulk operators and OPE coefficients of local operators on the defects,
 there are also bulk-to-defect couplings\footnote{
There is also the interesting possibility of introducing an expansion of the defect itself in terms of local  bulk
operators as proposed in 
\cite{Berenstein:1998ij},
and recently investigated in \cite{Gadde:2016fbj}.
We will not consider such expansions in this paper.
}. 
 The latter control the convergent expansion of  local operators in the bulk in terms of local operators on the defect.
 The increase in data is balanced by the fact that there are also more constraints. Two-point functions in the presence of the defect exemplify this feature. In figure \ref{fig:crossing} there is a pictorial representation of ``crossing symmetry'' for this system. 
\begin{figure}[h!]
\label{fig:crossing}
\centering
\includegraphics[scale=0.35]{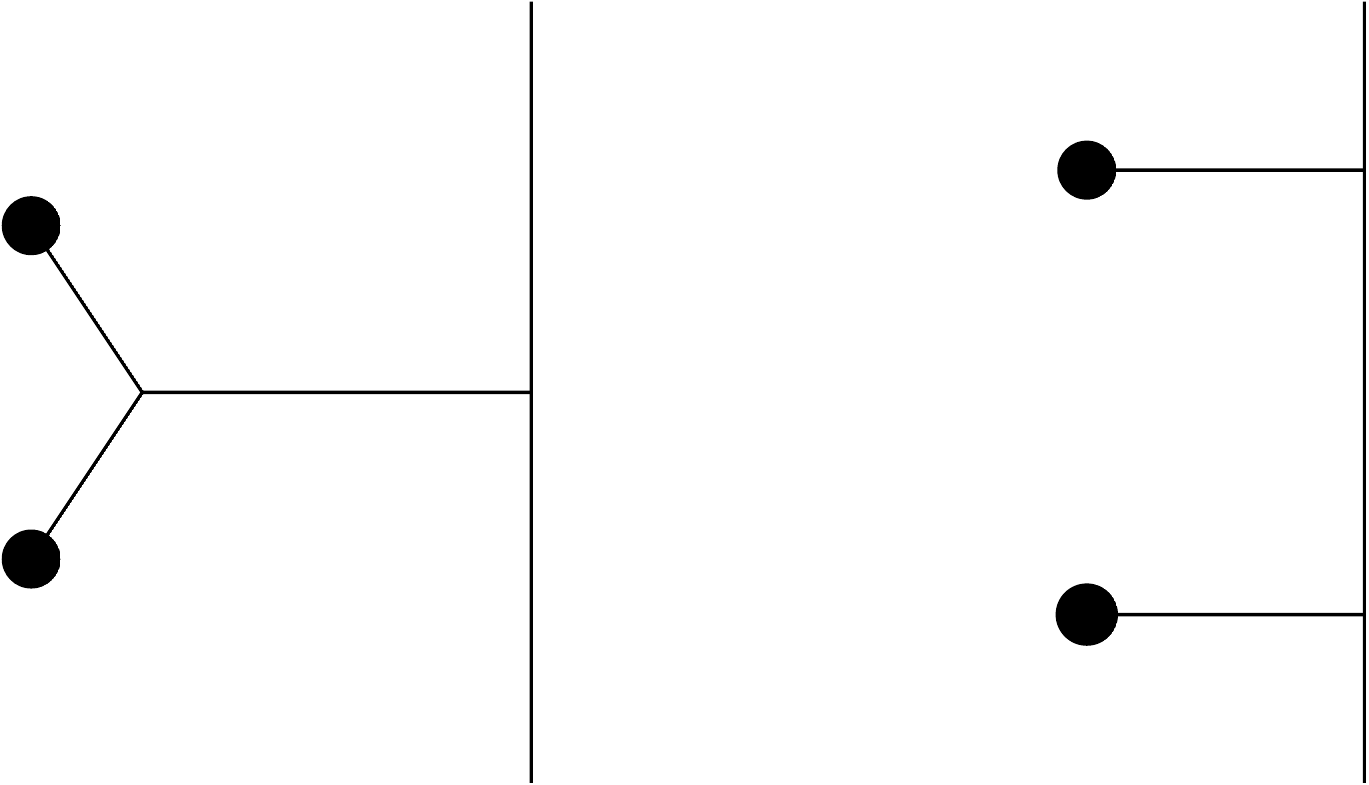}      
\put(-75,37){$=$} 
\put(-165,37){\LARGE{$\sum$}} 
\put(-55,37){\LARGE{$\sum$}} 
\caption{Crossing symmetry for two-point functions in the presence of a defect. The channel on the left represents the standard bulk OPE between local operators.
 The channel on the right is the bulk-to-defect OPE in which each local operator can be written as a convergent sum of defect operators.}
\label{fig:crossing}
\end{figure}
 
The channel on the left depicts the standard bulk OPE followed by taking the expectation value in the presence of the defect.
Because the full conformal symmetry is partially broken,
 certain bulk local operators can have a  non-zero one-point function. 
  The CFT data for this channel are thus the standard three-point couplings times the one-point function coefficients. 
  The channel on the right corresponds to expanding each bulk local operator in terms of local operators on the boundary.
  The resulting two-point function of boundary operators is then fixed by the defect conformal symmetry.
This channel thus involves only the new bulk-to-defect couplings, a subset of which is given by the
one-point function coefficients.
  It should be noticed that, given CFT data for the  theory in the bulk and for the theory restricted to the boundary,
the  crossing equations   in figure \ref{fig:crossing} do not imply by themselves that the full boundary CFT is consistent.
They should be supplemented with the generalization of the case presented 
in figure \ref{fig:crossing}, when an arbitrary local operator on the boundary is added.
Note that even when the external operators are identical, the coefficients of the conformal block expansion do not exhibit any positivity property.
The numerical bootstrap program was applied to defect CFTs in \cite{Liendo:2012hy,Gaiotto:2013nva} using the method of \cite{Rattazzi:2010yc}, and in \cite{Gliozzi:2015qsa,Gliozzi:2016cmg} using the method of \cite{Gliozzi:2013ysa}.

In this work we will be interested in flat defect CFTs that also exhibit supersymmetry. In particular, $\Nm=4$ SYM with $\tfrac{1}{2}$-BPS boundary conditions. This set of boundary conditions was extensively studied by Gaiotto and Witten in \cite{Gaiotto:2008sa}\footnote{
A class of 
 $\frac{1}{2}$-BPS interfaces in $\mathcal{N}=4$  SYM were  constructed in \cite{D'Hoker:2006uv}.
 In the context of integrability, there have been perturbative calculations of one-point functions in this setup,
see  \cite{deLeeuw:2015hxa,Rapcak:2015lhn,Buhl-Mortensen:2016pxs,deLeeuw:2016vgp} and references therein.
Defect CFTs  have also been studied from the point of view of holography, see
 \cite{Karch:2000gx,DeWolfe:2001pq,Erdmenger:2002ex,Constable:2002xt} and references thereafter.}.
 Even though this is an interesting system on its own, a particular motivation for our work is to present a set of bootstrap equations that might be tractable analytically. The analytical understanding of the bootstrap has improved significantly in recent years, however, much remains to be done. Two-point functions in the presence of a flat defect are a promising arena: the blocks depend on only one cross-ratio and are therefore relatively simple. The addition of supersymmetry makes the setup even more attractive. Supersymmetry is a powerful tool that gives good analytic control on certain quantities which are otherwise hard to study. If a non-trivial analytic solution to the bootstrap equations is within reach, the equations presented in this paper are a promising candidate for it. 

The system we discuss possesses the three-dimensional superconformal symmetry $OSP(4|4)$.
As explained in \cite{Chester:2014mea,Beem:2016cbd}, 
 any three-dimensional $\mathcal{N}=4$ superconformal theory 
contains a closed subsector of operators 
  whose correlators are described by a one-dimensional topological theory.
It was also mentioned in \cite{Beem:2016cbd} that such considerations extend to the case in which the
three-dimensional theory lives on the boundary of a $d=4$, $\mathcal{N}=4$ theory.
The construction consists in truncating the system by restricting to the cohomology of a 
  special supercharge inside $\mathfrak{osp}(4|4)$. 
 We will therefore refer to this subsector as the ``cohomological sector''. The approach championed in the original papers \cite{Beem:2013sza,Beem:2014kka,Chester:2014mea,Beem:2016cbd} is that the superconformal bootstrap can be implemented as a two-step process. First, we solve for the truncated cohomological sector using analytical tools, and then we proceed to the harder task of studying non-protected quantities, maybe resorting to numerics\footnote{The bootstrap for $\Nm=4$ SYM without defects was studied in \cite{Beem:2013qxa}.}.

In this paper we set the stage for this two-step program. We start by obtaining the full superconformal block expansion for two-point functions of $\tfrac{1}{2}$-BPS operators.
 This expansion, when restricted to the cohomological sector, implies an infinite set of polynomial equations relating bulk and boundary data. We present a preliminary study of solutions to this truncated system, leaving a more complete analysis for future work. We also explore the full bootstrap equations with the goal of eventually using modern numerical techniques to study their dynamics.

It should be pointed out that all the results we present for the codimension one $\frac{1}{2}$-BPS defect in
 a $d=4$, $\mathcal{N}=4$ superconformal theory, automatically extend to the case of a 
 codimension three $\frac{1}{2}$-BPS defect. The connection is spelled out in the main text and
 includes the superconformal blocks and the existence of a 
 cohomological sector.
 Moreover, and somewhat surprisingly, we find that 
 superconformal blocks for four-point functions in 
 $d=3$, $\mathcal{N}=4$ theories and $OSP(4^*|4)$ superformal quantum mechanics with no defect,
 are essentially equal to the defect superblocks we determine. 
 The latter superblocks have also not appeared in the literature yet, and we will therefore solve four different systems
 in one blow.

The outline of the paper is as follows. In section \ref{sec:superspace} we 
study one- and two-point functions using a novel superspace setup.
 For two-point functions of $\tfrac{1}{2}$-BPS operators we obtain the corresponding Ward identities, which capture in an elegant and compact form the constraints of superconformal invariance. In section \ref{sec:superblocks} we present the complete solution of the Ward identities in the form of a superconformal block expansion for the correlator. In section \ref{sec:bootstrap} we initiate the study of solutions to the bootstrap equations concentrating mostly in the restricted cohomological sector. We conclude with section \ref{sec:conclusions} and gather several technical details in the appendices.

%% file: sections/2_superspace.tex

\section{Correlation functions in superspace}
\label{sec:superspace}

Let us start by introducing the superspace we will use to describe correlation functions 
in a $d=4$, $\Nm=4$ superconformal theory in the presence of a flat  $\frac{1}{2}$-BPS defect. 
It is particularly useful for correlators of $\tfrac{1}{2}$-BPS operators and will allow us to write the Ward identities in a very compact form. Similar superspaces have already been used in the literature to study $\tfrac{1}{2}$-BPS correlators in different superconformal setups \cite{Dolan:2004mu,Doobary:2015gia,Liendo:2015cgi}. 

\subsection{Superspace setup}
\label{sec:superspacesetup}

The $\mathcal{N}=4$ superconformal group in four dimensions is $PSL(4|4)$\footnote{For the  discussion in this section 
 groups and coordinates are complexified.}.   
The four-dimensional Minkowski space can be extended to a superspace with coordinates
 \be \label{Xdef}
X= \left(X^{\mathsf{A}\dot{\mathsf{A}}}\right)\,=\,
\begin{pmatrix}
 x^{\alpha\dot{\alpha}} & \lambda^{\alpha\dot{a}}\\
\pi^{a\dot{\alpha}} & y^{a\dot{a}}
\end{pmatrix}\,,
\ee
where $\alpha\in\{1,2\}$, $\dot{\alpha}\in\{\dot{1},\dot{2}\}$,
$a\in\{1,2\}$, $\dot{a}\in\{\dot{1},\dot{2}\}$.
The R-symmetry coordinate $y$  can be considered as parameterizing 
a second copy of Minkowski space.
 The remaining variables $\lambda,\pi$ are fermionic.
The action of $GL(4|4)$ is given by
\be\label{GL44action}
g\circ X\,=\,
\left(A\,X+B\right)
\left(C\,X+D\right)^{-1}\,,
\qquad
g\,=\,\begin{pmatrix}
A & B \\
C & D
\end{pmatrix}\,\in\,GL(4|4)\,.
\ee
Due to the projective nature of these transformations, only $PGL(4|4)$ acts non-trivially on this set of coordinates.
The symmetry group  $PSL(4|4)$ corresponds to the elements in  $PGL(4|4)$ with unit superdeterminant.
The group element 
\be\label{gpsidef}
g_{\psi}\,:=\,
\begin{pmatrix}A_{\psi}&0\\0&A_{\psi}\end{pmatrix}
\in
 PGL(4|4)\,,\qquad
 A_{\psi}=\begin{pmatrix}\psi^{+1}1_2&0\\0&\psi^{-1}1_2\end{pmatrix}\,,
 \ee
  generates an outer automorphism of  $PSL(4|4)$.
This superspace goes under a variety of names (analytic/projective) and is particularly useful to describe correlation functions.

We will now turn  to the discussion of the symmetry group preserved by the presence of a $\frac{1}{2}$-BPS
 boundary, namely, the three-dimensional $\mathcal{N}=4$ superconformal group $OSP(4|4)$.

\paragraph{The supergroup $OSP(4|4)$.}
We define the  orthosymplectic group  as 
\be\label{OSPgroupdef}
OSP(4|4)\,=\,
\Big{\{}
g\,\in\,GL(4|4)\,\,
\text{such that }\,
g^{st}\,\eta\, g\,=\,\eta
\Big{\}}\,.
\ee
In the equation above $st$ denotes super-transposition (an operation with the properties 
$(AB)^{st}=B^{st} A^{st}$ and $(A^{st})^{st}=\Pi\,A\,\Pi$ where $\Pi$ is the 
super-parity matrix which acts as $+1$ on bosons and $-1$ on fermions)  and $\eta$ is a supersymmetric matrix, 
i.e.~$\eta=\eta^{st}\Pi=\Pi\,\eta^{st}$.
We choose conventions as
\be
\label{eq:defPiandSigma}
\Pi\,=\,
\begin{pmatrix}
\Sigma& 0 \\
0  &\Sigma
\end{pmatrix}\,,
\qquad
\eta\,=\,
g_{\psi}\,
\begin{pmatrix}
0 & 1_4 \\
\Sigma  &0
\end{pmatrix}
\,g_{\psi}\,,
\ee
where
 $\Sigma=\left(\begin{smallmatrix}-1_2&0\\0&+1_2\end{smallmatrix}\right)$
 and $g_{\psi}$ is defined in \eqref{gpsidef}.
Notice that $\psi$ parametrizes inequivalent embeddings $OSP(4|4)\subset GL(4|4)$, 
see e.g.~\cite{Gaiotto:2008sa}.
The $U(1)_Y$ outer automorphism, 
which is not a symmetry of the $\mathcal{N}=4$ $d=4$ superconformal theory \cite{Intriligator:1998ig},
 changes the value of the embedding parameter.
Finally, 
we define super-transposition as $(A^{st})_{ij}:=(-1)^{(|i|+1)|j|}\,A_{ji}$, 
where $(-1)^{|i|}:=\Pi_{ii}$. Notice that this definition can be applied to square matrices as well as to rectangular ones. 

\paragraph{The superalgebra $\mathfrak{osp}(4|4)$.}
The even and odd parts of the superalgebra $\mathfrak{g}=\mathfrak{osp}(4|4)$ are given by
\be\label{algebra}
\mathfrak{g}_{\underline{0}}\,=\,\mathfrak{sp}(4)\oplus  \mathfrak{su}(2)_+\oplus  \mathfrak{su}(2)_-\,,
\qquad
\mathfrak{g}_{\underline{1}}\,=\,(\mathbf{4},\mathbf{2},\mathbf{2})\,,
\ee
where the latter notation indicates that the fermionic generators transform in the tri-fundamental representation
of the bosonic subalgebra $\mathfrak{g}_{\underline{0}}$.
This super-algebra possess a $\mathbb{Z}_2$ outer automorphism $\mathbf{M}$ 
that exchanges $ \mathfrak{su}(2)_+$ with $\mathfrak{su}(2)_-$ and the fermionic generators accordingly\footnote{The choice of  letter $\mathbf{M}$ 
is motivated by the fact  that this automorphism is related to mirror symmetry.}.

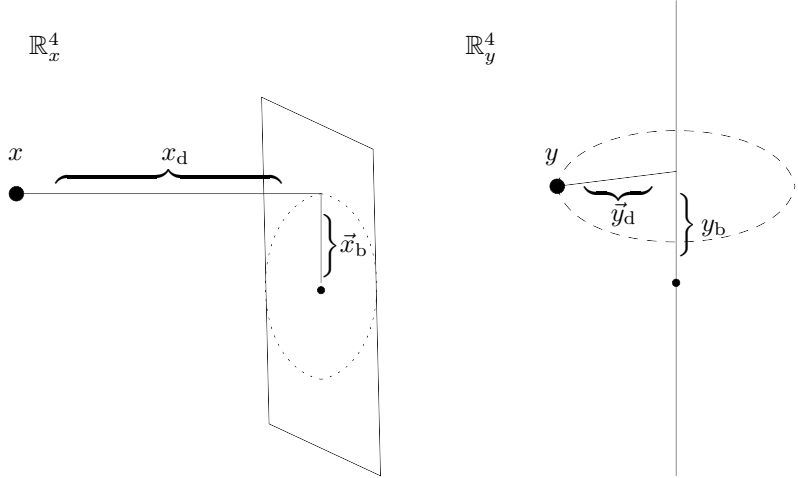
\begin{figure}[h!]
\centering
\includegraphics[scale=0.35]{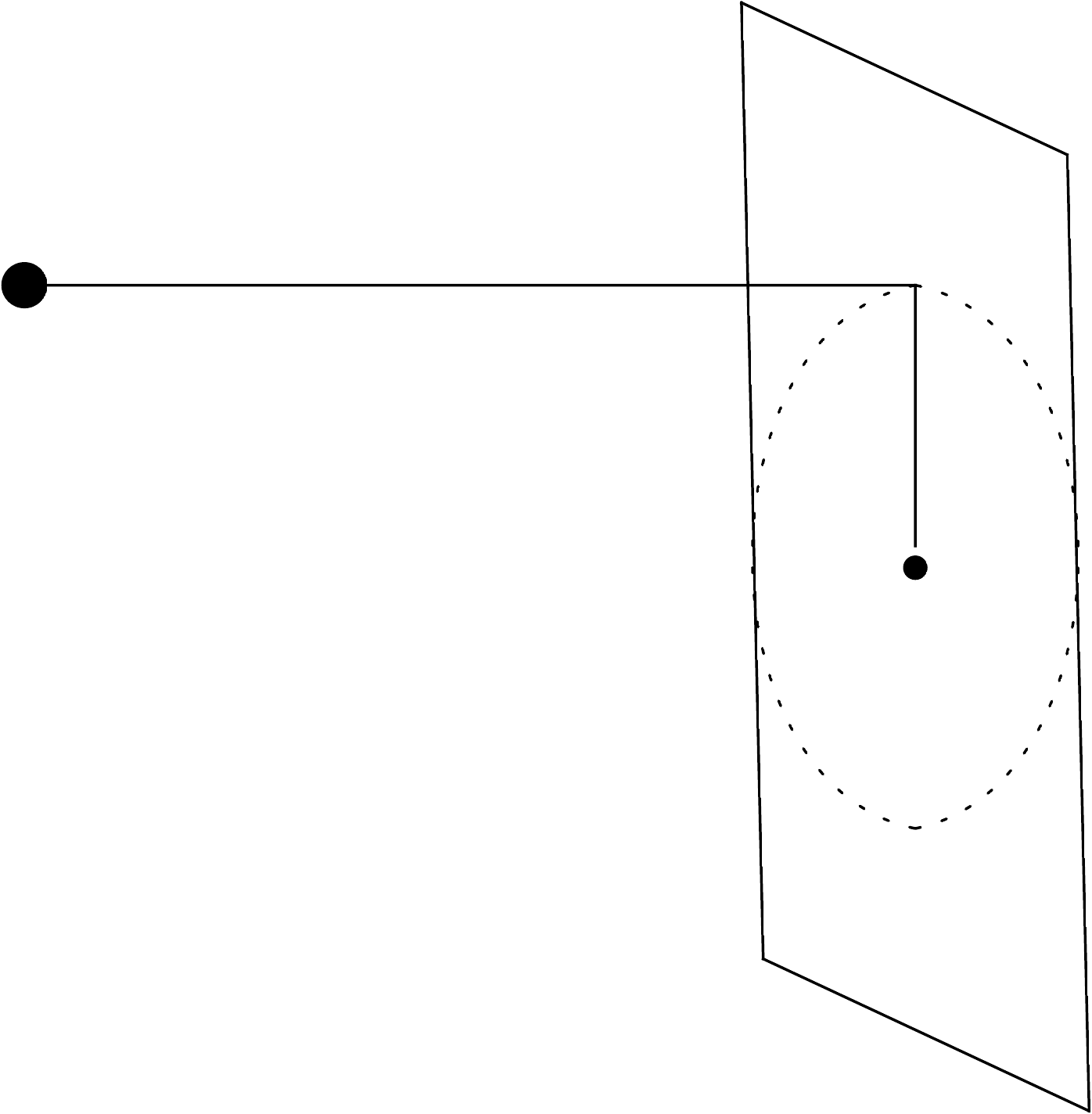}
\qquad
\qquad
\qquad
\includegraphics[scale=0.35]{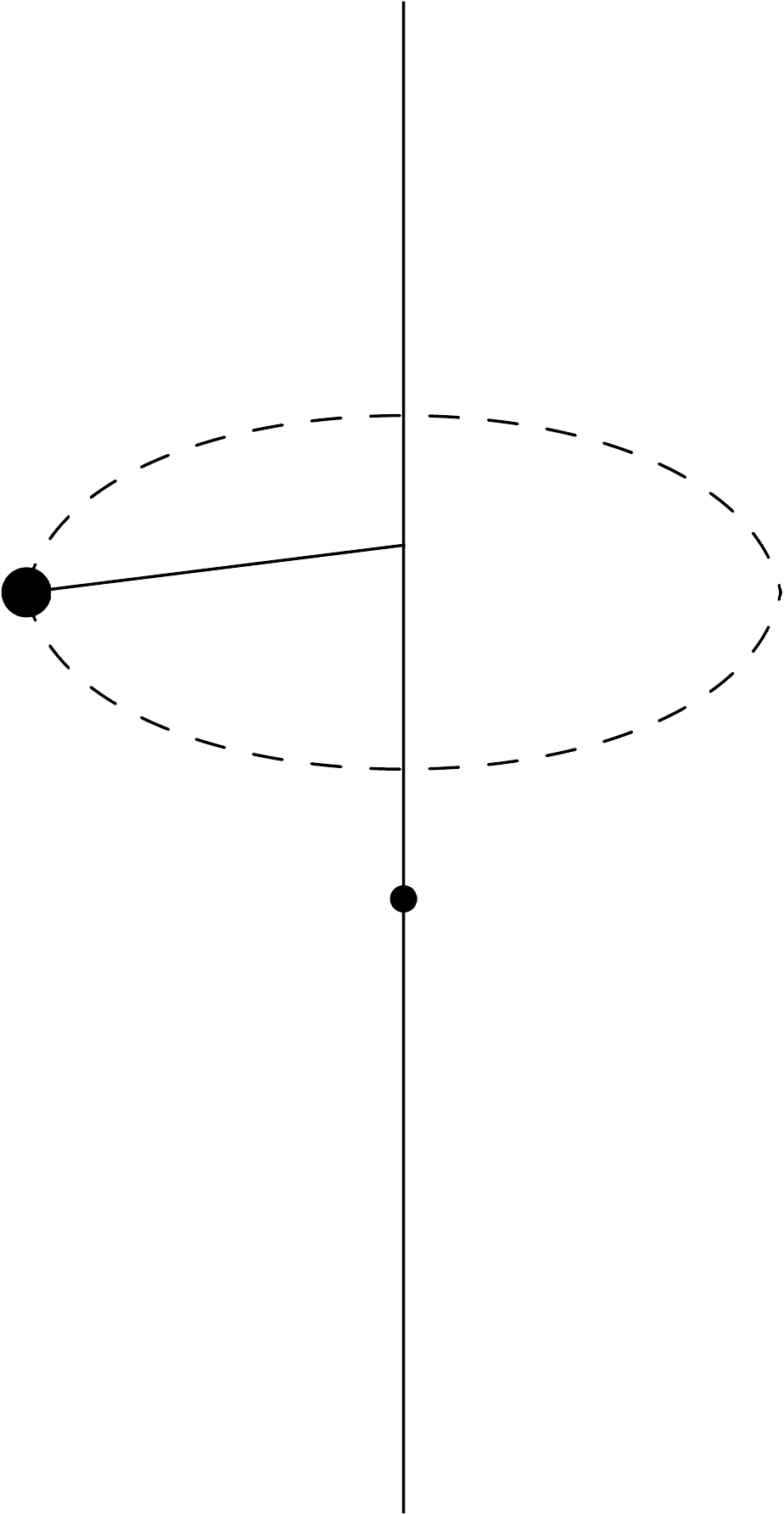}        
\put(-290,160){$\mathbb{R}^4_x$} 
\put(-240,120){$x_{\text{d}}$}  
\put(-298,120){$x$} 
\put(-280,110){$\overbrace{\hspace{3cm}} $}
\put(-180,85){$ \bigg{\}}\vec{x}_{\text{b}}$}
\put(-125,160){$\mathbb{R}^4_y$}
\put(-45,93){$ \bigg{\}}\,y_{\text{b}}$}
\put(-80,111){$\underbrace{\hspace{0.9cm}} $}
\put(-70,97){$ \vec{y}_{\text{d}}$}
\put(-95,120){$y$} 
\caption{On the left we have the configuration space for the spacetime coordinates and on the right for the R-symmetry coordinates.
The subgroup of the conformal group unbroken in the presence of the defect is clear from the picture.
On the left is the group of conformal transformations of the three-dimensional boundary $SO(3,2)$.
On the right is the product of rotations $SO(3)$ in the space orthogonal to the line with the 
$SL(2,\mathbb{R})$ conformal transformations on the line.}
\label{fig:spaces}
\end{figure}

\paragraph{Superspace coordinates and their transformations.}
The next step is to decompose the coordinate \eqref{Xdef} of the 
 four-dimensional superspace in boundary coordinates and distance coordinates as follows
\be\label{XbdfromX}
 X=A_{\psi}^{-1}\left(X_{\text{b}}+X_{\text{d}}\right)\,A_{\psi}^{}\,,
\,
\qquad
X_{\text{b}}^{st}\,=\,-\Sigma\,X_{\text{b}}^{}\,,
\qquad
X_{\text{d}}^{st}\,=\,+\Sigma\,X_{\text{d}}^{}\,.
\ee
where the matrix $A_{\psi}$ is defined in \eqref{gpsidef}
and the subscripts ${\text{b}}$ and ${\text{d}}$ stand for  boundary and distance respectively.
More explicitly
\be
\label{eq:coordexpl}
X^{}_{\text{b}}=\left(X_{\text{b}}^{\mathsf{A}\mathsf{B}}\right)\,=\,
\begin{pmatrix}
x_{\text{b}}^{\alpha\beta} & \,\theta^{a\beta} \\
\theta^{b\alpha} & \epsilon^{ab} y_{\text{b}}
\end{pmatrix}\,,
\qquad
X^{}_{\text{d}}=\left(X_{\text{d}}^{\mathsf{A}\mathsf{B}}\right)\,=\,
\begin{pmatrix}
\epsilon^{\alpha\beta}\,x_{\text{d}} & \,\chi^{a\beta} \\
-\chi^{b\alpha} & y_{\text{d}}^{ab}
\end{pmatrix}\,,
\ee
where  $ x_{\text{b}}^{\alpha\beta}= x_{\text{b}}^{\beta\alpha}$ and $ y_{\text{d}}^{ab}= y_{\text{d}}^{ba}$. 
When the fermions are zero, the geometric interpretation of these coordinates is given in figure \ref{fig:spaces}.

It should be noticed that  the parameter $\psi$, which parametrizes different embeddings of  $OSP(4|4)\subset PSL(4|4)$ affects only the relation between fermionic coordinates as
\be
\theta\,=\,\tfrac{1}{2}\left(\psi^{+2}\,\lambda\,+\,\psi^{-2}\,\pi\right)\,,
\qquad
\chi\,=\,\tfrac{1}{2}\left(\psi^{+2}\,\lambda\,-\,\psi^{-2}\,\pi\right)\,.
\ee
From now on we will set $\psi=1$.  The general case can be easily recovered using \eqref{XbdfromX}\footnote{
The factor $\psi$ is important if we want to extract component-correlators from a superfield correlator.
For example, it corresponds to a one parameter family of boundary conditions in the free $U(1)$ theory,
 see  \cite{Gaiotto:2008sd,Mikhaylov:2014aoa}.
}.
The transformation properties of these coordinates under $OSP(4|4)$ follow from \eqref{GL44action}, \eqref{OSPgroupdef} and  \eqref{XbdfromX}. 
More explicitly, 
super-translations and super-rotations  are given by 
\be\label{supertraandrot}
(X_{\text{b}},X_{\text{d}})\,\mapsto \,(X_{\text{b}}+B,X_{\text{d}})\, ,\qquad
(X_{\text{b}},X_{\text{d}})\,\mapsto \,(A\,X_{\text{b}}\,A^{st},A\,X_{\text{d}}\,A^{st})\,,
\ee
where  $B^{st}=-\Sigma\,B$ and $A\in GL(2|2)$.
Special superconformal transformations mix boundary and distance coordinates and act as follows
\be\label{specCONF}
(X_{\text{b}},X_{\text{d}})\,\mapsto \,K\,(X_{\text{b}}-X\,C\,\Sigma\,X^{st},X_{\text{d}})\,K^{st}\, ,
\qquad
K=(1+X\,C)^{-1}\, ,
\ee
where $C^{st}\,=\,-C\,\Sigma$.
It is worth remarking that on the boundary these transformations reduce to Special superconformal transformations of the three-dimensional boundary theory, namely
\be
(X_{\text{b}},0)\,\mapsto \,((1+X_{\text{b}}\,C)^{-1}\,X_{\text{b}},0)\,.
\ee
The element $\eta$ given in \eqref{eq:defPiandSigma} generates the superconformal inversion 
$\eta \circ X\,=\,A_{\psi}^2(X^{-1}\Sigma)A_{\psi}^{-2}$.

A  remark concerning the action of R-symmetry on the $y$ coordinates is in order.
The  R-symmetry preserved by the boundary conditions is $\mathfrak{o}(4)\simeq \mathfrak{su}(2)\oplus \mathfrak{su}(2)$.
One $\mathfrak{su}(2)$ acts linearly on the indices $a,b,\dots$
and will be denoted by $\mathfrak{su}(2)_+$, the second $\mathfrak{su}(2)$
 acts projectively on the R-symmetry boundary coordinate $y_{\text{b}}$
 and will be denoted by $\mathfrak{su}(2)_-$.
This observation can be better understood by noticing that the superspace setup just described can be applied to the case of a 
$\frac{1}{2}$-BPS \emph{line defect} in $\mathcal{N}=4$ as well.
The only difference is that we reinterpret the coordinate $y$ as spacetime coordinates and $x$ as R-symmetry coordinates, 
see figure \ref{fig:spaces}.
At the complexified level both the codimension one and codimension three cases preserve an $OSP(4|4)$ embedded in the bulk symmetry as described by \eqref{OSPgroupdef}.
At the level of real forms, in the codimension one case the symmetry is  
$OSP(4|4,\mathbb{R})\subset PSU(4|2,2)$,
while in the codimension three case  one has 
$OSP(4^*|4)\subset PSU(2,2|4)$.
The fact that a $\tfrac{1}{2}$-BPS Wilson line
in $\mathcal{N}=4$ SYM is symmetric under $OSP(4^*|4)$ was 
first observed in
 \cite{Bianchi:2002gz}, the $SP(4)\simeq SO(5)\subset SO(6)$ is the symmetry that survives after fixing 
 a non-null direction in the space of six scalars.
\paragraph{Superspace description of codimension two defects.}

Four dimensional $\mathcal{N}=4$ superconformal theories also admit codimension two superconformal defects.
Let us present how this set up can be realized in superspace in the case in which the defect preserves half of the supersymmetries.
As before one splits the superspace coordinate in coordinates on the surface and coordinates perpendicular to the surface
$X=X_{\text{S}}+X_{\perp}$ where 
\be\label{Xsurfacedefect}
X_{\text{S}}\,=\,
\mathpzc{x}{}_{\text{S}}\otimes
 \left(\begin{smallmatrix}1& 0 \\  0&0
\end{smallmatrix}\right)\,+\,
\bar{\mathpzc{x}}{}_{\text{S}}\otimes
 \left(\begin{smallmatrix}0& 0 \\  0&1
\end{smallmatrix}\right)\,,
\qquad
X_{\perp}\,=\,
\mathpzc{x}{}_{\perp}\otimes
 \left(\begin{smallmatrix}0& 1 \\  0&0
\end{smallmatrix}\right)\,+\,
\bar{\mathpzc{x}}{}_{\perp}\otimes
 \left(\begin{smallmatrix}0& 0 \\  1&0
\end{smallmatrix}\right)\,,
\ee
where $\mathpzc{x}{}_{\text{S}}$,
 $\bar{\mathpzc{x}}{}_{\text{S}}$, 
$ \mathpzc{x}{}_{\perp}$ and 
$\bar{\mathpzc{x}}{}_{\perp}$ are $(1|1)\times (1|1)$ supermatrices.
The subgroup of the conformal group preserved by  an $\frac{1}{2}$-BPS 
surface defect is $P(SU(2|2)\times SU(2|2))\subset PSU(2,2|4)$, see e.g.~\cite{Drukker:2008wr}, 
corresponding to elements of the form
\be\label{surfacedefectsubgroup}
\begin{pmatrix}
A & B \\
C & D
\end{pmatrix}\,=\,
 \left(\begin{smallmatrix}a& b \\  c&d
\end{smallmatrix}\right)\,\otimes\,
 \left(\begin{smallmatrix}1& 0 \\  0&0
\end{smallmatrix}\right)\,
+
 \left(\begin{smallmatrix}\bar{a}& \bar{b} \\ \bar{ c}&\bar{d}
\end{smallmatrix}\right)\,\otimes\,
 \left(\begin{smallmatrix}0& 0 \\  0&1
\end{smallmatrix}\right)\,,
\ee
with $\text{sdet} \left(\begin{smallmatrix}\bar{a}& \bar{b} \\ \bar{ c}&\bar{d}
\end{smallmatrix}\right)=
\text{sdet} \left(\begin{smallmatrix}a& b \\  c&d
\end{smallmatrix}\right) =1$. 
The projective condition corresponding to the letter ``$P$" is automatic 
in the representation  \eqref{GL44action}.
The action of the subgroup \eqref{surfacedefectsubgroup}
 on \eqref{Xsurfacedefect} follows from \eqref{GL44action}.
 For convenience we spell it out here. Super-translations and  super-rotation 
 respectively act as
 \begin{align}
\big( \mathpzc{x}{}_{\text{S}},\bar{\mathpzc{x}}{}_{\text{S}}
\big)&\mapsto
\big( \mathpzc{x}{}_{\text{S}}+b,\bar{\mathpzc{x}}{}_{\text{S}}
+\bar{b}
\big)\,,
\qquad \quad\,\,\,\,
\big( \mathpzc{x}{}_{\perp},\bar{\mathpzc{x}}{}_{\perp}
\big)\mapsto
\big( \mathpzc{x}{}_{\perp},\bar{\mathpzc{x}}{}_{\perp}
\big)\,,\\
\big( \mathpzc{x}{}_{\text{S}},\bar{\mathpzc{x}}{}_{\text{S}}
\big)&\mapsto
\big( a\,\mathpzc{x}{}_{\text{S}}\,d^{-1},\bar{a}\,\bar{\mathpzc{x}}{}_{\text{S}}\,\bar{d}^{-1}
\big)\,,
\qquad
\big( \mathpzc{x}{}_{\perp},\bar{\mathpzc{x}}{}_{\perp}
\big)\mapsto
\big( a\,\mathpzc{x}{}_{\perp}\,\bar{d}^{-1},\bar{a}\,\bar{\mathpzc{x}}{}_{\perp}\,d^{-1}
\big)\,.
 \end{align}
 Concerning special superconformal transformations, for $\bar{c}=0$
 they take the form
 \begin{align}
 \big( \mathpzc{x}{}_{\text{S}},\bar{\mathpzc{x}}{}_{\text{S}}
\big)&\mapsto
 \big( \mathpzc{x}{}_{\text{S}}(1+c\, \mathpzc{x}{}_{\text{S}})^{-1},
 \bar{\mathpzc{x}}{}_{\text{S}}
 -\bar{\mathpzc{x}}{}_{\perp}(1+c\, \mathpzc{x}{}_{\text{S}})^{-1}c\,\mathpzc{x}{}_{\perp}
\big)\,,\\
\big( \mathpzc{x}{}_{\perp},\bar{\mathpzc{x}}{}_{\perp}
\big)&\mapsto
\big( (1+\mathpzc{x}{}_{\text{S}}\,c)^{-1}\mathpzc{x}{}_{\perp},\bar{\mathpzc{x}}{}_{\perp}(1+c\, \mathpzc{x}{}_{\text{S}})^{-1}
\big)\,.
 \end{align}
 The transformations for $c=0$ are obtained from the above by changing barred and unbarred quantities.
 Notice that on the surface, where $ \mathpzc{x}{}_{\perp}=\bar{\mathpzc{x}}{}_{\perp}=0$,
the coordinates $ \mathpzc{x}{}_{\text{S}}$, $\bar{\mathpzc{x}}{}_{\text{S}}$ transform independently.
We remark that both boundary defect and surface defect preserve an $\mathfrak{su}(2)\otimes \mathfrak{su}(2)$
subgroup of the R-symmetry algebra $\mathfrak{su}(4)$, but with different embeddings.
In the surface case the embedding is specified by the decomposition of the fundamental representation 
$[1,0,0]\rightarrow (\tfrac{1}{2},0)\oplus  (0,\tfrac{1}{2})$, while in the boundary case
$[1,0,0]\rightarrow (\tfrac{1}{2},\tfrac{1}{2})$.
It follows that in the  case of a $\frac{1}{2}$-BPS surface defect
\be
[0,p,0]\rightarrow \bigoplus_{f=0}^p\,m_{p,f}\,(\tfrac{f}{2},\tfrac{f}{2})\,,
\qquad
m_{p,f}\,=\,p-f+1\,,
\ee
in particular the singlet appears $p+1$ times.

\subsection{Correlation functions}
\label{sec:correlation}

We now turn our attention to correlation functions in the presence 
of a $\frac{1}{2}$-BPS boundary or interface. In the first part of this section we will classify 
which bulk operators can have a non-zero one-point function,
and which boundary operators can have a non-zero two-point function with a $\frac{1}{2}$-BPS  bulk operator.
The second part of the section contains a derivation of the so-called superconformal Ward identities (WI) for
two-point functions of $\frac{1}{2}$-BPS bulk operators. These identities follow from the requirement that the correlation function is free of spurious singularities in the R-symmetry coordinates, a condition that goes under the name of superspace analyticity.
\paragraph{Supermultiplets/Superfields.}
Let us start by reviewing the relevant representations of superconformal algebras in three and four dimensions
and set up the notation.
A unitary highest weight representation $\chi_{\text{blk}}$ of $\mathfrak{psu}(2,2|4)$ is usually 
identified using the  Dynkin labels $\{\Delta,(\ell,\bar{\ell}),[q,p,r]\}$, where $\ell,\bar{\ell}$ and $[q,p,r]$ 
are spin and $\mathfrak{su}(4)$ labels respectively.
For the discussion of correlation functions it is  convenient to reorganize these labels into a representation 
$\mathcal{R}_{\text{blk}}$ of 
$\mathfrak{ps}(\mathfrak{u}(2|2)\oplus \mathfrak{u}(2|2))$, see \cite{Doobary:2015gia}.
Similarly, 
a unitary highest weight representation $\chi_{\text{bdy}}$ of $\mathfrak{osp}(4|4)$ can be 
identified using the  Dynkin labels $\{\delta,s,(k_+,k_-)\}$.
Where $\delta$ is the dilatation weight and we use conventions with  
$s,k_{\pm}\in\, \tfrac{1}{2}\mathbb{Z}_{\geq0}$.
For the discussion of correlation functions it is  convenient to reorganize these labels into a representation
 $\mathcal{R}_{\text{bdy}}$ of 
$\mathfrak{u}(2|2)$. Notice that the bosonic subalgebra of this $\mathfrak{u}(2|2)$ is $\mathfrak{su}(2)_+\oplus\mathfrak{su}(2)_{\text{Lorentz}}\oplus \mathfrak{u}(1)_-\oplus\mathfrak{u}(1)_{D}$,
where $ \mathfrak{u}(1)_-$ corresponds to the  Cartan generator of $\mathfrak{su}(2)_-$  and 
$\mathfrak{u}(1)_{D}$ corresponds to dilatations in spacetime.
The list of the unitary irreducible highest weight representations is given in Appendix \ref{App:supermultiplets}.

The $\frac{1}{2}$-BPS operators in the bulk have Dynkin labels
$\{p,(0,0),[0,p,0]\}$ and will be denoted by $\mathcal{W}_p(X)$. 
The $\frac{1}{2}$-BPS operators in the boundary come in two families $(B,+)_k$ and $(B,-)_k$
 with Dynkin labels given by $\{k,0,(0,k)\}$ and
$\{k,0,(k,0)\}$ respectively. Their superspace description is not symmetrical. The $(B,+)_k$ supermultiplets 
correspond to the one-dimensional representation of the $\mathfrak{u}(2|2)$ introduced above and are denoted by
$\widehat{\varphi_{+ ,k}}(X_{\text{b}})$. The  $(B,-)_k$ supermultiplets  
correspond to the rank $2k$ totally symmetric  representation of the same $\mathfrak{u}(2|2)$, they are denoted by
 $\widehat{\varphi_{-,k}}(X_{\text{b}},V)$, where $V^{\mathsf{A}}$ is an auxiliary variable with $V^{\alpha}$ fermionic and $V^{a}$ bosonic and   $\widehat{\varphi_{-,k}}(X_{\text{b}},\lambda V)= \lambda^{2k}\,\widehat{\varphi_{-,k}}(X_{\text{b}},V)$.
 
 The description of superfields just provided is actually oversimplified.
 What we did not specify are certain  analyticity constraints  on the superfields.
 These enforce the requirement that null states are actually zero
  (like the familiar $\partial^\mu\partial_\mu \varphi=0$ for a massless scalar) and 
 include the condition  that only finite dimensional  representation of R-symmetry, 
 which is non-linearly realized in our setup, can appear.
While such  constraints are known in various examples, see e.g.~\cite{Eden:2001ec},   
we feel that they have not been fully explored in the current literature.
A complete classification is left to the future.

\paragraph{Covariance properties of correlation functions.}
Correlation functions of super-primary fields are superconformally covariant in the following sense:
\begin{align}\label{cov}
\langle \mathcal{O}_1(X'_1)\dots  \mathcal{O}_n(X'_n)\,
\widehat{\mathcal{O}}_{n+1}(X'_{\text{b},n+1})\dots \,
 \widehat{\mathcal{O}}_{n+m}(X'_{\text{b},n+m)})
 \rangle\,=\,&\\
 \mathscr{M}(\underline{X},\underline{X}_{\text{b}},g)\,\,
 \langle \mathcal{O}_1(X_1)\dots  \mathcal{O}_n( X_n)\,
\widehat{\mathcal{O}}_{n+1}( X_{\text{b},n+1})\dots\, &
 \widehat{\mathcal{O}}_{n+m}(X_{\text{b},n+m)})
 \rangle\, ,
\end{align}
for any $g\,\in\,OSP(4|4)$. In the equation above $X'=g\circ X$ and the conformal factor is given by
\be
 \mathscr{M}(\underline{X},\underline{X}_{\text{b}},g)\,=\,
\Bigg(\prod_{i=1}^n \mathcal{R}_{\text{blk},i}(\Omega_{g,X_i})\Bigg)
\Bigg(\prod_{j=n+1}^{n+m} \mathcal{R}_{\text{bdy},j}(\omega_{g,X_{\text{b},i}})\Bigg)\,,
\ee
where $\Omega_{g,X}\in\,PS(GL(2|2)\times GL(2|2))$ and $\omega_{g,{X_{\text{b}}}}\in GL(2|2)$ are given in appendix \ref{App:conformalfactors}. Notice that while $\mathcal{R}_{\text{blk},i}$ is defined for any $\Omega_{g,X}$, the restriction to 
$g\,\in\,OSP(4|4)$ identifies a $GL(2|2)\subset PS(GL(2|2)\times GL(2|2))$, see \eqref{Omegafactors}.

\paragraph{One-point functions of bulk operators.}
Since the full $\mathcal{N}=4$ superconformal symmetry of the four-dimensional theory is reduced to 
 $OSP(4|4)\subset PSL(4|4)$ by the presence of the boundary, certain bulk 
operators can have non-vanishing one-point function.
This follows form the covariance property \eqref{cov} that we will now analyze.
Due to super-translation invariance \eqref{supertraandrot} 
the one-point function can depend only on the distance coordinate 
$X_{\text{d}}$, so we introduce the notation  $G_{\mathcal{O}}(X_{\text{d}}):=\langle \mathcal{O}(X)\rangle$.
For given $X$, one can restrict attention to $g\in\,OSP(4|4)$ that leave $X$ invariant, i.e.~such that
 $X'=g\,\circ \,X=X$. We call this stability group $S_X$. It follows from \eqref{cov} that
 \begin{align}\label{first_for_onept}
G_{\mathcal{O}}(X_{\text{d}})\,&=\,
 \mathcal{R}_{\text{blk}}(\Omega_{g,X_{\text{d}}})\,
 G_{\mathcal{O}}(X_{\text{d}})\,,
  \qquad\,\,\,\,\,\,\,\,\,\,\,\,
  \forall\,\, g\,\in\,S_{X_{\text{d}}}\,,\\
  \label{second_for_onept}
G_{\mathcal{O}}(AX_{\text{d}}A^{st})\,&=\,
 \mathcal{R}_{\text{blk}}(
 A,(A^{st})^{-1}%
)\,
G_{\mathcal{O}}(X_{\text{d}})\,,
 \qquad
  \, A=
   \left(\begin{smallmatrix}\alpha^{}\, 1_2
&0\\0&\alpha^{-1} 1_2\end{smallmatrix}\right)\,\in\,GL(2|2)\,.
 \end{align}
In the second line we added the covariance properties with respect to the relevant
  $GL(1)\subset GL(2|2)$.
In general the two conditions \eqref{first_for_onept} and \eqref{second_for_onept} should be supplemented by the requirement of superspace analyticity\footnote{
An illustrative example is the case of the three-point function of $\frac{1}{2}$-BPS operators. The object
\be\label{3ptfootnote}
\langle \mathcal{W}_{p_1}(X_1)\mathcal{W}_{p_2}(X_2)\mathcal{W}_{p_3}(X_3)\rangle_0\,\propto\,
\frac{1}
{\text{sdet}(X_{12})^{\frac{p_1+p_2-p_3}{2}}\text{sdet}(X_{13})^{\frac{p_1+p_3-p_2}{2}}\text{sdet}(X_{23})^{\frac{p_2+p_3-p_1}{2}}}\,,
\ee
where $\langle\dots \rangle_0$ means that there is no defect, is superconformally covariant for any $p_1,p_2,p_3$. The requirement of
analyticity gives the condition
 \begin{equation}
p_1\,\in\,\{|p_2-p_3|,|p_3-p_3|+2,\dots,p_2+p_3\}\,.
\end{equation}
} discussed below.
It is not hard to verify that the $S_X$ is isomorphic to $OSP(2|2)\times OSP(2|2)$, see appendix \ref{App:stabilityalgebra}.
The first condition \eqref{first_for_onept}  implies that for the one-point function to be non-vanishing
the representation $\mathcal{R}_{\text{blk}}$ has to contain a singlet with respect to the decomposition
$\mathfrak{osp}(2|2)\oplus \mathfrak{osp}(2|2)\subset \mathfrak{ps}(\mathfrak{u}(2|2)\oplus\mathfrak{u}(2|2))$.
Such representations are classified in appendix \ref{App:branching}. The second condition \eqref{second_for_onept} then fixes the one-point function uniquely up to a multiplicative constant.
Let us illustrate this point in the simple example of the one-point function of a $\frac{1}{2}$-BPS operator $\mathcal{W}_p(X)$. In this case the representation $\mathcal{R}_{\text{blk}}$ is trivial and the condition  \eqref{first_for_onept}  is automatically satisfied. The second condition implies that 
\be
\langle \mathcal{W}_p(X)\rangle\,=\,
\delta_{p,\,\text{even}}\,
\frac{\mathsf{a}_p}{\left(\text{sPf}\,X_{\text{d}}\right)^{p}}\,,
\ee
where the super-Pfaffian is defined in \eqref{sPfdef} and the constant 
$\mathsf{a}_p$, which is left undetermined by conformal symmetry, encodes dynamical information.
Notice that the condition that $p$ is even comes from the requirement of superspace analyticity. 
The case of a generic bulk operator is similar, but we will not give nor use the explicit expression for their one-point function here.
Using the criterion above together with the requirement of superspace analiticity we arrive at the following 

\vspace{0.5cm}
\noindent
\emph{Classification:}
the only operators, apart from the identity, that can have non-zero one-point function in the presence of an $OSP(4|4)$ symmetric codimension one defect 
 are
 \begin{itemize}
 \item 
 $\frac{1}{2}$-BPS  representations $\mathcal{B}_{[0,2n,0]}$ with $n\,\in\,\mathbb{Z}_{>0}$, 
 \item 
 $\frac{1}{4}$-BPS  representations $\mathcal{B}_{[2 m,2n,2m]}$  with $m,n\,\in\,\mathbb{Z}_{>0}$, 
 \item
 Long representations $\mathcal{A}^{\Delta}_{[2m,2n,2\bar{m}],(0,0)}$  with $\bar{m},m,n \,\in\,\mathbb{Z}_{>0}$.
 \end{itemize} 
Notice that  the one-point function of a bulk operator is non-zero if and only if  the top component of the supermultiplet, obtained by setting all the Grassmann variables to zero, has a non-zero one point function.
The latter can be studied using 
the bosonic symmety, i.e.~three-dimensional  conformal and R-symmetry.
The first symmetry implies that only  scalar operators, i.e.~representations $\chi_{\text{blk}}$
with $\ell=\bar{\ell}=0$ can have non-zero one-point function,  the second  that only representations $[q,p,r]$ with $q,p,r$ all even can have non-zero one-point function. This follows by looking at the decomposition of $[p,q,r]$ in 
representations of $\mathfrak{o}_R(4)$. This condition is not sufficient, as can be seen in the case of the so-called semi-short multiplets $\mathcal{C}$, see appendix \ref{App:supermultiplets}.

It is gratifying  that the same classification can be obtained independently by solving the superconformal Ward identities for bulk superconformal blocks. This is done in section \ref{sec:superblocks}.

\paragraph{Bulk-boundary two-point functions.}
While the one-point function of a bulk operator in the presence of a defect can be considered as the analogue 
of the two-point function without defect,
the two-point function of one bulk and one boundary operator in the presence of a defect is the analogue of the 
three-point function in the CFT without defect.  Generically, it is a linear combination of a finite number of superconformal structures.
Given a bulk point $X_1$ and a point on the boundary $X_{2,\text{b}}$, it is convenient to construct the combination
\be\label{Xtilde1def}
X_{1,\widehat{2}}\,:=\,
\Sigma\,
\left(X_{1,\text{d}}+X_{12,\text{b}}\right)^{st}X_{1,\text{d}}^{-1}\left(X_{1,\text{d}}+X_{12,\text{b}}\right)\,,
\qquad
X_{1,\widehat{2}}^{st}\,=\,+\Sigma\,X_{1,\widehat{2}}^{}\,.
\ee
This combination is like a distance coordinate, see \eqref{XbdfromX}, and 
transforms covariantly with respect with the boundary point  $X_{2,\text{b}}$. In particular
$X_{1,\widehat{2}}\mapsto (1+X_{2,\text{b}}C)^{-1}X_{1,\widehat{2}}(1+CX_{2,\text{b}})^{-1}$
under special superconformal transformations. 
One can  perform a super-translation to set $X_{2,\text{b}}=0$
followed by a special superconformal transformation, see \eqref{specCONF},  to set 
 $(X_{1,\text{b}},X_{1,\text{d}})=(0,X_{1,\widehat{2}})$.
  In this frame the definition \eqref{Xtilde1def} reduces to an identity.
Let
\be\label{GOOdef}
G_{\mathcal{O}\widehat{O}}(X_{1,\widehat{2}})\,:=\,
\langle
\mathcal{O}(X_{1,\widehat{2}})\widehat{O}(0)
\rangle=\mathcal{R}_{\text{blk},1}(\Omega^*)
\langle
\mathcal{O}(X_{1})\widehat{O}(X_{2,\text{b}})
\rangle\,,
\ee
where $\Omega^*=\left(1-X^{}_{12,\text{b}}X^{-1}_{1,\text{d}},1-X^{-1}_{1,\text{d}}X^{}_{12,\text{b}}\right)$.
The last equality, which follows from \eqref{cov},
allows to reconstruct in a simple way the bulk-boundary two-point function from $G_{\mathcal{O}\widehat{O}}$.
We can now write equations analogous to 
\eqref{first_for_onept} and \eqref{second_for_onept} in the case of
 one bulk and one boundary operator:
 \begin{align}\label{first_for_onept}
G_{\mathcal{O},\widehat{O}}(X_{1,\widehat{2}})\,&=\,
 \mathcal{R}_{\text{blk},1}(\Omega_{g,X_{1,\widehat{2}}})\,
 \mathcal{R}_{\text{bdy},2}(\omega_{g,0})\,
 G_{\mathcal{O},\widehat{O}}(X_{1,\widehat{2}})\,,
  \qquad\,\,\,\,\,\,\,\,\,\,\,\,
  \forall\,\, g\,\in\,S_{X_{1,\widehat{2}},0}\,,\\
  \label{second_for_onept}
G_{\mathcal{O}}(AX_{1,\widehat{2}}A^{st})\,&=\,
 \mathcal{R}_{\text{blk},1}(
 A,(A^{st})^{-1}%
)\,
\mathcal{R}_{\text{bdy},2}(A)\,
G_{\mathcal{O},\widehat{O}}(X_{1,\widehat{2}})\,,
 \qquad
  \, A=  \left(\begin{smallmatrix}\alpha^{}\, 1_2
&0\\0&\alpha^{-1} 1_2\end{smallmatrix}\right)
 \,\in\,GL(2|2)\,.
 \end{align}
 In the equation above $S_{X_1,X_{2,\text{b}}}$ denotes the stability group of one bulk point and one boundary point, specialized to  $S_{X_{1,\widehat{2}},0}$ by the choice of frame \eqref{GOOdef}.
This stability group is the subgroup of the stability group of one bulk point 
$S_{X=X_{1,\widehat{2}}}\simeq OSP(2|2)\times OSP(2|2)$ 
that leaves the second boundary point $X_{2,\text{b}}=0$ fixed.
It is not hard to see that this is the diagonally embedded $OSP(2|2)$. 
In summary, the elements of $S_{X_{1,\widehat{2}},0}$ are
 super-rotations, see \eqref{supertraandrot}, such that 
 $A\,X_{1,\widehat{2}}\,A^{st}=X_{1,\widehat{2}}$.
We conclude that superconformal structures for the two-point function \eqref{GOOdef}  
corresponds to $\mathfrak{osp}(2|2)$ invariant states in the triple tensor product  of the  
$\mathfrak{sl}(2|2)$ representations
 $\mathcal{R}^{\text{left}}_{\text{blk},1}$,  $\mathcal{R}^{\text{right}}_{\text{blk},1}$
, $\widetilde{\mathcal{R}}_{\text{bdy},2}$, see tables 
 \ref{tab:nonlongrepsosp4NEW} and \ref{tab:repsPSU224NEW},
  regarded as reducible
$\mathfrak{osp}(2|2)$ representations.
Among these, the superconformal structures that are not superspace analytic should be discarded as an allowed
bulk-boundary two-point function.

The $\frac{1}{2}$-BPS bulk operators correspond to  $\mathcal{R}^{\text{left/right}}_{\text{blk}}$
being the trivial representation. Using the result of appendix \ref{App:branching}
and by 
looking at table   \ref{tab:nonlongrepsosp4NEW} 
one immediately concludes that the only boundary operators that  can appear in the boundary OPE of 
a $\frac{1}{2}$-BPS bulk operator
are the ones given in  \eqref{changeofnames1}, \eqref{changeofnames2}.
The range of the R-symmetry labels $(k_+,k_-)$ is dictated by superspace analyticity.
We conclude that

\vspace{0.5cm}
\noindent
\emph{Classification:}
the only boundary operators $\widehat{\mathcal{O}}$, apart from the identity, that can have a non-zero two point function with a 
$\frac{1}{2}$-BPS bulk  operator in the presence of a $OSP(4|4)$ symmetric codimension one defect are 
\begin{itemize}
\item $\frac{1}{2}$-BPS operators $(B,\pm)_{k}$, $\,\,k\in \mathbb{Z}_{>0}$.
\item $\frac{1}{4}$-BPS operators $(B,1)_{(k_+,k_-)}$, $\,\,k_\pm\in \mathbb{Z}_{>0}$, $\,\,k_+k_-\neq 0$.
\item Long operators $L^{\delta}_{(k_+,k_-)}$, $\,\,k_\pm\in \mathbb{Z}_{>0}$.
\end{itemize}
Where we used the notation \eqref{changeofnames1}, \eqref{changeofnames2}.
The precise range of representation labels for the operators appearing in the bulk-boundary OPE  of a given 
 $\frac{1}{2}$-BPS bulk  operator $\mathcal{W}_p(X)$ can in principle be derived by imposing the requirement of harmonic analyticity. 
Let us illustrate the derivation in the simplest case of $\frac{1}{2}$-BPS boundary operators. 
The general case will be analyzed in a different way using the superconformal block expansion 
of the two-point function of $\frac{1}{2}$-BPS bulk operators in section \ref{sec:superblocks}, 
see equations \eqref{bdyOPEsummary}, \eqref{OPEbulksummary}.
The explicit form of the simplest  bulk-boundary two-point function is
\be\label{12BPSbulkbdy}
\langle \mathcal{W}_p(X_1)\,\widehat{\varphi_{+,k}}(X_{\text{b},2})\rangle\,=\,
\frac{\mu^{+}_{p,k}}{\text{sPf}(X_{1,\text{d}})^{p}\,\text{sPf}(X_{1,\widehat{2}})^{k}}\,=\,
\frac{\mu^{+}_{p,k}}{\text{sDet}(X_{12,\text{b}}+X_{1,\text{d}})^{k}\,\text{sPf}(X_{1,\text{d}})^{p-k}}
\,.
\ee
The middle expression in \eqref{12BPSbulkbdy} should be compared to \eqref{GOOdef},
the second equality follows from the definition \eqref{Xtilde1def}.
The constants $\mu^{+}_{p,k}$ are not fixed by superconformal symmetry and encode dynamical information
about the boundary conditions.
Analyticity in the  R-symmetry coordinates implies that
$p-k \,\in\,2\,\mathbb{Z}_{\geq 0}$.
As already discussed, the choice of superspace is not symmetric in the descriptions 
of $\frac{1}{2}$-BPS boundary operators $(B,+)$ and $(B,-)$. In the latter case the relevant two point function is given by 
\be
\langle \mathcal{W}_p(X_1)\,\widehat{\varphi_{-,k}}(X_{\text{b},2},V)\rangle\,=\,
\frac{\mu^{-}_{p,k}}{\text{sPf}(X_{1,\text{d}})^{p}}
\left(V\,X^{-1}_{1,\widehat{2}}V\right)^k
\,.
\ee
%
We will not consider more general bulk-boundary two-point functions 
in this paper as they are not directly relevant for the study of the bootstrap equations
for $\frac{1}{2}$-BPS operators.
 Nevertheless we stress that the criterion given above works in general.
 
\subsubsection{Correlation functions in the presence of a $\frac{1}{2}$-BPS line defect and other examples}

\paragraph{$\frac{1}{2}$-BPS line defect.}
As already pointed out in section \ref{sec:superspacesetup},
 and summarized in figure \ref{fig:spaces}, the superspace setup we introduced can be 
 directly applied to the study of correlation functions in the presence of a $\frac{1}{2}$-BPS line defect.
 The only modification is that one should exchange the spacetime coordinates $x$ with the R-symmetry coordinates $y$
 and change the analyticity conditions accordingly. 
 In this case,
  an $OSP(4^*|4)$ superconformal quantum mechanics lives on the defect\footnote{In principle, 
 there are many superalgebras that can serve as  symmetries of superconformal quantum mechanics, 
 see table 6 in  \cite{VanProeyen:1999ni}.
 The one living on $\frac{1}{2}$-BPS line defects of a four-dimensional $\mathcal{N}=4$ theory of course belongs to this list.}.
 
 Repeating the analysis of section \ref{sec:correlation},
  one concludes that the only bulk operators that can have non-zero one-point functions
   in the presence of a  $\frac{1}{2}$-BPS line defect are $\frac{1}{2}$-BPS, $\frac{1}{4}$-BPS
    and  long multiplets of the type
$\mathcal{B}_{[0,n,0]}$, 
$\mathcal{C}_{[0,n,0],(\ell,\ell)}$ and 
$\mathcal{A}^{\Delta}_{[0,n,0],(\ell,\bar{\ell})}$.
See \eqref{ABCd4} for the notation. 
This classification corresponds to the supermultiplets of $\mathfrak{psu}(2,2|4)$ 
for which the inducing  representations
 $\widetilde{\mathcal{R}}_{\text{blk}}^{\text{left}}$ and 
 $\widetilde{\mathcal{R}}_{\text{blk}}^{\text{right}}$ 
 are either atypical representations whose Young diagram
  is a single row of even length, or long representations of the form $[0,2n]_{\gamma}$.
 This follows from the analysis in Appendix \ref{App:branching} after a rotation by $90$ degrees.
  
  In order to discuss the classification of boundary operators that can have a non-vanishing two-point function 
  with a $\frac{1}{2}$-BPS bulk operator, it is necessary to first discuss some representation theory of
    $OSP(4^*|4)$ superconformal quantum mechanics. The relevant representations are reviewed in appendix
    \ref{App:supermultiplets}. The local operators living on the line defect that
    can have a non-vanishing two-point function with a $\frac{1}{2}$-BPS 
    bulk operator are listed in \eqref{unitaryrepsSCQM}.

  \paragraph{Correlators in $d=3$, $\mathcal{N}=4$ superconformal theories.}  
These considerations are relevant for the discussion in the next section.
Using the same ideas as in section \ref{sec:correlation},
one can determine  
the structure of boundary three-point functions. 
The stability group of three points on the boundary is $OSP(2|2)$ where
$O(2)\subset OSP(2|2)$ is a subgroup of the conformal group.
It follows that superconformal structures for three boundary operators
 are in one-to-one correspondence with $OSP(2|2)$ invariant states in the triple tensor product 
 $\widetilde{\mathcal{R}}_{\text{bdy},1}\otimes \widetilde{\mathcal{R}}_{\text{bdy},2}\otimes \widetilde{\mathcal{R}}_{\text{bdy},3}$.
It follows from this observation, consulting table \ref{tab:nonlongrepsosp4NEW}, that the 
  only operators that can appear in the OPE of $(B,+)_{k_1}$ with $(B,+)_{k_2}$ are 
   $\frac{1}{2}$-BPS, $\frac{1}{4}$-BPS
  and  long multiplets respectively
  of the type 
   $(B,+)_k$,  $(A,+)^{s}_k$ and 
   $L[2s]_{\delta}^{(2k;0)}$.
  If $k_1=k_2$ also higher-spin conserved currents of even spin $A_1[2s]_{s+1}^{(0;0)}$, where the lowest value $s=0$ corresponds
  to the stress-tensor supermultiplet, can appear.
  This OPE will be further discussed in section \ref{sec:3dN=41dQM}. 
  More general three-point functions can be determined in a similar way.
 
  \paragraph{Codimension two defect.}  
  In this case the one-point functions of bulk operators contain more than one covariant structure.
  Even in the simplest example of $\frac{1}{2}$-BPS operators one has
  \be\label{1ptsurface}
  \langle\mathcal{W}_{p}(X)\rangle_{\text{S}}\,=\,
  \sum_{n=0}^p\,
  \frac{\mathsf{a_{p,n}^{(\text{S})}}\,\,\,}{
  \left(\text{sdet}\,\mathpzc{x}{}_{\perp}\right)^n
  \left(\text{sdet}\,\bar{\mathpzc{x}}{}_{\perp}\right)^{p-n}
  }\,,
  \ee
  where the perpendicular supercoordinates are defined in 
\eqref{Xsurfacedefect}. 
The selection rule that determines which supermultiplets can have a non-zero one-point function is
found by looking at the stability group for one bulk point, which in this case is $SU(2|2)$. 
The class of operators that can have a non-vanishing one-point function is thus much larger than 
in the codimension one and three cases.

\subsubsection{The super-displacement operator}

In any defect CFT there is a distinguished boundary operator known as the displacement operator.
We will denote it by $\widehat{D}(x_{\text{b}})$.
It is associated to the breaking of the translation symmetry in the direction perpendicular to the defect.
In the case of a CFT in $d$ dimensions with a $d-1$ dimensional defect, it is a scalar of dimension 
$\delta=d$.
The goal of this section is to determine which three-dimensional $\mathcal{N}=4$
superconformal multiplet contains the displacement operator.
The main condition, apart from the fact that the supermultiplet should contain a conformal primary with the correct
quantum numbers, is that it preserves supersymmetry. This is translated into the requirement that
\be\label{Qdisplacement}
Q_{\text{3d}}\, \widehat{D}(x_{\text{b}})\,=\,\frac{\partial}{\partial  x_{\text{b}}}(\dots)\,,
\qquad
\forall\,\,\,\text{d=3, $\mathcal{N}=4$ Poincar\'e supersymmetry $Q_{\text{3d}}$}\,,
\ee
in other words, $Q_{\text{3d}}\, \widehat{D}$ is a conformal descendant.
This condition can be understood by recalling that
\be\label{displacementinsertion}
\int d^3x_{\text{b}}\,\langle  \widehat{D}(x_{\text{b}})\mathcal{O}_1(x_1)\dots \mathcal{O}_n(x_n)\rangle\,=\,
-\sum_{k=1}^n\,\frac{\partial}{\partial x_{k,\text{d}}}\,\langle\mathcal{O}_1(x_1)\dots \mathcal{O}_n(x_n)\rangle\,.
\ee
The condition 
\eqref{Qdisplacement} is strong enough to leave only two possible super-displacement operators.
They correspond to the $\frac{1}{2}$-BPS $\mathfrak{osp}(4|4)$ supermultiplets $(B,\pm)_2$.
By comparison recall that flavor currents sit in  $(B,\pm)_1$ supermultiplets. 
The structure of the $(B,+)_2$ supermultiplet can be found e.g.~in equation (2.5) of \cite{Beem:2016cbd}
with $r=2$. The case of
$(B,-)_2$ is obtained by applying the mirror automorphism
$\mathbf{M}$ defined below \eqref{algebra}.
The supermultiplets contain also other defect primaries
corresponding to the bulk conserved currents that are broken by the defect.
Recall that the bulk R-symmetry current, which is in the adjoint of $\mathfrak{su}(4)$, decomposes as
$(1,0)\oplus (0,1)\oplus (1,1)$  with respect to the relevant embedding of the defect R-symmetry $\mathfrak{su}(2)\oplus \mathfrak{su}(2)\subset \mathfrak{su}(4)$. 
The supermultiplets $(B,\pm)_2$ contain the conformal primary associated to the broken R-symmetry
which is a scalar of dimension $\delta=3$ and R-symmetry representation $(1,1)$.
Finally, there are operators associated to the breaking of supersymmetry with dimension $\delta=3+\tfrac{1}{2}$
spin $\tfrac{1}{2}$ and R-symmetry representation $(\tfrac{1}{2},\tfrac{1}{2})$.
As expected they are part of the $(B,+)_2$ and $(B,-)_2$ supermultiplets.

\subsection{Ward identities}

In the presence of a superconformal defect, the functional form of the two-point function of bulk operators is not fixed by 
the unbroken superconformal symmetry: there are certain combinations of coordinates that are invariant 
under its action. The first step is to determine such ``cross ratios".
The second step is to impose that the two-point function is free from certain superspace singularities. 
This requirement produces very powerful constraints known as superconformal Ward identities (WI).
While these identities are new in this specific context, identical identities have appeared before, 
see e.g.~\cite{Dolan:2004mu,Liendo:2015cgi}  .
This analogy suggests a deeper connection revealed in section \ref{sec:3dN=41dQM}.

\paragraph{Superconformal invariants made of two bulk points.}
The set of eigenvalues of the super-matrix
\be\label{Zdef}
\mathcal{Z}\,:=\,\Sigma\,
\left(X_{2,\text{d}}^{-1}\,X_{12}^{}\right)^{st}\,\Sigma\,
\left(X_{12}^{}\,X_{1,\text{d}}^{-1}\right)\,=\,(1-\mathcal{Y}^{+1})(\mathcal{Y}^{-1}-1)\,,
\ee
is $OSP(4|4)$ invariant. This statement can be verified using the transformation properties
 \eqref{supertraandrot} and \eqref{specCONF}.
The second equality in \eqref{Zdef} defines the matrix $\mathcal{Y}$  up to 
$\mathcal{Y}\,\mapsto \mathcal{Y}^{-1}$.
It turns out that the three distinguished eigenvalues\footnote{
The fact that the $4\times4$ matrix $\mathcal{Z}$ has only three distinguished eigenvalues can be 
verified by setting the fermionic coordinates to zero (this can be achieved with a superconformal transformation).
}
 of $\mathcal{Y}$ are convenient variables in which to
write the superconformal Ward identities.
A possible way to visualize the invariants above is to use 
 superconformal transformations\footnote{
Explicitly, this can be achieved by  performing a super-translation 
\eqref{supertraandrot} followed by a special superconformal transformations \eqref{specCONF} with 
$C\Sigma=\left[(X_1+B)^{-1}\right](X_{1,\text{b}}+B)\left[(X_1+B)^{-1}\right]^{st}=
\left[(X_2+B)^{-1}\right](X_{2,\text{b}}+B)\left[(X_2+B)^{-1}\right]^{st}$.
The fact that a graded-symmetric matrix $B$ that satisfies the last equality exists is not obvious but true.
}
  to  choose a frame in which
\be\label{frameforY}
(X_{1,\text{b}},X_{1,\text{d}})\,=\,(0,\widetilde{X}_{1,\text{d}})\, ,
\qquad
(X_{2,\text{b}},X_{2,\text{d}})\,=\,(0,\widetilde{X}_{2,\text{d}})\, ,
\ee
this frame is still acted upon by $GL(2|2)$.
The only invariants under this action are the eigenvalues of the matrix 
$\widetilde{X}_{2,\text{d}}^{}\widetilde{X}_{1,\text{d}}^{-1}$. It is not hard to verify that
\be\label{fromYtoeigs}
\mathcal{Y}\big{|}_{\text{frame \eqref{frameforY}}}
=\widetilde{X}_{2,d}^{}\widetilde{X}_{1,d}^{-1}\,\sim\, \text{diag}(z,z,w_1,w_2)\,.
\ee
The fact that $\mathcal{Y}$ is defined up to inversion is translated to the requirement of Bose symmetry 
when changing the first and the second point.
\paragraph{A useful $\mathbb{Z}_2$ action.}
It follows from the definition of $z,w_1,w_2$ 
that these variables are defined up to three involutions: 
\be\label{threeinvolutions}
(z,w_1,w_2)\mapsto (z,w_2,w_1)\,,
\quad
(z,w_1,w_2)\mapsto (z^{-1},w_1,w_2)\,,
\quad
(z,w_1,w_2)\mapsto (z,w_1^{-1},w_2^{-1})\,.
\ee
The first involution corresponds to the standard action of the symmetric group on the eigenvalues of a graded matrix.
The second and third follow from the fact that the matrix $\mathcal{Y}$ is defined by \eqref{Zdef} up to inversion\footnote{
More concretely, $z,w_1,w_2$ are defined by the three equations
\be
\text{Str}(\mathcal{Z}^n)\,=\,
((1-z)(z^{-1}-1))^n-\sum_{k=1}^2((1-w_k^{})(w_k^{-1}-1))^n\,,
\qquad n=1,2,3\,.
\ee
These equations are invariant under the three involutions \eqref{threeinvolutions}.}.
There is an extra $\mathbb{Z}_2$ that acts on these variables as 
\be\label{extraZ2}
(z,w_1,w_2)\mapsto (z,w_1,w_2^{-1})\,.
\ee

\paragraph{Superconformal Ward Identities.}
It follows from superconformal invariance and the previous discussion that 
\be\label{2ptfun}
\langle \mathcal{W}_{p_1}(X_1)\mathcal{W}_{p_2}(X_2)\rangle\,=\,
\frac{F_{p_1,p_2}(z,w_1,w_2)}{\left(\text{sPf}\,X_{\text{d},1}\right)^{p_1}\,
\left(\text{sPf}\,X_{\text{d},2}\right)^{p_2}}\,,
\ee
where  $F_{p_1,p_2}(z,w_1,w_2)=F_{p_1,p_2}(z,w_2,w_1)=
F_{p_1,p_2}(z^{-1},w_1^{},w_2^{})=F_{p_1,p_2}(z^{},w_1^{-1},w_2^{-1})$.
While \eqref{2ptfun} transforms covariantly  under superconformal transformations
for any function $F_{p_1,p_2}$,
 in general it does not possess the correct analyticity properties. 
 If one expands the eigenvalues of the supermatrix \eqref{Zdef} in Grassmann coordinates, one encounters 
 poles when the R-symmetry cross ratios are equal to the spacetime one.
 The vanishing of the residue at these spurious poles translates into the equations
\be\label{WI}
\left(\partial_{w_1}+\tfrac{1}{2}\partial_{z}\right)F_{p_1,p_2}(z,w_1,w_2)\big{|}_{w_1=z}\,=\,0\,,
\qquad
\left(\partial_{w_2}+\tfrac{1}{2}\partial_{z}\right)F_{p_1,p_2}(z,w_1,w_2)\big{|}_{w_2=z}\,=\,0\,.
\ee
See \cite{Dolan:2004mu,Liendo:2015cgi}  for more details. 
In particular, \eqref{WI} imply that $F_{p_1,p_2}(t,t,t)$ is a constant.
These equations will play a crucial role in the derivation of superconformal blocks in the next section.
It should be noticed that $F_{p_1,p_2}$ is different from zero only when $p_1$ and $p_2$ have the same parity
 (both even or both odd).
The structure of $F_{p_1,p_2}$ depends only on $M:=\text{min}(p_1,p_2)$ and is given by
\be\label{formFROMRstructure}
F_{p_1,p_2}(z,w_1,w_2)\,=\,
\sum_{a,b=0}^{M}\,\,w_1^{a-\tfrac{M}{2}}\,w_2^{b-\tfrac{M}{2}}\,A_{ab}(z)\,.
\ee
Invariance under \eqref{threeinvolutions} implies that $A_{ab}(z)=A_{ab}(1/z)$ and the various $A_{ab}(z)$ are not independent.
The number of independent functions depends on whether $M$ is even or odd. If we rewrite $M=2n+\varepsilon$ with 
$\varepsilon\in\{0,1\}$ the number of independent functions is $(n+1)(n+\varepsilon+1)$, which coincides with the number
 of R-symmetry channels as discussed in appendix \ref{app:Rchannells}. 
 It turns out that the general solution of the  Ward identities is given in terms of two constants and $n(n+\varepsilon)$ single variable functions.
  This is what we will now explain.
\paragraph{Solving the WI in two examples.}
Let us first see how the superconformal Ward identities can be solved explicitly 
in the two examples $(p_1,p_2)=(1,1)$ and $(p_1,p_2)=(2,2)$.
The general solution is discussed below. 
The goal is to solve the equations \eqref{WI} imposing the form \eqref{formFROMRstructure}.
It is not hard to find that in the case  $(p_1,p_2)=(1,1)$ the WI fix the two-point function uniquely up to two constants
\be\label{F11solveWI}
F_{1,1}(z,w_1,w_2)\,=\,
\mathsf{A}\,\Omega_++\mathsf{B}\,\Omega_-\,,
\qquad
\Omega_{\pm}:=\prod_{i=1}^2
\left(\frac{w^{+\frac{1}{2}}_i\mp w_i^{-\frac{1}{2}}}{z^{+\frac{1}{2}}\mp z^{-\frac{1}{2}}}\right)\,.
\ee
This is not surprising as in this case the external operators correspond to $\mathcal{N}=4$ free fields.
We will come back to a discussion of these constants in section \ref{sec:exampleboot:free}.
The solution in the $(p_1,p_2)=(2,2)$ is more interesting. 
It is not hard to show that the most general solution to the WI in this case is given by
\be\label{22factor}
  F_{2,2}(z,w_1,w_2)\,=\,\mathsf{C}_+\,+\,\kappa\,\mathsf{C}_-\,+\,\mathbb{D}\,H(z)\,,
\ee
where
\be\label{introduced}
  \kappa\,:=\,\frac{w^{}_1-w_1^{-1}}{z^{}-z^{-1}}
\frac{w^{}_2-w_2^{-1}}{z^{}-z^{-1}}\,,\qquad
 \mathbb{D}:=(g_1+g_2)+\frac{g_1g_2\,\,}{z-z^{-1}}\,z\partial_z\,,
  \qquad
 g_a:=\Big( 1-\frac{w_a}{z}\Big)\Big( z-\frac{1}{w_a}\Big)\,,
\ee
Notice that, given a solution of the WI, like a superconformal block, the coefficients $\mathsf{C}_{\pm}$
 and the function $H(z)$ can be extracted unambiguously.
  In this case,
all the dynamics is contained in the two constants $\mathsf{C}_{\pm}$
and the single-variable function $H(z)$.
It should be noticed that since the structure of $F_{p_1,p_2}$ depends only on $\text{min}(p_1,p_2)$,
the results \eqref{F11solveWI}, \eqref{22factor} can be applied to the cases of 
$F_{1,2n+1}$ and  $F_{2,2n+2}$ with $n>0$ as well.

 \paragraph{The general solution of the WI.}
 It is not hard to convince oneself that any function of the form \eqref{F11solveWI} subject to 
 \eqref{threeinvolutions}, can be rewritten with the help of the quantities introduced in \eqref{introduced} as 
  \be\label{Fingform}
\Omega_-^{\delta_{\varepsilon,1}}
\left( \sum_{t=0}^n \sum_{s=0}^t\,(g_1g_2)^s (g_1+g_2)^{t-s}\,f^{(0)}_{s,t}(z)\right)+ 
\Omega_-^{-\delta_{\varepsilon,1}}\kappa
\left( \sum_{t=0}^{n+\varepsilon-1}\sum_{s=0}^t\,(g_1g_2)^s (g_1+g_2)^{t-s}\,f^{(1)}_{s,t}(z)\right)\,,
 \ee
 where $\text{min}(p_1,p_2)=2n+\varepsilon$ with 
$\varepsilon\in\{0,1\}$.
 Compare to \eqref{projectedTENSORprod}. 
 Notice that the part multiplied by $\kappa$ is odd under \eqref{extraZ2} while the remaining part is even.
 Using the rewriting \eqref{Fingform} it is not hard to show that the general solution of the WI takes the form
    \be\label{FingformBIS}
\Omega_-^{\delta_{\varepsilon,1}}\,\mathsf{C}_1\,+\,
\Omega_-^{-\delta_{\varepsilon,1}}\kappa\,\mathsf{C}_2
+
\Omega_-^{\delta_{\varepsilon,1}}
G_{n}^{\text{even}}(z,w_1,w_2)+ 
\Omega_-^{-\delta_{\varepsilon,1}}\kappa\,
G_{n+\varepsilon-1}^{\text{odd}}(z,w_1,w_2)\,,
 \ee
  where $\mathsf{C}_1$, $\mathsf{C}_2$ are constants and 
  \be
G_{N}^{\text{even}/\text{odd}}(z,w_1,w_2)\,=\,
   (g_1g_2)^2\,F_N^{\text{even}/\text{odd}}(z,w_1,w_2)+\sum_{i=1}^{N}\,(g_1+g_2)^{i-1}\,\mathbb{D}f^{\text{even}/\text{odd}}_i(z)\,,
  \ee
  where the degree of $F_N^{\text{even}/\text{odd}}$ in $w_1,w_2$ is lowered by the presence of the factor $(g_1g_2)^2$.
  This function should be invariant under  \eqref{threeinvolutions}, so that it depends on $\tfrac{1}{2}N(N-1)$ functions of $z$
  individually invariant under $z\rightarrow z^{-1}$.
  The functions $G_{N}^{\text{even}/\text{odd}}$ are thus specified in terms of $\tfrac{1}{2}N(N+1)$ single variable functions,
  moreover they vanish when evaluated at $z=w_1=w_2$ and its images under  \eqref{threeinvolutions}.
  Notice that in order to extract $f_n(z)$ and $F(z,w_1,w_2)$, 
  it is convenient to first evaluate the quantity above at the special kinematical point $w_1=z$ so that $g_1=0$.
We should remark that while \eqref{FingformBIS} is the most general solution to the Ward identities, it is not clear whether more
convenient parameterizations exist.

%% file: sections/3_superblocks.tex

\section{Superconformal blocks}
\label{sec:superblocks}

In this section we present the superconformal blocks
for the two-point function of $\tfrac{1}{2}$-BPS operators, obtained by solving the Ward identities derived in the previous section. Each superconformal block captures the contribution coming from the exchange of a particular superconformal multiplet in the relevant OPE, see figure \ref{fig:crossing}.
 A superconformal block is therefore a sum of products of spacetime and R-symmetry blocks, 
 whose relative coefficients are fixed by superconformal symmetry.

A straightforward way to proceed is to consider the most general linear combination of spacetime and R-symmetry blocks 
dictated by the operator content of the exchanged  supermultiplet.
 The latter can be extracted, for example, using superconformal characters. 
 Notice that out of the full content of the supermultiplets only a small fraction of  spacetime and R-symmetry multiplets can appear due to the bosonic symmetry, e.g.~conformal representations with non-zero spin are excluded.
Once the ansatz is written, we can plug it in the WI and look for a solution for the unknown coefficients. 
In this section we systematically scan over all possible supermultiplets
 both in the boundary and bulk channels. 
 If the WI have no solution, it indicates that the corresponding supermultiplet cannot be exchanged in the given channel.
 On the other hand, if the  supermultiplets can be exchanged, the WI can be solved and the solution is unique, up to normalization. The results of this section will confirm the superspace analysis of section \ref{sec:superspace}, and further refine it by providing the range of representation labels for the exchanged operators.
 
An alternative way to proceed is by defining bulk and boundary  superblocks as eigenfunctions of the Casimir operator,
supplemented by certain boundary condition reflecting the OPE behavior of the correlator.
In the boundary channel, the relevant operator is the $\mathfrak{osp}(4|4)$ Casimir acting on one of the two points,
in the bulk channel,  the relevant operator is the $\mathfrak{psu}(2,2|4)$ Casimir acting on both points.
The method we used here to determine the superblocks is to solve the Casimir equations only for the bosonic subalgebras and then impose the Ward identities. This method turned out to be sufficient in this case, but in more general situations it is not\footnote{Two examples are $\Nm=3$ theories in $4d$ and $\Nm=6$ theories in $3d$ \cite{Lemos:2016xke,LLMM}.}.

To conclude the section we will present an interesting connection between several systems that exhibit $\mathfrak{osp}(4|4)$ symmetry. It turns out that from the superblocks of the $\frac{1}{2}$-BPS codimension one case,
 one can also obtain the superblocks relevant for the codimension three case, as well as 
the superblocks for four-point functions of $\frac{1}{2}$-BPS operators in  $d=3$ $\mathcal{N}=4$ and $d=1$ $OSP(4^*|4)$ theories with no defect.

\subsection{Boundary channel}

Let us write the boundary superblock as 
\be\label{superblockBULKfromblocks}
\mathfrak{F}^{\text{bdy}}_{\chi_{\text{bdy}}}(z,w_1,w_2)\,=\,
\sum_{L,(k_+,k_-)}\,c_{\delta,(k_+,k_-)}(\chi_{\text{bdy}})\,
\mathfrak{h}_{k_+}^{\text{bdy}}(w_+)\,
\mathfrak{h}_{k_-}^{\text{bdy}}(w_-)\,
\mathfrak{f}_{\delta}^{\text{bdy}}(z)\,,
\ee
where $w_\pm^2:=w_1w_2^{\pm1}$,
$\chi_{\text{bdy}}$ is a representation of $\mathfrak{osp}(4|4)$,
$(k_+,k_-)$ is a representation of $\mathfrak{su}(2)\oplus \mathfrak{su}(2)$,
and $\delta$ denotes the dimension of the three-dimensional operator.
The precise range of the finite summation in \eqref{superblockBULKfromblocks} depends on the 
$\mathfrak{sp}(4)\oplus\mathfrak{su}(2)\oplus \mathfrak{su}(2)$ content of the exchanged 
supermultiplet $\chi_{\text{bdy}}$ and can be extracted using the characters of \cite{Dolan:2008vc}.
An efficient way to solve for the spacetime and R-symmetry blocks is to use the bosonic Casimir equation. 
The spacetime blocks  $\mathfrak{f}_{\delta}^{\text{bdy}}(z)$  have already been obtained in the literature 
\cite{McAvity:1995zd,Liendo:2012hy}. In four dimensions they take the form
\be
\mathfrak{f}^{\text{bdy}}_{\delta}(z)
:=(4\,\xi)^{-\delta}\,\,{}_2F_{1}\left(\delta,\delta-1;2\delta-2;-\xi^{-1}\right)\,,
\qquad
\xi\,=\,\frac{(z-1)^2}{4\,z}\,.
\ee
Notice that the boundary block is independent of the dimensions of the external operators $p_1,p_2$.
The normalization has been chosen so that $\mathfrak{f}^{\text{bdy}}_{\delta}(z)\sim z^{\delta}$
in the boundary  OPE channel, corresponding to $z$ close to  zero.
The R-symmetry boundary block solves two 
Casimir equations corresponding to the two $\mathfrak{su}(2)$ R-symmetry factors and therefore they have a factorized form.
The relevant  Casimir equation for each factor is
\be 
\hat{C}_2\, \mathfrak{h}_{k}^{\text{bdy}}(w) = k(k+1)\,\mathfrak{h}_{k}^{\text{bdy}}(w)\,,
\qquad
\hat{C}_2\,=\,w^2\partial_w^2+\frac{2\,w^2}{w-w^{-1}}\partial_w\,,
\ee
as can be derived by acting with the $\mathfrak{su}(2)$ Casimir on one of the two points in R-symmetry space. 
The solution with the right asymptotics is given by 
\be
\mathfrak{h}_{k}^{\text{bdy}}(w)\,=\,w^{-k}\,{}_{2}F_{1}(\tfrac{1}{2},-k;\tfrac{1}{2}-k;w^2)\,=\,
\sqrt{\pi}\frac{\Gamma(k+1)}{\Gamma(k+\tfrac{1}{2})}\,C_k^{(\frac{1}{2})}(\cos \phi)\,.
\ee
Above,  $k$ is a non-negative integer and we emphasize that, up to normalization, the R-symmetry blocks 
$\mathfrak{h}_{k}^{\text{bdy}}(w)$ are  Gegenbauer polynomials with argument $\cos \phi$ where $w=e^{i \phi}$.
Their asymptotic behavior is given by $\mathfrak{h}_{k}^{\text{bdy}}(w)\sim w^{-k}$ for $w\sim 0$.
Notice that only integer spin representations of $\mathfrak{su}(2)$ appear\footnote{In our
 conventions the dimension of the $\mathfrak{su}(2)$ representation is $2k+1$.
}, as can be understood by looking at the branching ratios \eqref{branchinRsymm}.

We will now proceed to  fix the coefficients in \eqref{superblockBULKfromblocks} by solving the Ward 
identities \eqref{WI}. In the following we will only present the non-zero solutions. 
These solutions correspond to the $\mathfrak{osp}(4|4)$
supermultiplets that can have a non-vanishing two-point function with a $\frac{1}{2}$-BPS bulk operator.
As we will see, the results of this section fully agree with the analysis in section  \ref{sec:correlation}.

\subsubsection{Boundary superblock for $\chi_{\text{bdy}}=(B,\pm)$}
\label{sec:boundblocksBPS}

The simplest superblocks correspond to short multiplets of the type $(B,\pm)$ being exchanged. 
The expansion reads
\be
\mathfrak{F}^{\text{bdy}}_{(B,+)_{k}}=
\mathfrak{h}_{k}^{\text{bdy}}(w_+)\,\mathfrak{f}_{k}^{\text{bdy}}(z)-
\,c_1(k)\,
\mathfrak{h}_{k-1}^{\text{bdy}}(w_+)\,\mathfrak{h}_{1}^{\text{bdy}}(w_-)\,\mathfrak{f}_{k+1}^{\text{bdy}}(z)+
\,c_2(k)\,
\mathfrak{h}_{k-2}^{\text{bdy}}(w_+)\,\mathfrak{f}_{k+2}^{\text{bdy}}(z)\,,
\ee
where
\be
c_1(k)\,=\,
\frac{2\,k}{2k-1}\,,
\qquad
c_2(k)\,=\,
\frac{16(k-1)^2k(k+1)}{(2k-1)^2(2k-3)(2k+1)}\,.
\ee
Notice that we suppressed the dependence on $z,w_+,w_-$ from the left-hand side.
The superblocks $\mathfrak{F}^{\text{bdy}}_{(B,-)_{k}}$ are obtained from $\mathfrak{F}^{\text{bdy}}_{(B,+)_{k}}$ by the replacement $w_+\leftrightarrow w_-$.

\subsubsection{Boundary superblock for $\chi_{\text{bdy}}=(B,1)$}

Next in line are the blocks for multiplets of type $(B,1)$, the most general ansatz consistent with the character is
\begin{align}
\begin{split}
\mathfrak{F}^{\text{bdy}}_{(B,1)_{(k_+,k_-)}} = & \mathfrak{h}_{k_+}^{\text{bdy}}\,\mathfrak{h}_{k_-}^{\text{bdy}} \mathfrak{f}_{\delta}^{\text{bdy}}
+\left(c_{\delta+1,(k_+-1,k_--1)} \mathfrak{h}_{k_+-1}^{\text{bdy}}\,\mathfrak{h}_{k_--1}^{\text{bdy}}\right.
\\
&
\left.+c_{\delta+1,(k_++1,k_--1)}\mathfrak{h}_{k_++1}^{\text{bdy}}\,\mathfrak{h}_{k_--1}^{\text{bdy}}
+c_{\delta+1,(k_+-1,k_-+1)} \mathfrak{h}_{k_+-1}^{\text{bdy}}\,\mathfrak{h}_{k_-+1}^{\text{bdy}}
\right)
\mathfrak{f}_{\delta+1}^{\text{bdy}}
\\
&
+\left(c_{\delta+2,(k_+,k_--2)} \mathfrak{h}_{k_+}^{\text{bdy}}\,\mathfrak{h}_{k_--2}^{\text{bdy}}
+ c_{\delta+2,(k_+-2,k_-)} \mathfrak{h}_{k_+-2}^{\text{bdy}}\,\mathfrak{h}_{k_-}^{\text{bdy}}\right.
\\
&
\left. + c_{\delta+2,(k_+,k_-)} \mathfrak{h}_{k_+}^{\text{bdy}}\,\mathfrak{h}_{k_-}^{\text{bdy}}\right)
\mathfrak{f}_{\delta+2}^{\text{bdy}}
+ c_{\delta+3,(k_+,k_-)} \mathfrak{h}_{k_+-1}^{\text{bdy}}\,\mathfrak{h}_{k_--1}^{\text{bdy}}
\mathfrak{f}_{\delta+3}^{\text{bdy}}\, ,
\end{split}
\end{align}
where we have suppressed the coordinate dependence to avoid cluttering and $\delta=k_++k_-$. By solving the Ward identities one finds
\begin{align}
\label{B1coeffs}
\begin{split}
c_{\delta+1,(k_+-1,k_-+1)} & = \frac{2 k_+}{1-2 k_+}\, ,
\qquad
c_{\delta+1,(k_++1,k_--1)} = \frac{2 k_-}{1-2 k_-}\, ,
\\
c_{\delta+1,(k_+-1,k_--1)} & = -\frac{32 k_+^2 k_-^2 (\delta+1)}{(2 k_+-1) (2 k_++1) (2 k_--1) (2 k_-+1) (2\delta+1)}\, ,
\\
c_{\delta+2,(k_+-2,k_-)} & = \frac{16 (k_+-1)^2 k_+ (\delta+1)}{(1-2 k_+)^2 (2 k_+-3) (2 \delta+1)}\, ,
\\
c_{\delta+2,(k_+,k_--2)} & = \frac{16 (k_--1)^2 k_- (\delta+1)}{(1-2 k_-)^2 (2 k_--3) (2 \delta+1)}\, ,
\\
c_{\delta+2,(k_+,k_-)} & = \frac{16 k_+ k_- (\delta-1) (\delta+1)}{(2 k_+-1) (2 k_--1) (2 \delta-1) (2 \delta+1)}\, ,
\\
c_{\delta+3,(k_+,k_-)} & = -\frac{32 k_+ k_- (\delta) (\delta+1) (\delta+2)}{(2 k_+-1) (2 k_--1) (2 \delta+1)^2 (2 \delta+3)}\, .
\end{split}
\end{align}
\subsubsection{Boundary superblock for $\chi_{\text{bdy}}=L^{\delta}$}

For the long blocks the solutions are a quite involved, but the procedure is the same as before. We start with the most general ansatz consistent with the content of the supermultiplet and the bosonic symmetries and fix the relative coefficients using \eqref{WI}. 
We have written the full solution in appendix \ref{app:boundarylongblocks}.
It is interesting to note that the $(B,1)$ blocks presented above can be obtained as a special limit of the long block:
\be
\label{polebdylong}
\lim_{\delta\rightarrow k_++k_-+1}\left[(k_++k_-+1-\delta) \mathfrak{F}^{\text{bdy}}_{L^{\delta}_{(k_+,k_-)}}\right]\,=\,
 \mathfrak{F}^{\text{bdy}}_{(B,1)_{(k_++1,k_-+1)}}\,.
\ee

\subsection{Bulk channel}

Now we calculate the superblocks for the bulk channel. Recall that, unlike $F_{p_1,p_2}(z,w_1,w_2)$ in the boundary channel, here $\Omega^{-\frac{p_1+p_2}{2}}F_{p_1,p_2}(z,w_1,w_2)$  is the quantity to be expanded in bulk superblocks, where
\be\label{Omegadef}
\Omega\,:=\,
\frac{(\text{sPf}\,X_{\text{d},1})\,(\text{sPf}\,X_{\text{d},2})}{\text{sdet}(X_{1}-X_2)}
\,=\,
\frac{\xi_R}{\xi}
\,=\,\prod_{i=1}^2
\left(\frac{w^{+\frac{1}{2}}_i-w_i^{-\frac{1}{2}}}{z^{+\frac{1}{2}}-z^{-\frac{1}{2}}}\right)\,,
\ee
and $\xi_R$ and $\xi$ are given in \eqref{cosphixiR} and \eqref{xireview}.
The origin of this factor is clear by looking at \eqref{2ptfun}.

The general form of the superblock in the bulk OPE channel is given by
\be\label{superblockBDRfromblocks}
\mathfrak{F}^{\text{blk}}_{\chi_{\text{blk}}}(z,w_1,w_2)\,=\,\sum_{\Delta,R}\,c_{\Delta,R}(\chi_{\text{blk}})\,
\mathfrak{h}_{R}^{\text{blk}}(w_1,w_2)\,
\mathfrak{f}_{\Delta}^{\text{blk}}(z)\,,
\ee
where $\chi_{\text{blk}}$ is a representation of $\mathfrak{psu}(2,2|4)$ for which the corresponding operator has non-zero one-point function, and $R$ is a representation of $\mathfrak{su}(4)$ with Dynkin label $[2m,2n,2\bar{m}]$, 
$n,m,\bar{m}\in \mathbb{Z}_{\geq 0}$. 
In the following we will restrict our attention to the case in which the external operators have the same 
quantum numbers $p_1=p_2$ so that only the case $m=\bar{m}$ is relevant.
Again, the most efficient way for obtaining the bosonic spacetime $\mathfrak{f}_{\Delta}^{\text{blk}}(z)$  and R-symmetry blocks $\mathfrak{h}_{R}^{\text{blk}}(w_1,w_2)$ is by solving the Casimir equation.  In this case the relevant operator is the Casimir of the bulk symmetry
acting on both  points  in the two-point function. 
As in the boundary channel, the spacetime blocks are already given in the literature \cite{McAvity:1995zd,Liendo:2012hy}:
\be
\mathfrak{f}^{\text{blk}}_{\Delta}(z)
:=(4\,\xi)^{\frac{\Delta}{2}}\,\,{}_2F_{1}\left(\tfrac{1}{2}\Delta,\tfrac{1}{2}\Delta;\Delta-1;-\xi\right)\,,
\qquad
\xi\,=\,\frac{(z-1)^2}{4\,z}\,.
\ee
The normalization has been chosen so that $\mathfrak{f}^{\text{blk}}_{\Delta}(z)\sim (z-1)^{\Delta}$
 in the bulk OPE channel that corresponds to 
$z$  close to one.
Similarly, the R-symmetry bulk blocks can be defined as polynomial solutions to certain second and fourth order differential equations
coming from the Casimir, see \eqref{App:casimirRsymm} for the explicit form.  
The normalization is chosen so that in the bulk OPE limit $w_i\sim 1$ we have 
\be\label{asymRsymm}
\mathfrak{h}_{[2m,2n,2m]}^{\text{blk}}(w_1,w_2)\,\sim_{\epsilon\sim 0}\,
(w_1-1)^{-(n+m)}(w_2-1)^{-(n+m)}\,P_{m}(\cos\phi)\,,
\qquad
w_{1,2}=1+\epsilon\,e^{\pm i\frac{\phi}{2}}\, ,
\ee
and $P_m(x)$ are Legendre polynomials normalized as $P_{m}(1)=1$.
It is also useful to observe that
\be
\mathfrak{h}_{[2m,2n,2m]}^{\text{blk}}(w_1^{-1},w_2^{})\,=\,(-1)^n\,\,
\mathfrak{h}_{[2m,2n,2m]}^{\text{blk}}(w_1^{},w_2^{})\,,
\ee
while they are invariant under \eqref{threeinvolutions}.

We found it convenient in our analysis of the Casimir equation to introduce the following variables
\be\label{tifromwi}
t_i:=\frac{w_i+1}{w_i-1}\,,
\ee
which seem to play a special role in the mathematical literature \cite{Koor:1978}\footnote{An interesting connection between conformal blocks and Koornwinder polynomials was observed in \cite{Isachenkov:2016gim}.}.
The advantage of the variables  $t_i$ is that we managed to write the $R=[0,2n,0]$ and $R=[2m,0,2m]$
blocks in closed form, see \eqref{Rsymmclosedform}.
It turns out that this change of variables brings a pleasant surprise: the R-symmetry block  
$\mathfrak{h}_{[2q,2k,2q]}^{\text{blk}}$ can be considered as the analytic continuation in the representation labels 
of the standard four-point function $d=3$ bosonic block
of a CFT without defects\footnote{
We are very grateful to M. Isachenkov and V. Schomerus for pointing this out to us.
} . The precise correspondence is obtained by comparing the Casimir equations and is given in 
section \ref{sec:3dN=41dQM}
 and appendix \ref{APP:Rsymmblocksdetails}.
 For now let us notice the relation between representation labels and 
cross-ratios
\be\label{3dRsymmannounce}
\{\Delta^{3d},s\}\,=\,\{\tfrac{1}{2}-k-q,q\}\,,
\qquad
x=t_1^{-2}\,,
\qquad
\bar{x}=t_2^{-2}\,.
\ee
It is worth mentioning that identical external operators in the defect block correspond to non-identical external operators
in the  $d=3$ picture: 
$\Delta_{12}=\Delta_{34}=-\tfrac{1}{2}$ 
where $\Delta_{ij}:=\Delta_{i}-\Delta_{j}$.
It turns out that such a relation can be extended to relations between full superblocks and is explained with more detail in section \ref{sec:3dN=41dQM}.

As for the superblocks in the defect OPE channel,
 the basic idea is that the Ward identities \eqref{WI} fix the coefficients $c_{\Delta,R}(\chi_{\text{blk}})$
 uniquely up to an overall normalization. We fix the normalization by 
  $c_{\Delta_{\text{min}},R}(\chi_{\text{blk}})=1$.
 If the representation $\chi_{\text{blk}}$ is such that superconformal symmetry implies 
 $\langle\mathcal{O}_{\chi_{\text{blk}}}\rangle=0$, see the classification  in section \ref{sec:correlation},
  then the corresponding Ward identity has no solution.
This provides a cross-check of our analysis.

\subsubsection{Bulk superblock for $\chi_{\text{blk}}=\mathcal{B}_{[0,2n,0]}$}

Apart from the identity superblock which is just $1$, the next simplest superblock correspond to the
 $\frac{1}{2}$-BPS supermultiplet  $\mathcal{B}_{[0,p,0]}$. The corresponding block is non-trivial only for 
 $p=2n$ even.
Looking at table I in  \cite{Gunaydin:1984fk} we see which Lorentz scalar conformal multiplets are in the 
$\mathcal{B}_{[0,2n,0]}$ supermultiplets.  One should further restrict to operators with zero $U(1)_Y$  charge,
see below \eqref{eq:defPiandSigma} for the definition. 
This leads to  an ansatz for the superblock of the form
\be\label{12BPSexpansion}
\mathfrak{F}^{\text{blk}}_{\mathcal{B}_{[0,2n,0]}}\,=\,
\mathfrak{h}_{[0,2n,0]}^{\text{blk}}\,\,
\mathfrak{f}_{2n}^{\text{blk}}+
\,c_1(n)\,\,
\mathfrak{h}_{[2,2n-4,2]}^{\text{blk}}\,\,
\mathfrak{f}_{2n+2}^{\text{blk}}+
\,c_2(n)\,\,
\mathfrak{h}_{[0,2n-4,0]}^{\text{blk}}\,\,
\mathfrak{f}_{2n+4}^{\text{blk}}\,.
\ee
Notice that we suppressed the dependence on $z,w_1,w_2$.
The  coefficients are fixed by the Ward identities \eqref{WI} to be 
\be
c_1(n)\,=\,
\frac{(n-1)^2n}{2^4(2n-3)(2n-1)^2}\,,
\qquad
c_2(n)\,=\,
\frac{(n-1)n^2(n+1)}{2^{12}(2n-3)(2n-1)(2n+1)^2}\,.
\ee
As an example, in the case $n=1$,
 which corresponds to the exchange of the  stress-tensor supermultiplet in the bulk channel,
the superblock takes the form
\be
\mathfrak{F}^{\text{blk}}_{{\mathcal{B}_{[0,2,0]}}}(z,w_1,w_2)\,=\,
\left[\prod_{i=1}^2\left(\frac{w_i+1}{w_i-1}\right)\right]
\mathfrak{f}^{\text{blk}}_{2}(z)\,=\,
\prod_{i=1}^2
\left[\left(\frac{w_i+1}{w_i-1}\right)\left(\frac{z-1}{z+1}\right)\right]
\,.
\ee
This expression is remarkably simple. It is useful to check its properties.
Firstly, it is invariant under the transformations 
\eqref{threeinvolutions}
 and picks up a sign under \eqref{extraZ2}.
Secondly, when specialized to $w_1=w_2=z$ it reduces to a constant.
In our normalizations this constant is one.
%
For $\Delta=2n$ with $n\in\mathbb{Z}_{>0}$ the spacetime part of the blocks takes the form
\be
\mathfrak{f}^{\text{blk}}_{2n}\,=\,
\left[
\frac{(n-1)}{(1+\xi)^{n}}\binom{2n-2}{n-1}{}_2F_1\left(n-1,n,1,\frac{1}{1+\xi}\right)
\right]\,
\log(1+\xi)\,+\,
\frac{Q_{n}(\xi)}{1+\xi}\,,
\ee
where $Q_{n}(\xi)$ and the quantity in square bracket are Laurent polynonials in $\xi$.
\subsubsection{Bulk channel superblocks for $\chi_{\text{blk}}=\mathcal{B}^{\Delta}_{[2m,2n,2m]}$ and $\chi_{\text{blk}}=\mathcal{A}^{\Delta}_{[2m,2n,2m]}$}

For the long block in the bulk channel we have collected the coefficients in appendix \ref{app:bulklongblocks}.
For the $\Bm_{[2m,2n,2m]}$ block, the most efficient way to calculate the coefficients is using a relation analogous to \eqref{polebdylong}.
Indeed, the long block has a simple pole at $\Delta=2+2n+4m$, which corresponds to the unitarity bound. 
The residue of this corresponds to the block associated to  $\Bm_{[2q,2n,2q]}$ multiplets. More precisely
\be
\label{polebulklong}
\lim_{\Delta\rightarrow  2+2n+4m}\,
\left[\frac{1}{c_{4,m+1,n}}\,
\mathfrak{F}_{\mathcal{A}^{\Delta}_{[2m,2n,2m]}}^{\text{blk}}
\right]\,=\,
\mathfrak{F}_{\mathcal{B}^{}_{[2(m+1),2n,2(m+1)]}}^{\text{blk}}\,.
\ee
\paragraph{Cohomological sector from superconformal blocks.}
It is instructive to see how the cohomological sector of \cite{Chester:2014mea,Beem:2016cbd}, which we will study with more detail in section \ref{sec:bootstrap},
is singled out by restricting the correlation functions to certain values of the cross-ratios.
 %
Superconformal blocks possess the following remarkable properties
\be\label{blocksatspcialpointsMEW}
\mathfrak{F}^{\text{blk}}_{\chi_{\text{blk}}}(t,t,t^{\pm 1})\,=\,
\sum_{n}\,(\pm1)^{n}\,\delta_{\chi_{\text{blk}},\mathcal{B}_{[0,2n,0]}}\,,
\qquad
\mathfrak{F}^{\text{bdy}}_{\chi_{\text{bdy}}}(t,t,t^{\pm 1})\,=\,
\sum_{k}\,\delta_{\chi_{\text{bdy}},(B,\pm)_{k}}\,,
\ee
where $n,k\in \mathbb{Z}_{>0}$.
This means that the complicated crossing equations of figure \ref{fig:crossing} reduce to identities involving a finite 
number of operators at a time.
%
The relations \eqref{blocksatspcialpointsMEW} can be verified using the explicit expressions for the 
superblocks, but it should also be possible to prove them directly from their definition.
The crucial input comes from the Ward identities \eqref{WI}. Indeed, they imply that
\be
\mathfrak{F}^{\text{blk}}_{\chi_{\text{blk}}}(t,t,t)\,=\,\text{Const}(\chi_{\text{blk}})\,,
\qquad
\mathfrak{F}^{\text{bdy}}_{\chi_{\text{bdy}}}(t,t,t)\,=\,\text{Const}(\chi_{\text{bdy}})\,.
\ee
The value of the constants corresponding to \eqref{blocksatspcialpointsMEW}
 can then be fixed by a careful analysis of the relevant OPE limit.

\subsection{OPE summary}

Here we give a summary of the selection rules obtained in the bulk and boundary OPE channels.
For the boundary channel we have
\be\label{bdyOPEsummary}
\left[\mathcal{B}_{[0,2n+\varepsilon,0]}\right]_{\text{\tiny{$\partial$OPE}}}\,
\simeq\,
\delta_{\varepsilon,0}\,\oplus\,
\bigoplus_{(k_+,k_-)\,\in\,S_{n,\varepsilon}}\,
B_{(k_+,k_-)}
\,
\oplus
\,
\bigoplus_{(k_+,k_-)\,\in\,S_{n-1,\varepsilon}}\,
L^{\delta}_{(k_+,k_-)}\,,
\ee
where $\varepsilon\,\in\{0,1\}$ and we define the set
$S_{n,\varepsilon}=
\bigcup_{s=0}^n
\bigcup_{a=0}^{2s+\varepsilon}
\big{\{}(2s+\varepsilon-a,a)\big{\}}$.
In the OPE \eqref{bdyOPEsummary} the short operators of $B$-type are further divided 
as $B_{(k,0)}=(B,+)_k$, $B_{(0,k)}=(B,-)_k$ and $B_{(k_+,k_-)}=(B,1)_{(k_+,k_-)}$
otherwise.

Concerning the bulk channel, the OPE 
of $\frac{1}{2}$-BPS operators
 is well known, 
see  \cite{Arutyunov:2001qw,Eden:2001ec,Dolan:2001tt},
and is independent of the presence of the defect.
 To determine which operators contribute to the two-point function under study,
  one should further project to the exchanged operators that can have a non-zero one-point function.
 The latter operation is denote by $\mathsf{Pr}$.
  Finally, it is convenient to split the operators appearing in the OPE into even and odd parts with respect to 
  \eqref{extraZ2}. In summary we obtain
\begin{align}\label{OPEbulksummary}
\mathsf{Pr}\left(\mathcal{B}_{[0,2n+\varepsilon,0]}\stackrel{\text{\tiny{OPE}}}{\times}
\mathcal{B}_{[0,2n+\varepsilon,0]}\right)\Big{|}_{\text{even}}&
\,\simeq\,
\mathcal{I}\,\oplus\,
\bigoplus_{R\,\in\,T^{\text{e}}_{n,\varepsilon}}
\mathcal{B}_{R}\,\,\,
\oplus\,
\bigoplus_{R\,\in\,T^{\text{e}}_{n-1,\varepsilon}}
\mathcal{A}^{\Delta}_{R}\,,\\
\mathsf{Pr}\left(\mathcal{B}_{[0,2n+\varepsilon,0]}\stackrel{\text{\tiny{OPE}}}{\times}
\mathcal{B}_{[0,2n+\varepsilon,0]}\right)\Big{|}_{\text{odd}}\,&
\,\simeq\,
\bigoplus_{R\,\in\,T^{\text{o}}_{n,\varepsilon}}
\mathcal{B}_{R}\,\,\,
\oplus\,
\bigoplus_{R\,\in\,T^{\text{o}}_{n-1,\varepsilon}}
\mathcal{A}^{\Delta}_{R}\,,
\end{align}
where
\be
T^{\text{e}}_{n,\varepsilon}\,=\!
\bigcup_{0\leq a\leq b\leq n} [2(b-a),4a,2(b-a)]\,,
\qquad
T^{\text{o}}_{n,\varepsilon}\,=\!\!\,\!\!\!\!
\bigcup_{0\leq a\leq b\leq n+\varepsilon-1} 
\!
[2(b-a),4a+2,2(b-a)]\,.
\ee
For simplicity, in \eqref{OPEbulksummary} 
we restricted to the case of identical operators but the general case is immediately obtained.
Summation over the values of  $\delta$  and $\Delta$ for the long representations  
$L^{\delta}_{(k_+,k_-)}$
and $\mathcal{A}^{\Delta}_{R}$ is understood.

\subsection{One block to rule them all}
\label{sec:analyticcont}

In this section we elaborate on the relation between the polynomial R-symmetry blocks and
 spacetime bosonic blocks.
  Thanks to this analysis we will be able to write, with minimal effort, the superblocks for three additional systems: 
 $d=4$,  $\Nm=4$ theories in the presence of a $\frac{1}{2}$-BPS line defect,
 $3d$ $\Nm=4$ superconformal theories, and $OSP(4^*|4)$ superconformal quantum mechanics. 
In the latter two cases the theories have no defect, and the blocks we obtain are relevant for the expansion of four-point functions of 
$\frac{1}{2}$-BPS operators. All these systems exhibit the same $\mathfrak{osp}(4|4)$ symmetry as $4d$ $\Nm=4$ theories in the presence of a flat codimension one defect.

It is clear from section \ref{sec:superspace}
that  the superblocks for the  codimension three defects can be obtained from the codimension one case
by a sort of analytic continuation in the representation labels.
This is a special feature of the $4d$ $\mathcal{N}=4$ superconformal setup, and is therefore not surprising that the two-point function superblocks are related to each other. On the other hand, the connection to four-point function superblocks in theories without defect is unexpected. It also appears to be more general and certainly warrants further study.
 
\subsubsection{Line defect superblocks}

The superconformal blocks for $\frac{1}{2}$-BPS codimension three defects 
in $4d$ $\mathcal{N}=4$ superconformal theories can be obtained in a simple way from the codimension one case presented above. This fact is manifest by comparing  the superspace setup for the two defects.

\paragraph{Superblock dictionary.}
The bulk channel superblocks are related as
\begin{align}\label{fromcodimonetocodim3}
\Big[\mathfrak{F}^{\text{blk}}_{\mathcal{B}_{[0,2n,0]}}(z;w_1,w_2)
\Big]_{\text{codim-$1$}}
\stackrel{\text{a.c.}}{=}&\,\,\,
\Big[\mathfrak{F}^{\text{blk}}_{\mathcal{B}_{[0,-2n,0]}}(w_1,w_2;z)
\Big]_{\text{codim-$3$}}\\
\Big[\mathfrak{F}^{\text{blk}}_{\mathcal{B}_{[2m,2n,2m]}}(z;w_1,w_2)
\Big]_{\text{codim-$1$}}
\stackrel{\text{a.c.}}{=}&\,\,\,
\Big[\mathfrak{F}^{\text{blk}}_{\mathcal{C}_{[0,-2(n+2m),0],(m-1,m-1)}}(w_1,w_2;z)
\Big]_{\text{codim-$3$}}
\\
\Big[\mathfrak{F}^{\text{blk}}_{\mathcal{A}^{\Delta}_{[2m,2n,2m],(0,0)}}(z;w_1,w_2)
\Big]_{\text{codim-$1$}}
\stackrel{\text{a.c.}}{=}&\,\,\,
\Big[\mathfrak{F}^{\text{blk}}_{\mathcal{A}^{-2(m+n)}_{[0,-\Delta,0],(m,m)}}(w_1,w_2;z)
\Big]_{\text{codim-$3$}}
\end{align}
the defect channel superblocks are related as
\begin{align}\label{fromcodimonetocodim3defectchannel}
\Big[\mathfrak{F}^{\text{bdy}}_{(B,+)_k}(z;w_1,w_2)
\Big]_{\text{codim-$1$}}
\stackrel{\text{a.c.}}{=}&\,\,\,
\Big[\mathfrak{F}^{\text{line}}_{(B^*,+)_{-k}}(w_1,w_2;z)
\Big]_{\text{codim-$3$}}\\
\Big[\mathfrak{F}^{\text{blk}}_{(B,1)_{(k_+,k_-)}}(z;w_1,w_2)
\Big]_{\text{codim-$1$}}
\stackrel{\text{a.c.}}{=}&\,\,\,
\Big[\mathfrak{F}^{\text{blk}}_{(B^*,1)_{(-k_+,k_-)}}(w_1,w_2;z)
\Big]_{\text{codim-$3$}}
\qquad m>0\\
\Big[\mathfrak{F}^{\text{blk}}_{L[0]_{\delta}^{(2k_+,2k_-)}}(z;w_1,w_2)
\Big]_{\text{codim-$1$}}
\stackrel{\text{a.c.}}{=}&\,\,\,
\Big[\mathfrak{F}^{\text{blk}}_{L^*[-k_1]_{k_2}^{[0,\delta]}}(w_1,w_2;z)
\Big]_{\text{codim-$3$}}
\end{align}
see \eqref{ABCd4}  and \eqref{unitaryrepsSCQM} for the definition of the relevant exchanged supermultiplets.
Above, ``a.c." refers to analytic continuation which is defined as follows. 
Spacetime and R-symmetry blocks are individually analytically continued by the requirement that they 
satisfy the same Casimir equations with the same boundary conditions, but continued labels (this point is further discussed below). The analytic continuation of the coefficients is obvious as they are rational functions of the representation labels.
Notice that in the boundary channel the analytic continuation of unitary representations of the $3d$ $\mathcal{N}=4$ 
superconformal algebra $\mathfrak{osp}(4|4)$ gives 
unitary representations of the $1d$ $\mathcal{N}=4$ 
superconformal algebra $\mathfrak{osp}(4^*|4)$. This is the superconformal quantum mechanics living on the line defect.  

\subsubsection{$3d$ $\Nm=4$ theories and $1d$ $OSP(4^*|4)$ quantum mechanics}
\label{sec:3dN=41dQM}

It was mentioned in the previous section that R-symmetry  blocks in the bulk OPE channel of a two-point function in 
the presence of a codimension one defect are essentially analytic continuation in the representation labels
 of standard $3d$ conformal blocks,
 see \eqref{3dRsymmannounce}. 
 This observation is best understood by comparing their definition via the Casimir equations.
We will now sharpen and extend this surprising observation.

\paragraph{Superblock dictionary: line defect/$3d$ $\mathcal{N}=4$ theory.}
Let us start by giving the dictionary between blocks
\begin{align}\label{fromlinetobulk}
\left(\frac{y^2}{x\bar{x}}\right)^{\frac{1}{4}}\,\mathfrak{G}^{3d,\mathcal{N}=4}_{(B,+)_{\text{\tiny{$\frac{p+1}{2}$}}}}(x,\bar{x};y)
=&\,\,\,
\Big[\mathfrak{F}^{\text{blk}}_{\mathcal{B}_{[0,p,0]}}(z_1,z_2;w)
\Big]_{\text{codim-$3$}}\, ,\\
\left(\frac{y^2}{x\bar{x}}\right)^{\frac{1}{4}}\,\mathfrak{G}^{3d,\mathcal{N}=4}_{(A,+)^{\ell}_{\text{\tiny{$\frac{p+1}{2}$}}}}(x,\bar{x};y)
=&\,\,\,
\Big[\mathfrak{F}^{\text{blk}}_{\mathcal{C}_{[0,p,0],(\ell,\ell)}}(z_1,z_2;w)
\Big]_{\text{codim-$3$}}\, ,
\\
\left(\frac{y^2}{x\bar{x}}\right)^{\frac{1}{4}}\,
\mathfrak{G}^{3d,\mathcal{N}=4}_{L[2\ell]^{(p+1;0)}_{\text{\tiny{$\frac{\Delta+1}{2}$}}}}(x,\bar{x};y)
=&\,\,\,
\Big[\mathfrak{F}^{\text{blk}}_{\mathcal{A}^{\Delta}_{[0,p,0],(\ell,\ell)}}(z_1,z_2;w)
\Big]_{\text{codim-$3$}}\, ,
\end{align}
where the cross-ratios are identified as
\be
x=\left(\frac{z_1-1}{z_1+1}\right)^2\,,
\qquad
\bar{x}=\left(\frac{z_2-1}{z_2+1}\right)^2\,,
\qquad
y=\left(\frac{w-1}{w+1}\right)^2\,.
\ee
The superblocks $\mathfrak{G}^{3d,\mathcal{N}=4}$ for the $3d$ $\mathcal{N}=4$ superconformal theory are relevant for the expantion of the four-point function
\be
\langle (B,+)_{k_1}\,(B,+)_{k_1+\frac{1}{2}}\,(B,+)_{k_2}\,(B,+)_{k_2+\frac{1}{2}}\rangle\,.
\ee
The relations \eqref{fromlinetobulk} follow from the spacetime and R-symmetry block identities
\be\label{3dblockstoline}
(x\bar{x})^{-\frac{1}{4}}\mathfrak{g}^{3d}_{\frac{\Delta+1}{2},\ell}(x,\bar{x})
\,=\,
\left[\mathfrak{f}^{\text{blk}}_{\Delta,\ell}(z_1,z_2)\right]_{\text{line}}\,,
\qquad
\sqrt{y}\,
\mathfrak{g}^{\text{R-symm}}_{k}(y)
\,=\,
\left[\mathfrak{h}^{\text{blk}}_{k}(w)\right]_{\text{line}}\,,
\ee
together with the fact that both superblocks satisfy the same superconformal Ward identities, see 
\eqref{WI} and \cite{Dolan:2004mu,Liendo:2015cgi}.
Above, $\mathfrak{g}^{3d}_{\delta,\ell}$ are standard $3d$ blocks relevant for the expansion
of a four-point function of scalar operators with 
$\Delta^{3d}_{12}=\Delta^{3d}_{34}=-\tfrac{1}{2}$
where $\Delta_{ij}:=\Delta_{i}-\Delta_{j}$.  
Moreover, the representation label $k$ on the left hand side is an $\mathfrak{su}(2)$ labels,
while on the right-hand side is part of the $\mathfrak{usp}(4)$ Dynkin labels 
$[0,k]$.
See appendix \ref{APP:Rsymmblocksdetails} for more details.
It should be noticed that the first equation in \eqref{3dblockstoline}
 fills a gap in the literature. 
 The conformal blocks for the line defect in the bulk channel were not determined in  \cite{Billo:2016cpy}.
 Now,  thanks to \eqref{3dblockstoline}, we can say that they are
  as well understood as $3d$ bosonic blocks.

\paragraph{Superblock dictionary: boundary/$1d$ $OSP(4^*|4)$ quantum mechanics.}
In this case the dictionary between superblocks is given by
\begin{align}\label{frombdyoQM}
\left(\frac{x^2}{y\bar{y}}\right)^{\frac{1}{4}}\,\mathfrak{G}^{1d,OSP(4^*|4)}_{(B^*,+)_{\text{\tiny{$\frac{2n-1}{2}$}}}}(x;y,\bar{y})
=&\,\,\,
\Big[\mathfrak{F}^{\text{blk}}_{\mathcal{B}_{[0,2n,0]}}(w_1,w_2;z)
\Big]_{\text{codim-$1$}}\, ,\\
\left(\frac{x^2}{y\bar{y}}\right)^{\frac{1}{4}}\,\mathfrak{G}^{1d,OSP(4^*|4)}_{(B^*,1)_{\text{\tiny{$(\frac{2n-1}{2},m)$}}}}(x;y,\bar{y})
=&\,\,\,
\Big[\mathfrak{F}^{\text{blk}}_{\mathcal{B}_{[2m,2n,2m],(0,0)}}(w_1,w_2;z)
\Big]_{\text{codim-$1$}}\, ,
\qquad m>0\\
\left(\frac{x^2}{y\bar{y}}\right)^{\frac{1}{4}}\,
\mathfrak{G}^{1d,OSP(4^*|4)}_{L^*[\tfrac{1}{2}(\Delta-1)]_{\text{\tiny{$\frac{2n-1}{2}$}}}^{[0,m]}}
(x;y,\bar{y})
=&\,\,\,
\Big[\mathfrak{F}^{\text{blk}}_{\mathcal{A}^{\Delta}_{[2m,2n,2m],(0,0)}}(w_1,w_2;z)
\Big]_{\text{codim-$1$}}\,.
\end{align}
At the level of bosonic blocks, the identity reduces to an analytic continuation in the representation labels of \eqref{3dblockstoline}.
In particular, the spacetime part is realized by the identification
\be
 \sqrt{x}\,\,\mathfrak{g}^{1d}_{\frac{\Delta-1}{2}}(x)\,=\,
2^{-\Delta}\,\left[\mathfrak{f}_{\Delta}^{\text{blk}}(z)\right]_{\text{boundary}}\, ,
\ee
where $\mathfrak{g}^{1d}_{h}(x):=x^h{}_2F_1(h-h_{12},h+h_{34},2h,x)$ 
are the $1d$ blocks above specialized to $h_{12}=h_{34}=\tfrac{1}{2}$.

\vspace{0.5cm}
\noindent
\emph{Remark:} There is a relation between blocks for scalar four-point functions in $d$ dimensions,
and bulk channel blocks for scalar two-point functions in the same dimensionality $d$ in the presence of 
a codimension two defect \cite{Billo:2016cpy}. 
It is then natural to ask whether this statement can be extended to $\mathcal{N}=4$ superconformal blocks. 
An equality of blocks appears rather unlikely due to the fact that the relevant one-point functions contain multiple structures,
see \eqref{1ptsurface}. However, we still expect that the superconformal WI for the two-point function of bulk operators 
in the codimension two case take a similar form as the superconformal WI for the four-point function of $\frac{1}{2}$-BPS 
operators in $4d$, $\mathcal{N}=4$ theories. Clarifying this point remains an interesting problem for the future.

%% file: sections/4_bootstrap.tex

\section{Bootstrap equations}
\label{sec:bootstrap}

In this exploratory section we initiate the analysis of the dynamical constraints imposed by the bootstrap equations.
As discussed in the introduction, supersymmetry allows to divide the implementation of the bootstrap in two 
steps. Section \ref{sec:micro} concentrates on the cohomological sector while \ref{sec:macro} considers the full bootstrap equations.

\subsection{Microbootstrap}
\label{sec:micro}

We will now present the cohomologically  truncated bootstrap equations\footnote{
The term microbootstrap, introduced in \cite{LPR},
 is  justified by the fact that, from the point of view of the bulk theory, 
it corresponds to  a further reduction of the 
so called miniboostrap equations of \cite{Beem:2013sza}. %
}. 
The space  of  
$\frac{1}{2}$-BPS operators in a $4d$ $\mathcal{N}=4$ superconformal theory
is a graded vector space
  \begin{equation}\label{Vspace}
  \mathcal{V}\,=\,\bigoplus_{p=1}^{\infty}\,\mathcal{V}_p\,.
  \end{equation}
 For each element in $\mathcal{V}_p$ there is an associated superconformal primary 
 for the supermultiplets $\mathcal{B}_{[0,p,0]}$.
Since $\mathcal{B}_{[0,2,0]}$ contains the stress tensor, which one requires to be unique,
the space $\mathcal{V}_2$ is  one-dimensional.
The space $\mathcal{V}_1$ corresponds to massless representations and 
is expected to be part of a decoupled free theory\footnote{%
Notice that it cannot really be decoupled if the stress tensor is unique.
}. Each $\mathcal{V}_p$ is finite dimensional.
  The two-point function gives a non-degenerate pairing
   $\mathsf{G}:\mathcal{V}\times \mathcal{V}\rightarrow \mathbb{C}$ that respects the grading in \eqref{Vspace}. 
 The three-point function gives a trilinear map
 \be\label{trilinear}
 \mathsf{C}:\,\mathcal{V}\times \mathcal{V}\times \mathcal{V}\rightarrow \mathbb{C}\,,
 \ee
 which is invariant under permutation of the three vectors and due to the superconformal selection rules
  is non-vanishing only if $p_1\,\in\,S_{p_2,p_3}:=\{|p_2-p_3|,|p_3-p_3|+2,\dots,p_2+p_3\}$\,,
compare to \eqref{3ptfootnote}.
 The OPE is encoded in the bilinear map
 \be\label{OPEchat}
\widehat{ \mathsf{C}}:\,\mathcal{V}\times \mathcal{V}\rightarrow  \text{id}\,\oplus\,\mathcal{V}\,,
 \ee
 defined by the condition
  $\mathsf{C}(v_1,v_2,v_3)=\mathsf{G}\left(\widehat{\mathsf{C}}(v_1,v_2),v_3\right)$. 
  Above  $\text{id}$ corresponds to the identity operator. Notice that, as opposed to the full OPE of the 
  $4d$ $\mathcal{N}=4$ superconformal theory, the cohomologically truncated OPE \eqref{OPEchat}
  contains only finitely many terms on the right hand side.
  The crossing equations, i.e.~the requirement of associativity of the OPE, take the form 
  \be\label{cohomologicalcrossingfull}
  \mathsf{G}\left(\widehat{\mathsf{C}}(v_1,v_2),\widehat{\mathsf{C}}(v_3,v_4)\right)\,=\,
  \mathsf{G}\left(\widehat{\mathsf{C}}(v_1,v_3),\widehat{\mathsf{C}}(v_2,v_4)\right)\,.
  \ee
  As an example, let us consider the case in which all  $\mathcal{V}_p$ are one-dimensional.
  If we  choose an orthonormal basis $\{e_p\}$ 
 and define $\mathsf{C}_{p_1,p_2,p_3}:=\mathsf{C}(e_{p_1},e_{p_2},e_{p_3})$, the crossing equations read\footnote{Solutions to these equations were also explored in \cite{LPR}.}
 \be\label{CCbulk}
\sum_{k\,\in\,S_{p_1, p_2}}\,
\mathsf{C}_{p_1p_2 k}\,
\mathsf{C}_{p_3p_4 k}\,=\,
\sum_{k\,\in\,S_{p_1, p_3}}\,
\mathsf{C}_{p_1p_3 k}\,
\mathsf{C}_{p_2p_4 k}\,.
\ee
Notice that the summation is finite.
The simplest solution to these equations, which corresponds to 
$G_\text{gauge}=U(1)$,
 $\mathcal{N}=4$ SYM, is given by
\be\label{freeC}
\mathsf{C}_{abc}\,=\,
\frac{1}{\sqrt{a!\,b!\,c!}}\,
\sum_{n=0}^{a+b+c}\,
\delta_{a+b-2n,c}\,n!\,c!\,
{a \choose n}\,
{b \choose n}\,.
\ee
  The relations \eqref{CCbulk} together with the condition above correspond to the definition of an 
  infinite dimensional associative, commutative  algebra with some extra structure.
It thus makes sense to consider the subalgebra generated by the unique state in $\mathcal{V}_2$ which is the 
stress-tensor supermultiplet. This subalgebra contains a unique element for each $\mathcal{V}_{2n}$
and the relevant bootstrap equations take the form  \eqref{CCbulk} with $p_i$ even.
The extremal three-point structures $\mathsf{C}_{2n,2m,2(n+m)}$ 
in this case can be fixed by general principles \cite{Baggio:2015vxa} to be 
\be
\mathsf{C}_{2n,2m,2(n+n)}\,=\,
\sqrt{\frac{g_{n+m}}{g_n\,g_m}}\,,
\qquad 
g_n:=\frac{\Gamma(n+1)\Gamma(n+2\,c_{4d})}{\Gamma(2\,c_{4d})}\,,
\ee
where $c_{4d}$ is the four-dimensional central charge, which for a Lagrangian theory is
$c_{4d}=\tfrac{1}{4}\text{dim}(G_{\text{gauge}})$. 
Moreover, it follows from the conformal Ward identities for the stress tensor that, in our normalizations,
$\mathsf{C}_{2,2n,2n}=4 \sqrt{2c_{4d}}$.
It appears that, under these conditions and for any choice of the central charge $c_{4d}$,
 \eqref{CCbulk} possess a unique solution.
 For $c_{4d}=\tfrac{1}{4}$ this solution reduces to \eqref{freeC}.
As an example, for the next-to-extremal case we find
 \be
 \mathsf{C}_{2n,2m,2(n+n-1)}\,=\,
8\,mn\,c_{4d}\,
\frac{\Gamma(n+m)\Gamma(n+m-1+2\,c_{4d})}{\Gamma(2\,c_{4d})}\,
\sqrt{\frac{1}{g_n\,g_m\,g_{n+m-1}}}\,.
 \ee
We plan to report the general solution elsewhere.

Let us now introduce the boundary data.
The space of $\tfrac{1}{2}$-BPS operators in a $3d$ $\mathcal{N}=4$ superconformal theory splits
 into
 \begin{equation}\label{Vspace}
  \mathcal{U}\,=\,
    \mathcal{U}^{+}\oplus
      \mathcal{U}^{-}\,,
      \qquad  
      \mathcal{U}^{\pm}\,=\,\bigoplus_{k=1}^{\infty}\, \mathcal{U}_k^{\pm}\,,
  \end{equation}
   For each element in $\mathcal{U}^{\pm}_k$ there is an associated superconformal primary 
 for the supermultiplets $(B,\pm)_k$. Notice that we restricted to integer values of $k$ as they are the only ones entering  in the bulk-boundary OPE.
 The spaces $\mathcal{U}^{\pm}_{1}$ correspond to flavor symmetry currents and the spaces 
 $\mathcal{U}^{\pm}_2$ to the displacement operator. We thus require that 
 $\text{dim}(\mathcal{U}^+_2)+\text{dim}(\mathcal{U}^-_2)=1$ in the case of a boundary and 
 $\text{dim}(\mathcal{U}^+_2)=\text{dim}(\mathcal{U}^-_2)=1$ in the case of an interface.
 The boundary two-point function gives non-degenerate pairings
  $\mathsf{g}^{\pm}:\mathcal{U}^{\pm}\times \mathcal{U}^{\pm}\rightarrow \mathbb{C}$ 
  that respect the grading.
  The one-point function of bulk operators gives a map $\mathsf{a}:\mathcal{V} \rightarrow \mathbb{C}$ and
the bulk boundary two-point function defines two maps
 $\mu^{\pm}: \mathcal{V}\otimes 
 \mathcal{U}^{\pm}\rightarrow \mathbb{C}$.
 It follows from superconformal symmetry that they are non-vanishing only if 
  $p-k\in2\mathbb{Z}_{\geq 0}$.
 The bulk-boundary OPE   coefficients 
 \be
 \widehat{\mu}^{\pm}:\,\mathcal{V}\rightarrow
 \mathcal{U}^{\pm}\,,
 \ee
can then be obtained as $\mu^{\pm}(u,v)=\mathsf{g}^{\pm}(u,\widehat{\mu}^{\pm}(v))$. 
The bootstrap equations in figure \ref{fig:crossing}  take the form
\be
\mathsf{a}(\widehat{\mathsf{C}}(v_1,v_2))\,=\,
\mathsf{g}^{\pm}(\widehat{\mu}^{\pm}(v_1),\widehat{\mu}^{\pm}(v_2))\,.
\ee
In the case in which all the involved spaces are one-dimensional and choosing an orthonormal basis
 the equations above reduce to 
\be\label{rank1crossing}
\sum_{p\,\in\,S_{p_1, p_2}}\,
(\pm1)^{\frac{p_1+p_2+p}{2}}
\,\mathsf{C}_{p_1 p_2 p}\,
\mathsf{a}_p\,=\,\delta_{p_1,\text{even}}\,
\delta_{p_2,\text{even}}\,
\mathsf{a}_{p_1}\,\mathsf{a}_{p_2}+
\sum_{m=\varepsilon}^{\text{min}(p_1,p_2)}\,\mu^{\pm}_{p_1 m}\,\mu^{\pm}_{p_2 m}\,,
\ee
where in the sum on the right hand side $m\in\{\varepsilon,\varepsilon+2,\dots\}$ and $\varepsilon=1$ for $p_1,p_2$ odd and 
$\varepsilon=2$ for $p_1,p_2$ even.
One can study the equations \eqref{rank1crossing}
in the simplest example in which $\mathsf{C}_{p_1 p_2 p}$ are the bulk OPE coefficients corresponding to 
free $\mathcal{N}=4$ Maxwell theory given above. Notice that considering this equation makes sense also 
when $\text{dim}(\mathcal{U}^{\pm}_k)>1$, as the image of a one-dimensional space under 
$\widehat{\mu}^{+}$ or
$\widehat{\mu}^{-}$  is at most one dimensional.
We also remark that the equations  are invariant under $\mu^{\pm}_{p m}\mapsto \sigma_m\,\mu^{\pm}_{p m}$
where $\sigma_m^2=1$ and
that the two sets of equations map to each other under
 $\mathsf{a}_p\rightarrow (-1)^{\frac{p}{2}}\mathsf{a}_p$, $\mu^{\pm}\rightarrow (-1)^{\frac{p}{2}} \mu^{\mp}$.
The simplest solution is obtained by setting $\mu^{-}_{p,m}=0$ by hand. Then the solution of the remaining equations is
given by
\be\label{freeU1bdy}
\mathsf{a}_{2n}\,=\,2^{n}\,\frac{\Gamma(n+\tfrac{1}{2})}{\sqrt{\pi\,(2n)!}}\,,
\qquad
\mu^{+}_{p,k}=2^{k}\,\sqrt{{p}\choose{k}}\,\mathsf{a}_{p-k}\,.
\ee
One verifies that the displacement operator 
condition $\mu_{\mathcal{O}\widehat{D}}=\mathsf{M}\,\mathsf{a}_{\mathcal{O}}\,\Delta_{\mathcal{O}} $
holds with $\mathsf{M}=\sqrt{2}$. The result \eqref{freeU1bdy} can be confirmed by a relatively  
 simple calculation using Wick contractions
  and the definition of composite operators in the free theory by normal ordering. 
 It is interesting to observe that, after imposing 
 $\mu^{\pm}_{p,2}=\mathsf{M}^{\pm}\,\mathsf{a}_{p}\,p $, it appears that
 \eqref{rank1crossing} has a unique solution parametrized by 
 $(\mathsf{M}^+,\mathsf{M}^-)$.
  It is easy to generate the solution on a computer but we could not find a closed form expression.
 Let us also notice that 
the trivial defect corresponds to $\mathsf{a}_p=0$, $\mu^{\pm}\,=\,(\pm 1)^{\frac{p}{2}}\delta_{p,k}$, in this case $\mathsf{M}^{\pm} \rightarrow \infty$.

In order to have a sufficient set of conditions for the mircrobootstrap one needs to add the crossing equations for 
correlators involving two bulks and one boundary operator. The corresponding equation takes the form
\be
\mathsf{g}^{\pm}\big(\widehat{\mu}^{\pm}(\widehat{\mathsf{C}}(v_1,v_2)),u\big)
\,=\,
\mathsf{C}_{3d}\big(
\widehat{\mu}^{\pm}(v_1),\widehat{\mu}^{\pm}(v_2),u
\big)\,.
\ee

\vspace{0.5cm}
\noindent{\emph{Remark}:}
It is known that the  three-point functions encoded in the map $\mathsf{C}$ entering
\eqref{trilinear}, are independent of marginal deformations of the $\mathcal{N}=4$ four-dimensional theory.
This is a rather unique property of $4d$ $\mathcal{N}=4$, which is canonically equipped with a conformal manifold of complex dimension one. 
The boundary data encoded in $\mathsf{\mu}^{\pm}$ and $\mathsf{a}$, on the other hand, are expected to 
vary under such marginal deformations.
It should be possible to constraint their dependence on the marginal couplings 
by using superconformal perturbation theory.

\paragraph{Microbootstrap equations in the presence of a  $\frac{1}{2}$-BPS line defect.}
The  considerations above can be extended to the case of a $\frac{1}{2}$-BPS line defect.
In this case correlation functions of 
$\frac{1}{2}$-BPS local operator  can be related to certain quantities in 
$2d$ YM, see \cite{Giombi:2009ds,Giombi:2012ep}. In these examples, it should be possible to verify the microbootstrap equations directly.

\subsection{The full bootstrap}
\label{sec:macro}
In the remaining of the section we will take a closer look at the complete bootstrap equations in two 
 examples: the two-point function of massless representations and the two-point function of the super stress-tensor.
 
\subsubsection{Example: $\mathcal{N}=4$ Massless representation, $p=1$ }
\label{sec:exampleboot:free}
As we will see, for this simple example a solution of the  microbootstrap equations,
 gives automatically a solution of the full bootstrap equations.
In this case the bulk  and boundary OPE \eqref{OPEbulksummary}, \eqref{bdyOPEsummary} reduce to 
\be
\mathsf{Pr}\left(\mathcal{B}_{[0,1,0]}\stackrel{\text{\tiny{OPE}}}{\times}\mathcal{B}_{[0,1,0]}\right)\,\simeq\,
\mathcal{I}\,+\,\mathcal{B}_{[0,2,0]}\,,
\qquad
\left[\mathcal{B}_{[0,1,0]}\right]_{\text{\tiny{$\partial$OPE}}}\,\simeq\,
(B,+)_1\,+\,(B,-)_1\,,
\ee
where $\mathcal{B}_{[0,2,0]}$ is the $4d$ $\mathcal{N}=4$ stress-tensor supermultiplet and 
$(B,\pm)_1$ are $3d$ $\mathcal{N}=4$ multiplets corresponding to conserved currents.
As in the  non-supersymmetric case, see \cite{Dimofte:2012pd,Liendo:2012hy,Gliozzi:2015qsa},
 the crossing equation for massless representations is solved by a finite number of blocks  in both channels as
\be\label{lambdasol}
1+\lambda\,\mathfrak{F}^{\text{blk}}_{\mathcal{B}_{[0,2,0]}}\,=\,\Omega^{-1}\,\left((1+\lambda)
\mathfrak{F}^{\text{bdy}}_{(B,+)_{1}}-(1-\lambda)
\mathfrak{F}^{\text{bdy}}_{(B,-)_{1}}\right)\,.
\ee
In this case the full bootstrap equations are satisfied if the microbootstrap equations  \eqref{rank1crossing},
 that in this case read $(\mu^{\pm}_{1,1})^2=\pm1 +\mathsf{C}_{112}\mathsf{a}_2$, are.
 The parameter $\lambda$ interpolates between the trivial interface,
  corresponding to  $\lambda=0$, 
 and the boundary, i.e.~the interface with an empty theory on one side,
  corresponding to $\lambda=\pm1$.
 It should be possible to have a Lagrangian realization of all other values of the parameter 
 $\lambda$ as well.


\subsubsection{Example: $\mathcal{N}=4$ Stress tensor, $p=2$ }
\label{sec:exampleboot:stretttensor}

In this case the OPE  \eqref{OPEbulksummary}  and  \eqref{bdyOPEsummary} reduce to 
\begin{align}
\mathsf{Pr}\left(\mathcal{B}_{[0,2,0]}\stackrel{\text{\tiny{OPE}}}{\times}\mathcal{B}_{[0,2,0]}\right)\Big{|}_{\text{even}}&
\,\simeq\,
\mathcal{I}\,+\,\mathcal{B}_{[0,4,0]}
\,+\,\mathcal{B}_{[2,0,2]}
\,+\,\mathcal{A}^{\Delta}_{[0,0,0]}\,,\\
\mathsf{Pr}\left(\mathcal{B}_{[0,2,0]}\stackrel{\text{\tiny{OPE}}}{\times}\mathcal{B}_{[0,2,0]}\right)\Big{|}_{\text{odd}}\,&\,\simeq\,
\mathcal{B}_{[0,2,0]}\,,
\end{align}
for the bulk channel
and
\be\label{bdyOPEstresstensor}
\left[\mathcal{B}_{[0,2,0]}\right]_{\text{\tiny{$\partial$OPE}}}\,\simeq\,
1\,+\,(B,+)_2\,+\,(B,-)_2\,+\,(B,1)_{(1,1)}\,+\,A^{\delta}_{(0,0)}\,,
\ee
for the boundary channel.  This time the OPE contains infinitely many operators.
We will use the notation $\mathcal{T}=\mathcal{B}_{[0,2,0]}$ for the stress-tensor supermultiplet.
Recall that $(B,+)_2$ and $(B,-)_2$ are the correct representations to be the super-displacement operator.
The corresponding OPE coefficients $\mu^{\pm}_{2,2}$ should be considered on the same footing as the central charge $c_{4d}$.
%
In this case the microbootstrap equations read
\be\label{microbootstrapTT}
1\pm\mathsf{C}^{\text{blk}}_{\mathcal{T}\mathcal{T}\mathcal{T}}\mathsf{a}_{\mathcal{T}}\,+
\mathsf{\lambda}^{\text{blk}}_{\mathcal{B}_{[0,4,0]}}\,=\,
\mathsf{a}_{\mathcal{T}}^2
+
\mu^2_{\mathcal{T},(B,\pm)_2}\,.
\ee
The odd part of the full crossing equations 
\be\label{ODDTT}
\mathsf{C}^{\text{blk}}_{\mathcal{T}\mathcal{T}\mathcal{T}}\,\mathsf{a}_{\mathcal{T}}\,\,
\Omega^{2}\,\mathfrak{F}^{\text{blk}}_{\mathcal{T}}\,=\,
\frac{
(\mu_{\mathcal{T},(B,+)_2} )^2
-
(\mu_{\mathcal{T},(B,-)_2})^2 
}{2}\,
\left[\mathfrak{F}^{\text{bdy}}_{(B,+)_2}-
\mathfrak{F}^{\text{bdy}}_{(B,+)_2}\right]\,,
\ee
 is then automatically satisfied
 once \eqref{microbootstrapTT} holds.
As in the previous example, the crossing equation \eqref{ODDTT} involves only a finite number of blocks.
The fact that a solution of \eqref{ODDTT}  exists relies on the identity between superblocks
\be\label{blocksatspcialpoints}
\Omega^{2}\,\mathfrak{F}^{\text{blk}}_{\mathcal{T}}\,=\,
\mathfrak{F}^{\text{bdy}}_{(B,+)_{2}}-\mathfrak{F}^{\text{bdy}}_{(B,-)_{2}}\,,
\ee
where $\Omega$ is defined in \eqref{Omegadef}.
Let us turn to the  even part of the full crossing equation.
After rewriting bulk and boundary blocks in the form \eqref{22factor} and using \eqref{microbootstrapTT} 
to cancel the $\mathsf{C}_{\pm}$ contributions from the two sides of the equation, we are left with
\begin{align}
\mathfrak{H}^{\text{blk}}_{\mathcal{I}}+
\mathsf{\lambda}^{\text{blk}}_{\mathcal{B}_{[0,4,0]}}
\mathfrak{H}^{\text{blk}}_{\mathcal{B}_{[0,4,0]}}+
\mathsf{\lambda}^{\text{blk}}_{\mathcal{B}_{[2,0,2]}}
\mathfrak{H}^{\text{blk}}_{\mathcal{B}_{[2,0,2]}}+
\sum_{\Delta}\,
&\mathsf{\lambda}^{\text{blk}}_{\mathcal{A}^{\Delta}}\,
\mathfrak{H}^{\text{blk}}_{\mathcal{A}^{\Delta}}\,=\\
=\,\mathsf{R}\,
\left(\mathfrak{H}^{\text{bdy}}_{(B,+)_2}
+
\mathfrak{H}^{\text{bdy}}_{(B,-)_2}\right)&
+
\mu^2_{\mathcal{T},(B,1)_{(1,1)}} 
\mathfrak{H}^{\text{bdy}}_{(B,1)_{(1,1)}}
+\sum_{\delta}\,
\mu^2_{\mathcal{T},L^{\delta}} 
\mathfrak{H}^{\text{bdy}}_{L^{\delta}}\,.
\end{align}
where $
\mathsf{R}\,:=\,\tfrac{1}{2}(\mu_{\mathcal{T},(B,+)_2} )^2
+
(\mu_{\mathcal{T},(B,-)_2})^2$, we have defined
\be
\mathsf{\lambda}^{\text{blk}}_{\chi_{\text{blk}}}\,=\,
\sum_{\chi(\mathcal{O}_I)=\chi_{\text{blk}}}\,
\mathsf{C}^{\text{blk}}_{\mathcal{T}\mathcal{T}\mathcal{O}_I}\mathsf{a}_{\mathcal{O}_I}\,,
\ee
and used the shorthand notation 
$\mathcal{A}^{\Delta}=\mathcal{A}^{\Delta}_{[0,0,0]}$, $L^{\delta}=L^{\delta}_{(0,0)}$.
The reduced superblocks $\mathfrak{H}^{\text{blk/bdy}}$ are defined by writing 
$\mathfrak{F}^{\text{bdy}}$ and   $\Omega^{2}\mathfrak{F}^{\text{blk}}$
 in the form  \eqref{22factor}. Following this procedure one finds the following explicit expressions
 for the boundary channel contribution
 \be
 \mathfrak{H}^{\text{bdy}}_{\text{id}}=0,
 \,\,\,
   \mathfrak{H}^{\text{bdy}}_{(B,\pm)_{2}}=-\frac{1}{8}
\log(1+\xi^{-1})\,,
\quad
 \mathfrak{H}^{\text{bdy}}_{L^{\delta}}
=\,\frac{\Gamma(2\delta+2)}{1-\delta}\,(4\xi)^{-\delta-1}
{}_{2}F_{1}(\delta+1,\delta+1,2\delta+2,-\xi^{-1})\,,
 \ee
 and for the bulk channel 
 \be
 \mathfrak{H}^{\text{blk}}_{\mathcal{I}}=\frac{1}{4\xi},
 \,\,\,
 \mathfrak{H}^{\text{blk}}_{\mathcal{B}_{[0,2,0]}}\!=0,
\,\,\,
 \mathfrak{H}^{\text{blk}}_{\mathcal{B}_{[0,4,0]}}=-{}_2F_1(1,3,4,-\xi)\,,
 \quad
  \mathfrak{H}^{\text{blk}}_{\mathcal{A}^{\Delta}}=
  \frac{(4\,\xi)^{\beta-2}}{\beta-2}\,
{}_{2}F_{1}(\beta,\beta+1,2\beta+1,-\xi)
 \ee
 where $\beta=\tfrac{1}{2}\Delta+1$.
 Notice that strictly above the unitarity bound we have $\beta-2=\tfrac{1}{2}(\Delta-2)>0$.
 The blocks $\mathfrak{H}^{\text{bdy}}_{(B,1)_{(1,1)}}$ and
 $\mathfrak{H}^{\text{blk}}_{\mathcal{B}_{[2,0,2]}}$
  are immediately obtained using the relations 
  \eqref{polebdylong} and \eqref{polebulklong}.

\paragraph{A toy model two-point function.} 
We will now consider the superconformal block expansion of the toy model two-point
function for $\langle\mathcal{T}\mathcal{T}\rangle$ given by
\be
F_{2,2}(z,w_1,w_2)=\mathsf{C}_++\kappa\,\mathsf{C}_-+\,\mathbb{D}H(z)\,,
\qquad
H(z)\,=\,\frac{z}{(1-z)^2}-g\,\frac{z}{(1+z)^2}\,,
\ee
compare to \eqref{2ptfun} and  \eqref{22factor}, where $\mathsf{C}_{\pm}$ and $g$ are certain constants.
This three parameter family of functions can be obtained by postulating that 
$\langle\mathcal{T}\mathcal{T}\rangle$ is a linear combination of 
$\langle\mathcal{T}\rangle\langle\mathcal{T}\rangle$ and the square of the 
 two-point function of massless fields corresponding to \eqref{lambdasol}. 
 By expanding this function in superblocks in the two channels we extract the CFT data
in terms of the parameters $\mathsf{C}_{+},\mathsf{C}_-,g$
\be
a_{\mathcal{T}}^2\,=\,\mathsf{C}_+-1-g\,,
\qquad
\mu^2_{\mathcal{T},(B,\pm)_2}\,=\,1+g\pm\mathsf{C}_-\,,
\quad
\mu^2_{\mathcal{T},(B,1)_{(1,1)}}\,=\,2(g-1)\,,
\ee 
and
\be
\mu^2_{\mathcal{T},A^{\delta}}\,=\,\sqrt{\pi}\,(\delta-1)\,
\frac{\Gamma(\delta+1)}{\Gamma(\delta+\tfrac{1}{2})}\,
\left(1+(-1)^{\delta}g\right)\,,
\qquad
\delta\,\in\mathbb{Z}_{\geq 2}\,.
\ee
The data associated to the bulk channel are
\be
\mathsf{C}^{\text{blk}}_{\mathcal{T}\mathcal{T}\mathcal{T}}\mathsf{a}_{\mathcal{T}}\,=\,\mathsf{C}_-\,,
\qquad
\mathsf{\lambda}^{\text{blk}}_{\mathcal{B}_{[0,4,0]}}\,=\,\mathsf{C}_+-1\,,
\qquad
\mathsf{\lambda}^{\text{blk}}_{\mathcal{B}_{[2,0,2]}}\,=\,\tfrac{1}{3}(1-\mathsf{C}_+)+\tfrac{1}{2}\,g\,,
\ee
and
\be
\mathsf{\lambda}^{\text{blk}}_{\mathcal{A}^{\Delta}}\,=\,
2^{1-\Delta}
\Gamma^2(\tfrac{\Delta}{2}+2)\,\left(
(\tfrac{\Delta}{2}-1)(\tfrac{\Delta}{2}+1)(\mathsf{C}_+-1)+
\tfrac{\Delta}{2}(\tfrac{\Delta}{2}+2-(-1)^{\tfrac{\Delta}{2}})g\right)\,,
\ee
$\Delta=4,6,8,10,\dots$.
The trivial interface belongs to this class of examples and corresponds to $\mathsf{C}_-=0$,
$\mathsf{C}_+=1$ and $g=0$.
This is a very simple solution of the bootstrap equations but implies a remarkable identity
for the bulk superblock corresponding to the exchange of the identity operator $\mathscr{F}^{\text{blk}}_{\mathcal{I}}=1$, namely
\be
\Omega^2\,\mathscr{F}^{\text{blk}}_{\mathcal{I}}\,=\,
\mathscr{F}^{\text{bdy}}_{(B,+)_2}
+
\mathscr{F}^{\text{bdy}}_{(B,-)_2}-2\,
\mathscr{F}^{\text{bdy}}_{(B,1)_{(1,1)}}+\sum_{\delta=2}^{\infty}\,\sqrt{\pi}\,(\delta-1)\,
\frac{\Gamma(\delta+1)}{\Gamma(\delta+\tfrac{1}{2})}\,
\mathscr{F}^{\text{bdy}}_{A^{\delta}}\,.
\ee
Notice that this result works for any $\mathcal{N}=4$ theory in the case of a trivial interface.
It would be interesting to learn a general lesson from this example on how the exchange 
of the identity operators in the bulk channel is reproduced upon summing infinitely many contributions from the boundary channel,
along the lines of \cite{Komargodski:2012ek,Fitzpatrick:2012yx}.

%% file: sections/5_conclusions.tex

\section{Conclusions}
\label{sec:conclusions}
In this work we studied  $4d$ $\Nm=4$ superconformal theories in the presence of defects from the point of view of the conformal bootstrap. We obtained the full superconformal block expansion for the two-point function of $\tfrac{1}{2}$-BPS operators 
in the codimension one case and thanks to the analytic continuations explained in section \ref{sec:superblocks}, these results also apply to $4d$ $\Nm=4$ superconformal theories 
 in the presence of a line defect. Moreover, they also capture the block expansion
  of $3d$ $\Nm=4$ superconformal theories and $1d$ $OSP(4^*|4)$
superconformal quantum mechanics.

In the cohomological sector we presented an infinite set of polynomial equations that relate defect and bulk data. 
Apart from the equations coupling bulk and boundary,
one also needs to consider the truncated equations for pure bulk and pure boundary CFT data separately.
While solutions to the bulk equations correspond to a commutative algebra with certain properties, 
the associative algebra that describes the boundary data is not commutative \cite{Beem:2016cbd}.
We observed that the cohomological bulk bootstrap equations 
admit a truncation to  a subsector corresponding 
to the subalgebra generated by the super stress-tensor, whose structure constants 
are fully determined in terms of the central charge.
For this to be the case, we supplemented the equations with the knowledge of so-called extremal three-point couplings
which had been obtained from rather general principles in \cite{Baggio:2015vxa}.

Understanding the solutions of the truncated bootstrap equations is an important step in order to fully characterize the superconformal theory, and we leave a more thorough analysis of their solutions for the future. For example,
it would very interesting to map the space of solutions to the boundary conditions studied by Gaiotto and Witten in \cite{Gaiotto:2008sa}.
This set of superconformal $\frac{1}{2}$-BPS boundary conditions in $\mathcal{N}=4$ SYM is extremely rich, 
as they can be engineered 
by gauging a flavor symmetry of a generic three-dimensional
$\mathcal{N}=4$  SCFT.
It should be pointed out that defect configurations that are S-dual correspond to the same solution of the bootstrap equations.  
 
Another interesting question is to understand how deep is the connection between the four systems studied in this paper:
 $4d$ $\Nm=4$ superconformal theories with codimension one and three defects, 
 $3d$  $\mathcal{N}=4$ superconformal theories and $1d$ 
  $OSP(4^*|4)$ 
quantum mechanics. If the solutions of the truncated equations can be mapped to each other, it will bring the kinematic relation 
uncovered in this paper to the realm of dynamics.

Given a solution of the bootstrap equations in the cohomological sector, one would like to determine
a, or many, solutions to the full bootstrap equations. 
This is a complicated problem, in particular due to the fact that 
long operators with unknown conformal dimension appear.
The usual approach is to apply modern numerical techniques. 
The absence of positivity makes the original method of \cite{Rattazzi:2008pe} unsuitable, however, there are alternative techniques in the literature. Apart from the already mentioned Gliozzi approach, a new promising method is the one developed in \cite{El-Showk:2016mxr}.

Another interesting venue is the study  superconformal defects in superconformal theories in other dimensions.
Promising candidates are codimension one and two $\frac{1}{2}$-BPS defects in  $3d$ $\mathcal{N}=4$ theories 
\cite{Assel:2015oxa,Bullimore:2016nji} 
and  codimension two and four defects in  $6d$ $(2,0)$ theories, \cite{Bullimore:2014upa,Beem:2015aoa}. 
It would also be interesting to investigate whether certain bulk-boundary CFT data can be computed using localization methods.
This idea has been applied successfully to extract the one-point function of the super stress tensor 
for the case in which the defect is a Wilson line, see \cite{Lewkowycz:2013laa} and references therein.
For the case of a codimension one defect, the results of \cite{Gaiotto:2014ina,Gaiotto:2015una} 
may already contain the necessary information to extract some boundary CFT data.

%% file: sections/acknowledgments.tex

\acknowledgments

We have greatly benefited from discussions with F.~Bonetti, M.~G\"unaydin,
C.~Herzog, M.~Hogervorst, M.~Isachenkov, M.~Lemos, V.~Mitev, W.~Peelaers, L.~Rastelli, V.~Schomerus, and  K.~Zarembo.
The authors thank GGI Florence for hospitality during the workshop ``Conformal field theories and renormalization group flows in dimensions $d>2$''.
P.~L. thanks Nordita for hospitality during the workshop ``Holography and dualities 2016''.
P.~L. is supported by SFB 647 ``Raum-Zeit-Materie. Analytische und Geometrische Strukturen''.

%% file: sections/A_appendix.tex

\section{Some representation theory}

\subsection{Supermultiplets}
\label{App:supermultiplets}
We will now review aspects of representation theory of superconformal algebras relevant for the discussion in the main text.
The representation theory of the $4d$ $\mathcal{N}=4$ superconformal algebra 
is summarized in table \ref{tab:repsPSU224NEW}.
Certain representations of $\mathfrak{psu}(2,2|4)$ will play a distinguished role
\begin{align}\label{ABCd4}
\mathcal{B}_{[q,p,q]}&:=B_1\bar{B}_1[0;0]^{[q,p,q]}_{\Delta=p+2q}\,,\\
\mathcal{C}_{[0,p,0],(\ell,\ell)}&:=A_*\bar{A}_*[2\ell;2\ell]^{[0,p,0]}_{\Delta=2+2\ell+p}\,,\\
\mathcal{A}^{\Delta}_{[q,p,\bar{q}],(\ell,\bar{\ell})}&:=L\bar{L}[2\ell;2\bar{\ell}]^{[q,p,\bar{q}]}_{\Delta}\,.
\end{align}
where $A_*$ is either $A_1$ or $A_2$ depending on the value of $\ell$. 
\begin{table}
\centering
\renewcommand{\arraystretch}{1.7}
\begin{tabular}{%
| l
                |>{\centering }m{4.8cm}
             |>{\centering\arraybackslash}m{2.5cm}|
}
\hline
Supermultiplet
& Highest weight
& $\mathcal{R}_{\text{blk}}^{\text{left}}$
\\\hline
$L$
&$[2\ell,2\bar{\ell}]^{[q,p,\bar{q}]}_{\Delta}$\,
&
$[q,2\ell]_{\Delta-\gamma}$
\\\hline
$A_1$
&$[2\ell,2\bar{\ell}]^{[q,p,\bar{q}]}_{\Delta}$,\,\,\,\,
$\ell >0$\,
&
$ \text{\scriptsize{$q+1$}}\Big{\{}
\overbrace{\autoparbox{\young{& &&&\cr\cr\cr\cr}}}^{2\ell+2}\,$
\\\hline
$A_2$
&$[0,2\bar{\ell}]^{[q,p,\bar{q}]}_{\Delta}$
&
$ \text{\scriptsize{$q+1$}}\Big{\{}
\autoparbox{\young{&\cr\cr\cr\cr}}$
\\\hline
$B_1$
&$[0,2\bar{\ell}]^{[q,p,\bar{q}]}_{\Delta}$
&
$ \text{\scriptsize{$q$}}\Big{\{}
\autoparbox{\young{\cr\cr\cr\cr}}$
\\\hline
\end{tabular}
\renewcommand{\arraystretch}{1.0}
\caption{Unitary representations of the $4d$ $\mathcal{N}=4$ superconformal algebra in the notation of  
\cite{Cordova:2016xhm}, apart for the small modification $(R_1,R_2,R_3)$ replaced by $[q,p,\bar{q}]$
for Dynkin labels. This is just the chiral half of the conditions. 
The other half can be easily obtained. 
The labels $[a,b]_c$ denote  long representations of $\mathfrak{sl}(2|2)$ 
with dimension $16 (a+1)(b+1)$ and $c\in\mathbb{C}$
is the value of the central charge generators.
The rightmost entry of the table is part of the data of the inducing representation 
$\mathcal{R}_{\text{blk}}=\{\mathcal{R}_{\text{blk}}^{\text{left}},\mathcal{R}_{\text{blk}}^{\text{right}},Q\}$,
where $Q=\Delta-\ell-\bar{\ell}$ is the conformal twist,
see  \cite{Doobary:2015gia} and references therein for more details.
}
\label{tab:repsPSU224NEW}
\end{table}
For representations of the $3d$ $\mathcal{N}=4$ superconformal algebra see table \ref{tab:nonlongrepsosp4NEW}.
\begin{table}
\centering
\renewcommand{\arraystretch}{1.7}
\begin{tabular}{%
| l
                |>{\centering }m{2.8cm}
             |>{\centering\arraybackslash}m{3.5cm}
             |
}
\hline
Supermultiplet
&Unitarity bound
& $\widetilde{\mathcal{R}}_{\text{bdy}}$
\\\hline
$L[2s]^{(2k_+;\,2k_-)}_{\delta}$\,
& 
$\delta>s+k_++k_-+1$
&
$[2k_{\sigma},2s]_{\delta-k_{-\sigma}}$
\\\hline
$A_1[2s]^{(2k_+;\,2k_-)}_{\delta}$,\,\,\,\,
$s>0$\,
& 
$\delta=s+k_++k_-+1$
&
$ \text{\scriptsize{$2k_{\sigma}+1$}}\Big{\{}
\overbrace{\autoparbox{\young{& &&&\cr\cr\cr\cr}}}^{2s+2}\,$
\\\hline
$A_2[0]^{(2k_+;\,2k_-)}_{\delta}$
& 
$\delta=k_++k_-+1$
&
$ \text{\scriptsize{$2k_{\sigma}+1$}}\Big{\{}
\autoparbox{\young{&\cr\cr\cr\cr}}$
\\\hline
$B_1[0]^{(2k_+;\,2k_-)}_{\delta}$
& 
$\delta=k_++k_-$
&
$ \text{\scriptsize{$2k_{\sigma}$}}\Big{\{}
\autoparbox{\young{\cr\cr\cr\cr}}$
\\\hline
\end{tabular}
\renewcommand{\arraystretch}{1.0}
\caption{Unitary representations of the $3d$ superconformal algebra in the notation of \cite{Cordova:2016xhm},
$[2s]^{(2k_+;\,2k_-)}_{\delta}\,=\,\{\delta,s,(k_+,k_-)\}$ with $2s,2k_+,2k_-\in \mathbb{Z}_{\geq 0}$.
The sign $\sigma\in\{+,-\}$ denotes the choice of
 $\mathfrak{sl}(2|2)_{\pm} \supset \mathfrak{su}_{\text{Lorentz}}(2)\oplus \mathfrak{su}_{\pm}(2)$.
  The symbol  $\widetilde{\mathcal{R}}_{\text{bdy}}$ stands for a $\mathfrak{sl}(2|2)$  representation.
 The inducing representation of $\mathfrak{gl}(2|2)$
 is $\mathcal{R}_{\text{bdy}}=\{\widetilde{\mathcal{R}}_{\text{bdy}},b_{2|2}\}$, 
 where $b_{2|2}$ is a $U(1)$ quantum number. 
  Notice that $A_1[2s]^{(2k_+;0)}_{\delta}$ and their mirrors are somewhat special since 
  the $\mathfrak{sl}(2|2)$  inducing representation becomes a totally symmetric representation.
  Following  the Dolan classification  \cite{Dolan:2008vc} we denote  $(A,\pm)^s_{k_+}:=A_1[2s]^{(2k_+;0)}_{\delta}$.
Among these, the representations 
  $A_1[2s]^{(0;0)}_{\delta}$ are distinguished by the fact that 
 the $\mathfrak{sl}(2|2)$
   inducing representation becomes a totally symmetric representation with respect to each of the inducing factors. 
   These multiplets are referred to as conserved currents as they include higher-spin conserved currents;
    the $s=0$  corresponds to the stress-tensor supermultiplet.
  The $\mathfrak{sl}(2|2)$ central charge in our normalization is given by $\delta-k_{-\sigma}$, it is half the number of boxes in the 
  Young tableaux.
}
\label{tab:nonlongrepsosp4NEW}
\end{table}
We give the complete list of representations however, only a few of them will appear in the boundary OPE of a $\frac{1}{2}$-BPS bulk operator. We will therefore simplify the notation a bit following \cite{Dolan:2008vc}:
\be\label{changeofnames1}
(B,+)_k\,:=\,B_1[0]_{k}^{(2k,0)}\,,
\qquad
(B,-)_k\,:=\,B_1[0]_{k}^{(0,2k)}\,,
\ee
and
\be\label{changeofnames2}
(B,1)_{(k_+,k_-)}\,:=\,B_1[0]_{k_++k_-}^{(2k_+,2k_-)}\,,
\qquad
\text{cons}_s\,:=\,A_1[2s]_{\delta}^{(0;0)}\,,
\qquad
L^{\delta}_{(k_+,k_-)}\,:=\,L[0]_{\delta}^{(2k_+,2k_-)}\,.
\ee
We also recall that at the unitarity bound, long representation decompose as
\begin{align}\label{unitaritydecompositions}
\lim_{\delta\rightarrow k_++k_-+1}\, L[0]_{\delta}^{(2k_+,2k_-)}\,\,\,\sim&
\,\,
A_2[0]_{\delta}^{(2k_+,2k_-)}\,\,\,+\,
B_1[0]_{\delta+1}^{(2k_++1,2k_-+1)}\, , \\
\lim_{\delta\rightarrow s+k_++k_-+1}\, L[2s]_{\delta}^{(2k_+,2k_-)}\,\sim&
\,\,
A_1[2s]_{\delta}^{(2k_+,2k_-)}\,+\,
A_1[2s-1]_{\delta+\tfrac{1}{2}}^{(2k_++1,2k_-+1)}\, ,
\end{align}
where in the second line $s>0$.
These relations imply that certain boundary superblocks can be obtained as the residue of the pole 
in $\delta$ of the long block, see \eqref{polebdylong}.
We also remark that the decompositions \eqref{unitaritydecompositions} take place already at the level of inducing representation 
$\widetilde{\mathcal{R}}_{\text{bdy}}$ given in table \ref{tab:nonlongrepsosp4NEW}, see \cite{Beisert:2006qh}.

\paragraph{Representations of superconformal quantum mechanics.}
The relevant representation theory of  $OSP(4^*|4)$ superconformal quantum mechanics can essentially be extracted from 
\cite{Gunaydin:1990ag}.
In the following we will describe three classes of representations that are relevant for the discussion in the main text. The $d=1$ superconformal algebra $\mathfrak{osp}(4^*|4)$ contains 
$\mathfrak{sl}(2,\mathbb{R})\oplus\mathfrak{su}(2)\oplus \mathfrak{usp}(4)$ as bosonic subalgebra where $\mathfrak{sl}(2,\mathbb{R})$ plays the role of conformal algebra in one dimension. The operators with smallest value $\Delta_{1d}$ of the one dimensional dilatation 
generator form a finite dimensional irreducible 
representation of the R-symmetry $\mathfrak{su}(2)\oplus \mathfrak{usp}(4)$. 
We use Dynkin labels to  characterize such representation $\{n,[a,b]\}$, where $n$ labels the $n+1$
dimensional representation of $\mathfrak{su}(2)$ and $[a,b]$ are $\mathfrak{usp}(4)$ Dynkin labels, 
for example $[1,0]=\mathbf{4}$ and $[0,1]=\mathbf{5}$.
Representations of  $OSP(4^*|4)$ are then uniquely characterized by 
$\chi_{\text{SCQM}}=\{\Delta_{1d},n,[a,b]\}$. For special values of this labels the representation contains null states 
that have to be removed, these corresponds to so-called atypical representations.
We introduce the notation
\begin{align}\label{unitaryrepsSCQM}
(B^*,+)_{\frac{b}{2}}\,\,:&=\{\tfrac{b}{2},0,[0,b]\}\,,\\
(B^*,1)_{(\frac{b}{2},n)}:&=\{\tfrac{b}{2},n,[0,b]\}\,,\qquad n>0\\
L^*[\Delta_{1d}]_{n}^{[0,b]}:&=\{\Delta_{1d},n,[0,b]\}\,,
\end{align}
notice that there is no $(B,-)$ multiplet. 
Indeed the mirror automorphism is not compatible with the real form 
$OSP(4^*|4)$.
In these definition subtraction of null states is understood.

\subsection{Miscellanea}

\subsubsection{Details on symmetry and the mirror automorphism}
The R-symmetry subgroup of $OSP(4|4)$ can be defined in a similar way as \eqref{OSPgroupdef}
\be\label{Ogroupdef_APP}
O(4)\,=\,
\Big{\{}
g\,\in\,GL(4)\,\,
\text{such that }\,
g^{t}\,\eta_R\, g\,=\,\eta_R
\Big{\}}\,,
\qquad
\eta_R:=\begin{pmatrix}
0 & 1_2 \\
1_2 & 0
\end{pmatrix}\,.
\ee
At the lie algebra level, $\mathfrak{o}(4) \sim \mathfrak{su}(2)_+  \oplus \mathfrak{su}(2)_- $ is spanned by matrices
\be
\mathfrak{su}(2)_+ \,=\,\text{Span}\Big{\{}
\left(\begin{smallmatrix} a&0 & 0 & -b\\
0 & a &b &0\\
0 & c & -a & 0\\
-c & 0 & 0 & -a
\end{smallmatrix}\right)
\Big{\}}\,,
\qquad
\mathfrak{su}(2)_- \,=\,\text{Span}\Big{\{}
\left(\begin{smallmatrix} 
-\tilde{a}& -\tilde{c} & 0&0\\
-\tilde{b} & \tilde{a} &0 & 0\\
0 & 0 & \tilde{a} & \tilde{b}\\
0 & 0 & \tilde{c} & -\tilde{a}
\end{smallmatrix}\right)
\Big{\}}\,.
\ee
Notice that $\mathfrak{su}(2)_+$ acts projectively while $\mathfrak{su}(2)_-$ acts linearly on the R-symmetry 
coordinates.
Let us define
\be
\mathbf{T}_{R}\,:=\,
\left(\begin{smallmatrix} 
 0&0 & 0 & 1\\
1& 0 & 0&0\\
0 & 1& 0 & 0\\
0 & 0 & 1 & 0
\end{smallmatrix}\right)\,.
\ee
This element generates a (non-central) $\mathbb{Z}_4\subset SL(4)$. 
Moreover it generates the mirror automorphism $\mathbf{M}$ defined below \eqref{algebra}, 
which follows from  the identity
\be
\mathbf{T}_{R}^{}\,
\left(\begin{smallmatrix} a&0 & 0 & -b\\
0 & a &b &0\\
0 & c & -a & 0\\
-c & 0 & 0 & -a
\end{smallmatrix}\right)
\,\mathbf{T}_{R}^{-1}\,=\,
\left(\begin{smallmatrix} 
-a& -c & 0&0\\
-b & a &0 & 0\\
0 & 0 & a & b\\
0 & 0 & c & -a
\end{smallmatrix}\right)\,.
\ee
One should also notice that $\mathbf{T}_{R}^2\,=\,\eta_{R}$, $\mathbf{T}_{R}^{}\mathbf{T}_{R}^{\dagger}=1$,
$\det(\mathbf{T}_{R})=-1$.

\subsubsection{Conformal factors}
\label{App:conformalfactors}

The conformal factors are given by
\be\label{Omegafactors}
\Omega_{g,X}\,=\,
\begin{cases}
\Big(1,1\Big)&
g\,=\,\left(\begin{smallmatrix}1& B \\  0&1
\end{smallmatrix}\right)\,, \\
\Big(A, (A^{st})^{-1}\Big)\,,&
g\,=\,\left(\begin{smallmatrix}A& 0 \\  0&(A^{st})^{-1}
\end{smallmatrix}\right) \\
\Big((1+X\,C)^{-1}, (1+C\,X)\Big)\,,\qquad
&
g\,=\,\left(\begin{smallmatrix}1& 0 \\  C&1
\end{smallmatrix}\right) 
\end{cases}\,,
\ee
and
\be
\omega_{g,X_{\text{b}}}\,=\,
\begin{cases}
1\,,&
g\,=\,\left(\begin{smallmatrix}1& B \\  0&1
\end{smallmatrix}\right) \\
A\,,&
g\,=\,\left(\begin{smallmatrix}A& 0 \\  0&(A^{st})^{-1}
\end{smallmatrix}\right) \\
\left(1+ X_{\text{b}}\,C\right)^{-1}\,,\qquad&
g\,=\,\left(\begin{smallmatrix}1& 0 \\  C&1
\end{smallmatrix}\right) 
\end{cases}
\ee
The form for general $g$ follows by group composition law.
\subsubsection{Definitions}
We recall the definitions of  superdeterminant and super-Pfaffian
\be
\text{sdet}
\begin{pmatrix}
A & B \\
C & D
\end{pmatrix}\,\colonequals \,
\frac{\det(A-BD^{-1}C)}{\det(D)}
\,=\,
\frac{\det(A)}{\det(D-CA^{-1}B)}\,,
\ee
\be\label{sPfdef}
\text{sPf}
\begin{pmatrix}
A & sB \\
B^t & D
\end{pmatrix}\,\colonequals \,
\frac{\sqrt{\det(A)}}{\text{Pfaff}(D-s B^tA^{-1}B)}\,=\,
\frac{\sqrt{\det(A-sB D^{-1}B^t)}}{\text{Pfaff}(D)}\,.
\ee
\subsubsection{The stability group $S_X$}
\label{App:stabilityalgebra}

Without loss of generality we can take $X_{\text{b}}=0$, which can be obtained by 
performing an  $OSP(4|4)$ super-translation. 
From the infinitesimal transformation properties 
\be
\delta\big(0,X_{\text{d}}\big)\,=\,
\big(\beta- X_{\text{d}}\,\gamma\,X_{\text{d}} ,\alpha\,X_{\text{d}}+X_{\text{d}}\,\alpha^{st}\big)\,,
\ee
where close to the identity $\hat{g}:=g_{\psi}^{}g\,g_{\psi}^{-1}
=1+\left(\begin{smallmatrix}\alpha&\beta\\\gamma&-\alpha^{st}
\end{smallmatrix}\right)$,
one finds that the stability  group 
$S_{X}$ (with $X=A_{\psi}^{-1}X_{\text{d}}A_{\psi}^{}$) close to the identity looks like
\be
\hat{g}=1+
\begin{pmatrix}
\alpha & X_{\text{d}}\,\gamma\,X_{\text{d}} \\
\gamma & -\alpha^{st}
\end{pmatrix}+\dots
\qquad
\alpha\,X_{\text{d}}+X_{\text{d}}\,\alpha^{st}=0\,,
\quad
\gamma=-\gamma^{st}\,\Sigma\,.
\ee
It follows that
the stability algebra is spanned by 
\be\label{stabilityalgebraEQ}
m_{\pm}(\alpha)\,:=\,
\frac{1}{2}
\begin{pmatrix}
+\alpha & \pm \alpha\,X_{\text{d}}^{}\\
\pm X_{\text{d}}^{-1}\,\alpha& -\alpha^{st}
\end{pmatrix}\,,
\qquad
\alpha\,X_{\text{d}}+X_{\text{d}}\,\alpha^{st}=0\,.
\ee
Notice that the fact that $X_{\text{d}}$ is invertible is the statement that the configuration is generic.
It is a simple exercise to show that
\be
\left[ 
m_{\sigma_1}(\alpha_1), 
m_{\sigma_2}(\alpha_2)
\right]\,=\,
\delta_{\sigma_1,\sigma_2}\,m_{\sigma_1}([\alpha_1,\alpha_2])\,.
\ee
This proves that the stability algebra  is $\mathfrak{osp}(2|2)\oplus\mathfrak{osp}(2|2)$.

\subsection{Branching ratios and tensor products}

\subsubsection{R-symmetry channels}
\label{app:Rchannells}
For the boundary channel we use the  
$\mathfrak{su}(4)\rightarrow \mathfrak{su}(2)\oplus \mathfrak{su}(2)$ branching ratios
\be\label{branchinRsymm}
[0,2\ell+\varepsilon,0]\,\rightarrow\, \bigoplus_{s=0}^{\ell} \bigoplus_{a=0}^{2s+\varepsilon}\,\,(2s+\varepsilon-a,a)\,,
\qquad
\varepsilon\,\in\,\{0,1\}\,.
\ee
For the bulk channel we have
\be\label{projectedTENSORprod}
\mathsf{Pr}\Big([0,2\ell+\varepsilon,0]\otimes [0,2\ell+\varepsilon,0]\Big)\,=\,
\bigoplus_{0\leq m\leq n\leq \ell} [2(n-m),4m,2(n-m)]
\bigoplus_{0\leq m\leq n\leq \ell+\varepsilon-1} [2(n-m),4m+2,2(n-m)]\,,
\ee
where $\mathsf{Pr}$ denotes a projection into representations of the form $[2a,2b,2c]$.
Notice that the sum \eqref{projectedTENSORprod} splits into two sums, the blocks corresponding to each sum have definite $\mathbb{Z}_2$ transformation properties.
The number of multiplet is the same in bulk and boundary channel and it is equal to 
 $(\ell+1)(\ell+\varepsilon+1)$.

We also collect the defect channel R-symmetry OPE relevant for the line defect.
In this case one has $\mathfrak{su}(4)\rightarrow \mathfrak{sp}(4)$ branching ratios
\be
[0,p,0]\rightarrow \bigoplus_{d=0}^p\,[0,d]\,.
\ee
The $\mathfrak{sp}(4)$ Dynkin labels $[0,d]$
correspond to $\tfrac{1}{6}(d+1)(d+2)(2d+3)$ dimensional representations.
Recall that the representations of $\mathfrak{su}(4)$ that contains a $\mathfrak{sp}(4)$ 
singlet are the ones with Dynkin labels $[0,t,0]$.

\subsubsection{$\mathfrak{sl}(2|2)\rightarrow \mathfrak{osp}(2|2)$ branching ratios}
\label{App:branching}

In this appendix we determine which  $\mathfrak{sl}(2|2)$ representation listed in tables 
\ref{tab:nonlongrepsosp4NEW}
and \ref{tab:repsPSU224NEW} contains a state (vector) 
which is invariant under  $\mathfrak{osp}(2|2)\subset \mathfrak{sl}(2|2)$ where the choice of embedding is specified in 
Appendix \ref{App:stabilityalgebra}. It turns out that, apart from the case in which the  $\mathfrak{sl}(2|2)$ representation is trivial,
this invariant state will not appear as the trivial representation of $\mathfrak{osp}(2|2)$, but as part of the so called projective cover of the trivial representation denoted by $\mathcal{P}(0)$, see \cite{Gotz:2005jz}. We refer to Table \ref{tab:osp22reps} for the notation and a few facts about the relevant representations of  $\mathfrak{osp}(2|2) \simeq \mathfrak{sl}(2|1)$. Since we could not find the branching ratios  in the literature
we present them below.
\begin{table}
\centering
\renewcommand{\arraystretch}{1.6}
\begin{tabular}{%
 |l    |>{\centering }m{7.9cm}
             |>{\centering\arraybackslash}m{3.7cm}|
}
\hline 
symbol
&name
&dimension
\\\hline
$\{0\}$ &
trivial representation
&
$1$
\\\hline
$\{j\}_{\pm}$ &
atypical representation
&
$4j+1$
\\\hline
$\{b,j\}$ &
typical representation
&
$8j$
\\\hline
$\mathcal{P}(0)$ &
projective cover of the trivial representation
&
$8$
\\\hline
$\mathcal{P}(\pm j)$ &
projective cover of the atypical  representation $\{j\}_{\pm}$ 
&
$16j+4$
\\\hline
\end{tabular}
\renewcommand{\arraystretch}{1.0}
\caption{Brief summary of the $\mathfrak{osp}(2|2)$ representations appearing in the branching ratios considered here, see \cite{Gotz:2005jz} for a complete description of these representations. Above $2j\in\mathbb{Z}_{>0}$ and $b\neq \pm j$.}    
\label{tab:osp22reps}
\end{table}

The branching rules can be determined from the knowledge of tensor products in $\mathfrak{sl}(2|2)$ and $\mathfrak{osp}(2|2)$ together with the fact that the operations of taking tensor product and branching ratios commute and the tensor product is
 distributive with respect to  direct sums.
Let us illustrate this procedure in a simple example. The fundamental representation of $\mathfrak{sl}(2|2)$ decomposes as 
$ 
\autoparbox{\young{\cr}}\rightarrow \{0, \tfrac{1}{2}\}$. This branching ratio specifies the embedding  $\mathfrak{osp}(2|2)\subset \mathfrak{sl}(2|2)$.
Taking tensor products and branching ratios of the fundamental representation with itself gives
\be
\autoparbox{\young{\cr}}\otimes 
\autoparbox{\young{\cr}}\,=\,
\autoparbox{\young{&\cr}}\oplus
\autoparbox{\young{\cr\cr}}\,
\qquad
\rightarrow
\qquad
 \{0, \tfrac{1}{2}\}\otimes\{0, \tfrac{1}{2}\}\,=\, \{0, 1\}\oplus \mathcal{P}(0)\,,
\ee
see Table \ref{tab:osp22reps} for details about the representations on the right hand side.
Iterating this procedure, with extensive use of $\mathfrak{osp}(2|2) \simeq \mathfrak{sl}(2|1)$ tensor product decompositions from 
 \cite{Gotz:2005jz},  one arrives at the following  rules
  \be
  \underbrace{\autoparbox{\young{& &&&\cr}}}_s\,
  \rightarrow\,\{0,\tfrac{s}{2}\}\,,
  \ee
  \be
  \text{\scriptsize{$a$}}\left\{\autoparbox{\young{\cr\cr\cr\cr}}\right.\,
  \rightarrow\,
  \begin{cases}
\mathcal{P}(0)
\oplus
\bigoplus\limits_{b=-\text{\tiny{$\tfrac{a-1}{2}$}}}^{\text{\tiny{$\tfrac{a-1}{2}$}}}
{\vphantom{\bigoplus}}\!\!\!' \,\,\,
\{b,\tfrac{1}{2}\}\,,\quad & a=2+2g\,,\\
\bigoplus\limits_{b=-\text{\tiny{$\tfrac{a-1}{2}$}}}^{\text{\tiny{$\tfrac{a-1}{2}$}}}
\{b,\tfrac{1}{2}\}\,,
 & \text{otherwise}\,,
  \end{cases}
  \ee
    \be
 \text{\scriptsize{$a+1$}}\Big{\{}
\overbrace{\autoparbox{\young{& &&&\cr\cr\cr\cr}}}^{s+1}\,
  \rightarrow\,
  \begin{cases}
\mathcal{P}(\tfrac{s}{2})
\oplus 
\mathcal{P}(-\tfrac{s}{2})
\oplus
\bigoplus\limits_{b=-\text{\tiny{$\tfrac{a}{2}$}}}^{\text{\tiny{$\tfrac{a}{2}$}}}
{\vphantom{\bigoplus}}\!\!\!' \,\,\,
\{b,\tfrac{s+1}{2}\}\,\,
\oplus
\bigoplus\limits_{b=-\text{\tiny{$\tfrac{a-1}{2}$}}}^{\text{\tiny{$\tfrac{a-1}{2}$}}}
{\vphantom{\bigoplus}}\!\!\!' \,\,\,
\{b,\tfrac{s}{2}\}\,,\quad & a=s+1+2g\,,\\
\bigoplus\limits_{b=-\text{\tiny{$\tfrac{a}{2}$}}}^{\text{\tiny{$\tfrac{a}{2}$}}}
\{b,\tfrac{s+1}{2}\}\,\,
\oplus
\bigoplus\limits_{b=-\text{\tiny{$\tfrac{a-1}{2}$}}}^{\text{\tiny{$\tfrac{a-1}{2}$}}}
\{b,\tfrac{s}{2}\}\,,
 & \text{otherwise}\,,
  \end{cases}
  \ee
  \be
  [a,s]_{\gamma}\rightarrow
  \begin{cases}
  \mathcal{P}(0)\oplus\mathbf{S}_{0,0}\,, & s=0,\,\,a=0\,,\\
 \mathcal{P}(\tfrac{1}{2})
  \oplus   \mathcal{P}(-\tfrac{1}{2})
  \oplus \mathbf{S}_{1,1}\,, & s=1,\,\,a=1\,,\\
  \mathcal{P}(0)\oplus   \mathcal{P}(\tfrac{1}{2})
  \oplus   \mathcal{P}(-\tfrac{1}{2})
  \oplus \mathbf{S}_{0,2g+2}\,, & s=0,\,\,a=2g+2\,,\\
   \mathcal{P}(\tfrac{s}{2})
  \oplus   \mathcal{P}(-\tfrac{s}{2})
  \oplus
    \mathcal{P}(\tfrac{s+1}{2})
  \oplus   \mathcal{P}(-\tfrac{s+1}{2})
  \oplus \mathbf{S}_{a,s}\,, & s\geq 1,\,\,a=s+2g+2\,,\\
 \mathbf{S}_{a,s}\,, & \text{otherwise}\,,
  \end{cases}
  \ee
  where
  \be
   \mathbf{S}_{a,s}\,:=\,
   \bigoplus\limits_{b=-\text{\tiny{$\tfrac{a}{2}$}}}^{\text{\tiny{$\tfrac{a}{2}$}}}
   {\vphantom{\bigoplus}}\!\!\!' \,\,\,
\Big(\{b,\tfrac{s+2}{2}\}\oplus
\{b+\tfrac{1}{2},\tfrac{s+1}{2}\}\oplus
\{b-\tfrac{1}{2},\tfrac{s+1}{2}\}\oplus
\{b,\tfrac{s}{2}\}
\Big)\,.
  \ee
  Some remarks on the notation are in order: the sums $\bigoplus$ are integer spaced between the limits of summation,
  $\bigoplus'$  means that one should exclude from the sum terms that give contributions of the form $\{\pm j,j\}$, the latter are exactly the one that recombines into projective representations $\mathcal{P}$, finally $g\in \mathbb{Z}_{\geq 0}$.

The only $\mathfrak{osp}(2|2)$ representation appearing in the right hand side of the list above that contains an invariant state is
 $\mathcal{P}(0)$.
Notice that
\be
\mathcal{P}(0)\,\sim\, 2\{0\}\oplus\{\tfrac{1}{2}\}_+
\oplus\{\tfrac{1}{2}\}_-\,,
\qquad
\mathcal{P}(\pm\tfrac{1}{2})\,\sim\, 2\{\tfrac{1}{2}\}_{\pm}\oplus\{0\}
\oplus\{1\}_{\pm}\,,
\ee
where $\sim$ means that they have the same character. 
In both cases the trivial representation $\{0\}$ appears. 
In the first case it appears twice,
 only one of them is a true $\mathfrak{osp}(2|2)$ 
invariant state.
In the second case the state associated to $\{0\}$ by the character is not an invariant,
 see \cite{Gotz:2005jz}. 
This important difference would have been missed by a character analysis.

Looking at the branching ratios given above we conclude that,
apart from the trivial representation of  $\mathfrak{sl}(2|2)$, only the totally antisymmetric representations in $2g+2$ indices 
and  the long representations $[a,s]_{\gamma}=[2g,0]_{\gamma}$ with $g\in \mathbb{Z}_{\geq 0}$ contain an $\mathfrak{osp}(2|2)$
invariant state.

%% file: sections/B_appendix.tex

\section{Comparison of conformal blocks}
\label{APP:Rsymmblocksdetails}

In this short appendix we present the details of the comparison between our conformal blocks and the results already present in the literature.

\subsection{Casimir equation for the bulk R-symmetry blocks.}

The Casimir equation takes the form 
\be\label{App:casimirRsymm}
\left[\left(\sum_{i=1}^2\,w_i(w_i-1)^2\partial^2_{w_i}\right)+
k(w_1,w_2)\partial_{w_1}+k(w_2,w_1)\partial_{w_2}\right]\,
\mathfrak{h}^{\text{blk}}_{R}\,=\,C_{R}\,\mathfrak{h}^{\text{blk}}_{R}\,,
\ee
where
\be\label{kCdef}
k(w_1,w_2)=\left(\frac{w_1 \left(w_1-1\right)}{w_1-w_2}+\frac{w_1-1}{w_1 w_2-1}-2\right) \left(w_1-1\right)\,,
\,\,
C_{[2m,2n,2m]}\!=\!2 \left(n^2+2 n
   (m+1)+m (2 m+3)\right)\,.
\ee
The  change of variables \eqref{tifromwi} was motivated by the fact that 
we were able to find two special cases of R-symmetry blocks in closed form:
\begin{align}\label{Rsymmclosedform}
\mathfrak{h}^{\text{blk}}_{[0,2n,0]}&
= \frac{4^{-2 n} n!}{\left(\frac{1}{2}\right)_n}\sum_{a,b=0}^{n} \delta_{n-a,\text{even}}\delta_{n-b,\text{even}}
\frac{i^{b-a}(a+b)! \frac{a+k}{2}! \frac{b+k}{2}!}{i! j! \left(\frac{a+b}{2}!\right)^2 \frac{k-a}{2}! \frac{k-b}{2}!}  t_1^a t_2^b\,,
\\
\mathfrak{h}^{\text{blk}}_{[2m,0,2m]} &
=\frac{(-64)^{-m} (m!)^2}{\left(\frac{3}{4}\right)_m \left(\frac{5}{4}\right)_m}\sum_{a,b=0}^{m} 
\frac{(2 a)! (2 b)! (-1)^{a+b} 2^{-4 (a+b)-1} (2 (a+b+m+1))!}{(a!)^2 (b!)^2 ((a+b)!)^2 (-a-b+m)! (a+b+m+1)!}t_1^{2 a} t_2^{2 b}\,.
\end{align}
In the first equation $\delta_{a-b,\text{even}}=1$, if $a-b$ is even, and 0 otherwise.
For the special case $w_2=0$,$R= [0,2n,0]$ the expression above further simplifies to 
\be
\mathfrak{h}_{[0,2n,0]}^{\text{blk}}(w,0)\,=\,
\frac{(-1)^{n}(n+1)}{4^n\binom{2n}{n}}
{}_2F_{1}(-n,n+2,\tfrac{3}{2},\tfrac{1}{1-w})\,.
\ee
In the variables $t_i$, see \eqref{tifromwi}, the Casimir equation \eqref{App:casimirRsymm} takes the form 
,\be\label{App:casimirRsymmt}
\left[\left(\sum_{i=1}^2\,(t_i^2-1)\partial^2_{t_i}\right)+
\frac{1}{t_1^2-t_2^2}\left(\tilde{k}(t_1,t_2)\partial_{t_1}+\tilde{k}(t_2,t_1)\partial_{t_2}\right)\right]\,
\mathfrak{h}^{\text{blk}}_{R}\,=\,C_{R}\,\mathfrak{h}^{\text{blk}}_{R}\,,
\ee
where $\tilde{k}(t_1,t_2)=2 t_1 \left(2 t_1^2-t_2^2-1\right) $.
After performing the change of variables $t_1=\sqrt{x}$, $t_2=\sqrt{\bar{x}}$ and defining $\mathfrak{g}=(x\bar{x})^{\frac{1}{4}}\mathfrak{h}^{\text{bulk}}$, one finds that $\mathfrak{g}$ satisfies the Dolan-Osborn Casimir equations 
\cite{Dolan:2011dv} specialized to  $d=3$, upon identifying 
\be
\Delta_{3d}=\tfrac{1}{2}-n-m\,,
\qquad
\ell_{\text{\cite{Dolan:2011dv}}}=s=m\,.
\ee
These relations can be confirmed by comparing the asymptotic behavior in \eqref{asymRsymm}
with the one of \cite{Dolan:2011dv}. By looking at this asymptotic behavior we further conclude that
$J=2m$ and $\Delta=-2(m+n)$.

\subsection{Line defect from analytic continuation}

We will now show that the R-symmetry polynomials $\mathfrak{h}^{\text{blk}}$ are an analytic continuation 
of the spacetime blocks in the presence of a line defect by comparing with the analysis of  \cite{Billo:2016cpy}.
In the boundary channel the authors were able to find an explicit expression for the conformal blocks, 
concerning the bulk channel, \cite{Billo:2016cpy} contains the relevant Casimir equation without an explicit solution.
We somewhat fill this gap by relating it to the solution of the Casimir equation of the standard $3d$ four-point function blocks.

\paragraph{Comparison of cross-ratios.}
The first step is to relate the cross-ratios \eqref{fromYtoeigs}
to the cross-ratios in \cite{Billo:2016cpy}.
Let
\be
\cos \phi\,:=\,-\frac{1}{2}\frac{\mathrm{tr}(y_{1,\text{d}}\,\epsilon\, y_{2,\text{d}}\,\epsilon)}{\sqrt{\det(y_{1,\text{d}})\det(y_{2,\text{d}})}}\,,
\qquad
\xi_{R}\,:=\,\frac{1}{4}
\frac{\det(y_1-y_2)}{\sqrt{\det(y_{1,\text{d}})\det(y_{2,\text{d}})}}\,.
\ee
By going to the frame \eqref{fromYtoeigs} it is easy to verify that
\be\label{cosphixiR}
\cos \phi\,=\,\frac{1}{2}\frac{w_1+w_2}{\sqrt{w_1\,w_2}}\,=\,
\frac{\,w_-^{}+w_-^{-1}}{2}\,,
\qquad
\xi_{R}\,=\,\frac{1}{4}\,\prod_{i=1}^2\left(w_i^{+\frac{1}{2}}-w_i^{-\frac{1}{2}}\right)\,,
\ee
where  $w_\pm^2:=w_1w_2^{\pm1}$ have been defined below
\eqref{superblockBULKfromblocks}.
It is instructive to compare these cross-ratios with the spacetime
cross-ratio in the presence of a boundary 
\be\label{xireview}
\xi\,:=\,
\frac{\det(x_1-x_2)}{4\,x_{1,\text{d}}\,x_{2,\text{d}}}\,=\,
\frac{(z-1)^2}{4\, z}\,=\,
\frac{1}{4}\,\prod_{i=1}^2\left(z^{+\frac{1}{2}}-z^{-\frac{1}{2}}\right)\,.
\ee
Notice that $\xi_{R}=\tfrac{1}{4} \xi_{\text{\cite{Billo:2016cpy}}}$, while $\phi$ is the same, the additional variable 
$\chi = (4 \xi_R + 2 \cos \phi)=w_+^{}+w_{+}^{-1}$
is also used in \cite{Billo:2016cpy}.

\paragraph{Boundary channel.} The block in the defect channel take a factorized form.
With the  identifications above we obtain the following block identities valid for each factor,
\begin{align} 
w_{-}^{-k_-}{_2}F_1(\tfrac{1}{2},-k_-,\tfrac{1}{2}-k_-;w^2_{-}) 
&=  
\frac{\sqrt{\pi}\,\Gamma(1+s)}{\Gamma(\tfrac{1}{2}+s)}
{_2}F_1(\tfrac{3}{2}+\tfrac{s}{2}-1,-\tfrac{s}{2},\tfrac{3}{2}-1;\sin^2\phi)\,,
\\
w_{+}^{k_+}{_2}F_1(\tfrac{1}{2},-k_+,\tfrac{1}{2}-k_+;w^2_{+}) 
&=  \chi^{-\delta}\,{_2}F_1(\tfrac{\delta}{2}+\tfrac{1}{2},\delta+1-\tfrac{1}{2},;\tfrac{4}{\chi^2})\,.
\end{align}
In the equalities above, representation labels are identified as $s=k_-$ is the spin of the boundary operator 
appearing in this channel and $\delta=-k_+$ is its dimension.

\paragraph{Bulk channel} 
Using the change of variables \eqref{cosphixiR}
 it is a straightforward exercise to compare (4.12) of \cite{Billo:2016cpy} with our Casimir equation
 \eqref{App:casimirRsymm} for the bulk R-symmetry block. 
 The identification between the labels is
\be 
J(J+2)+\Delta(\Delta-4) = 2\,C_{[2m,2n,2m]}\,,
\ee
where $C_R$ is given in \eqref{kCdef} we have set $q=3$, $d=4$, and $\Delta_{12}=0$ in (4.12) of \cite{Billo:2016cpy}.

%% file: sections/C_appendix.tex

\section{Long blocks coefficients}

In this appendix we present the OPE coefficients for the long superblocks in the bulk and boundary channels.
   
\subsection{Boundary channel coefficients}
\label{app:boundarylongblocks}

As in \eqref{B1coeffs}, each coefficient $c_{\delta,(k_++i,k_-+j)}$ has a counterpart $c_{\delta,(k_++j,k_-+i)}$ where $k_{+}$ and $k_{-}$ are interchanged. Below we only list the ones with $i \geqslant j$.
{\footnotesize
\begin{align}
\begin{split}
c_{\delta+1,(k_{+}+1,k_{-}+1)} & = -\frac{\delta -k_{+}-k_{-}}{\delta -k_{+}-k_{-}-1}
\\
c_{\delta+1,(k_{+}-1,k_{-}-1)} & = -\frac{16 k_{+}^2 k_{-}^2 (\delta +k_{+}+k_{-}+2)}{(2 k_{+}-1) (2 k_{+}+1) (2 k_{-}-1) (2 k_{-}+1) (\delta +k_{+}+k_{-}+1)}
\\
c_{\delta+1,(k_{+}+1,k_{-}-1)} & = -\frac{4 k_{-}^2 (\delta -k_{+}+k_{-}+1)}{(2 k_{-}-1) (2 k_{-}+1) (\delta -k_{+}+k_{-})} 
\end{split}
\end{align}
}

{\footnotesize
\begin{align}
\begin{split}
c_{\delta+2,(k_{+}+2,k_{-})} & = \frac{(-\delta +k_{+}-k_{-}-1) (-\delta +k_{+}+k_{-})}{(-\delta +k_{+}-k_{-}) (-\delta +k_{+}+k_{-}+1)}
\\
c_{\delta+2,(k_{+},k_{-})} & = -\frac{8 k_{+} k_{-} (k_{+}+k_{-}-1) (2 k_{+}+2 k_{-}+3) (-\delta +k_{+}-k_{-}-1) (\delta +k_{+}-k_{-}+1) (\delta +k_{+}+k_{-}+2)}{(2 \delta -1) (2 \delta +3) (2 k_{+}-1)
   (2 k_{+}+1) (2 k_{-}-1) (2 k_{-}+1) (\delta +k_{+}+k_{-}+1)}
\\
& \frac{8 k_{+} (k_{-}+1) (2 k_{+}-2 k_{-}+1) (k_{+}-k_{-}-2) (-\delta +k_{+}+k_{-}) (\delta +k_{+}-k_{-}+1) (\delta +k_{+}+k_{-}+2)}{(2 \delta -1) (2 \delta +3) (2 k_{+}-1)
   (2 k_{+}+1) (2 k_{-}+1) (2 k_{-}+3) (\delta +k_{+}-k_{-})}
\\
&
\frac{8 (k_{+}+1) k_{-} (2 k_{+}-2 k_{-}-1) (k_{+}-k_{-}+2) (-\delta +k_{+}-k_{-}-1) (-\delta +k_{+}+k_{-}) (\delta +k_{+}+k_{-}+2)}{(2 \delta -1) (2 \delta +3) (2 k_{+}+1)
   (2 k_{+}+3) (2 k_{-}-1) (2 k_{-}+1) (-\delta +k_{+}-k_{-})} 
\\
&
-\frac{8 (k_{+}+1) (k_{-}+1) (k_{+}+k_{-}+3) (2 k_{+}+2 k_{-}+1) (-\delta +k_{+}-k_{-}-1) (-\delta +k_{+}+k_{-}) (\delta +k_{+}-k_{-}+1)}{(2 \delta -1) (2 \delta +3) (2
   k_{+}+1) (2 k_{+}+3) (2 k_{-}+1) (2 k_{-}+3) (-\delta +k_{+}+k_{-}+1)}
\\
c_{\delta+2,(k_{+},k_{-}-2)} & = \frac{16 (k_{-}-1)^2 k_{-}^2 (-\delta +k_{+}-k_{-}-1) (\delta +k_{+}+k_{-}+2)}{(2 k_{-}-3) (2 k_{-}-1)^2 (2 k_{-}+1) (-\delta +k_{+}-k_{-}) (\delta +k_{+}+k_{-}+1)} 
\end{split}
\end{align}
}

{\footnotesize
\begin{align}
\begin{split}
c_{\delta+3,(k_{+}+1,k_{-}-1)} & = -\frac{16 \delta  (\delta +2) k_{-}^2 (-\delta +k_{+}+k_{-}) (\delta -k_{+}+k_{-}+1) (\delta +k_{+}+k_{-}+2)}{(2 \delta +1) (2 \delta +3) (2 k_{-}-1) (2
   k_{-}+1) (-\delta +k_{+}+k_{-}+1) (\delta -k_{+}+k_{-}) (\delta +k_{+}+k_{-}+1)}
\\
c_{\delta+3,(k_{+}+1,k_{-}+1)} & = -\frac{4 \delta  (\delta +2) (\delta -k_{+}-k_{-}) (\delta +k_{+}-k_{-}+1) (\delta -k_{+}+k_{-}+1)}{(2 \delta +1) (2 \delta +3) (\delta -k_{+}-k_{-}-1)
   (\delta +k_{+}-k_{-}) (\delta -k_{+}+k_{-})}
\end{split}
\end{align}
}

{\footnotesize
\begin{align}
\begin{split}
c_{\delta+4,(k_{+},k_{-})} & = 
\frac{16 \delta  (\delta +1) (\delta +2) (\delta +3) (\delta -k_{+}-k_{-}) (\delta +k_{+}-k_{-}+1) (\delta -k_{+}+k_{-}+1) (\delta +k_{+}+k_{-}+2)}{(2
   \delta +1) (2 \delta +3)^2 (2 \delta +5) (\delta -k_{+}-k_{-}-1) (\delta +k_{+}-k_{-}) (\delta -k_{+}+k_{-}) (\delta +k_{+}+k_{-}+1)}  
\end{split} 
\end{align}
}

\subsection{Bulk channel coefficients}
\label{app:bulklongblocks}

Here we present the coefficients of the bulk channel.
{\tiny
\begin{align}
\begin{split}
c_{\Delta+2,[2m-2,2n,2m-2]} & =-\frac{(4 m-1)^2 (4 m+1)^2 (n+2 m)^2 (n+2 m+1)^2 (2 n+2 m+1)^2 (\Delta +2 n+4 m+6)}{2 m^3 (2 m+1) (n+m) (n+m+1) (2 n+4 m-1) (2 n+4 m+1)^2 (2 n+4 m+3)
   (\Delta +2 n+4 m+4)}
\\
c_{\Delta+2,[2m-2,2n+4,2m-2]} & =-\frac{(n+1)^2 (n+2)^2 (4 m-1)^2 (4 m+1)^2 (-\Delta +2 n-2)}{128 (2 n+1)^2 (2 n+3)^2 m^3 (2 m+1) (2 n-\Delta )}
\\ 
c_{\Delta+2,[2m,2n,2m]} & =\frac{1}{32} \left(\frac{\Delta +1}{\Delta -1}-\frac{(\Delta +5) (2 n+1)^2 (2 n+4 m+3)^2}{(\Delta +3) (2 n-1) (2 n+3) (2 n+4 m+1) (2 n+4 m+5)}\right)
\\
c_{\Delta+2,[2m+2,2n-4,2m+2]} & =-\frac{8 (2 n-3) (m+1)^5 (\Delta +2 n+4) (2 n+2 m+1)^2}{(2 n+1) (2 m+1) (4 m+3)^2 (4 m+5)^2 (\Delta +2 n+2) (n+m) (n+m+1)}
\\
c_{\Delta+2,[2m+2,2n,2m+2]} & =-\frac{(m+1)^5 (-\Delta +2 n+4 m)}{8 (2 m+1) (4 m+3)^2 (4 m+5)^2 (-\Delta +2 n+4 m+2)}
\end{split}
\end{align}
}
{\tiny
\begin{align}
\begin{split}
c_{\Delta+4,[2m-4,2n+4,2m-4]} & = \frac{(n+1)^2 (n+2)^2 (4 m-5)^2 (4 m-3)^2 (4 m-1)^2 (4 m+1)^2 (-\Delta +2 n-2) (n+2 m)^2 (n+2 m+1)^2 (\Delta +2 n+4 m+6)}{64 (2 n+1)^2 (2 n+3)^2 (m-1)^3 m^3
   (2 m-1) (2 m+1) (2 n-\Delta ) (2 n+4 m-1) (2 n+4 m+1)^2 (2 n+4 m+3) (\Delta +2 n+4 m+4)}
\\
c_{\Delta+4,[2m-2,2n,2m-2]} & = -\frac{(4 m-1)^2 (4 m+1)^2 (-\Delta +2 n-2) (\Delta +2 n+4) (n+2 m)^2 (n+2 m+1)^2 (2 n+2 m+1)^2 (\Delta +2 n+4 m+6)}{16 (\Delta +1) (\Delta +3) (2 n-1) (2
   n+3) m^3 (2 m+1) (n+m) (n+m+1) (2 n+4 m-1) (2 n+4 m+1)^2 (2 n+4 m+3) (\Delta +2 n+4 m+4)}
\\
c_{\Delta+4,[2m-2,2n+4,2m-2]} & = -\frac{(n+1)^2 (n+2)^2 (4 m-1)^2 (4 m+1)^2 (-\Delta +2 n-2) (-\Delta +2 n+4 m) (\Delta +2 n+4 m+6)}{1024 (\Delta +1) (\Delta +3) (2 n+1)^2 (2 n+3)^2 m^3 (2
   m+1) (2 n-\Delta ) (2 n+4 m+1) (2 n+4 m+5)}
\\
c_{\Delta+4,[2m,2n-4,2m]} & = \frac{(2 n-3) (\Delta +2 n+4) (n+2 m)^2 (n+2 m+1)^2 (2 n+2 m-1)^2 (2 n+2 m+1)^2 (\Delta +2 n+4 m+6)}{(2 n+1) (\Delta +2 n+2) (n+m-1) (n+m)^2 (n+m+1) (2 n+4
   m-1) (2 n+4 m+1)^2 (2 n+4 m+3) (\Delta +2 n+4 m+4)}
\\
c_{\Delta+4,[2m,2n,2m]} & = -\frac{(m+1) (-\Delta +2 n-2) (\Delta +2 n+4) (n+2 m+2)^2 (n+2 m+3) (2 n+2 m+3) (-\Delta +2 n+4 m)}{256 (\Delta +1) (\Delta +3) (2 m+1) (2 m+3) (n+m+1)
   (n+m+2) (2 n+4 m+5) (-\Delta +2 n+4 m+2)}
\\
& \frac{(n+1)^2 (n+2) m (-\Delta +2 n-2) (2 n+2 m+3) (-\Delta +2 n+4 m) (\Delta +2 n+4 m+6)}{256 (\Delta +1) (\Delta +3) (2 n+3) (2 m-1) (2 m+1) (2 n-\Delta )
   (n+m+1) (n+m+2)}
\\
& \frac{(n-1) n^2 (m+1) (\Delta +2 n+4) (2 n+2 m+1) (-\Delta +2 n+4 m) (\Delta +2 n+4 m+6)}{256 (\Delta +1) (\Delta +3) (2 n-1) (2 m+1) (2 m+3) (\Delta +2
   n+2) (n+m) (n+m+1)}
\\
& -\frac{m (-\Delta +2 n-2) (\Delta +2 n+4) (n+2 m) (n+2 m+1)^2 (2 n+2 m+1) (\Delta +2 n+4 m+6)}{256 (\Delta +1) (\Delta +3) (2 m-1) (2 m+1) (n+m) (n+m+1) (2
   n+4 m+1) (\Delta +2 n+4 m+4)}
\\
c_{\Delta+4,[2m,2n+4,2m]} & = \frac{(n+1)^2 (n+2)^2 (-\Delta +2 n-2) (-\Delta +2 n+4 m)}{4096 (2 n+1)^2 (2 n+3)^2 (2 n-\Delta ) (-\Delta +2 n+4 m+2)}
\\
c_{\Delta+4,[2m+2,2n-4,2m+2]} & = -\frac{(2 n-3) (m+1)^5 (\Delta +2 n+4) (2 n+2 m+1)^2 (-\Delta +2 n+4 m) (\Delta +2 n+4 m+6)}{(\Delta +1) (\Delta +3) (2 n+1) (2 m+1) (4 m+3)^2 (4 m+5)^2
   (\Delta +2 n+2) (n+m) (n+m+1) (2 n+4 m+1) (2 n+4 m+5)}
\\
c_{\Delta+4,[2m+2,2n,2m+2]} & = -\frac{(m+1)^5 (-\Delta +2 n-2) (\Delta +2 n+4) (-\Delta +2 n+4 m)}{64 (\Delta +1) (\Delta +3) (2 n-1) (2 n+3) (2 m+1) (4 m+3)^2 (4 m+5)^2 (-\Delta +2 n+4
   m+2)}
\\
c_{\Delta+4,[2m+4,2n-4,2m+4]} & = \frac{4 (2 n-3) (m+1)^5 (m+2)^5 (\Delta +2 n+4) (-\Delta +2 n+4 m)}{(2 n+1) (2 m+1) (2 m+3) (4 m+3)^2 (4 m+5)^2 (4 m+7)^2 (4 m+9)^2 (\Delta +2 n+2) (-\Delta
   +2 n+4 m+2)}
\end{split}
\end{align}
}

{\tiny
\begin{align}
\begin{split}
c_{\Delta+6,[2m-2,2n,2m-2]} & = -\frac{(\Delta +2) (\Delta +4) (4 m-1)^2 (4 m+1)^2 (-\Delta +2 n-2) (\Delta +2 n+4) (n+2 m)^2 (n+2 m+1)^2 (2 n+2 m+1)^2 (\Delta +2 n+4 m+6)}{512 (\Delta
   +3)^2 m^3 (2 m+1) (2 n-\Delta ) (\Delta +2 n+2) (n+m) (n+m+1) (2 n+4 m-1) (2 n+4 m+1)^2 (2 n+4 m+3) (\Delta +2 n+4 m+4)}
\\
c_{\Delta+6,[2m-2,2n+4,2m-2]} & = -\frac{(\Delta +2) (\Delta +4) (n+1)^2 (n+2)^2 (4 m-1)^2 (4 m+1)^2 (-\Delta +2 n-2) (-\Delta +2 n+4 m) (\Delta +2 n+4 m+6)}{32768 (\Delta +3)^2 (2 n+1)^2 (2
   n+3)^2 m^3 (2 m+1) (2 n-\Delta ) (-\Delta +2 n+4 m+2) (\Delta +2 n+4 m+4)}
\\
c_{\Delta+6,[2m,2n,2m]} & = \frac{(\Delta +2) (\Delta +4) (m+1) (-\Delta +2 n-2) (\Delta +2 n+4) (2 n+2 m+3) (-\Delta +2 n+4 m)^2 (\Delta +2 n+4 m+6)}{1024 (\Delta +1) (\Delta +3)^2
   (\Delta +5) (2 n-1) (2 n+3) (2 n-\Delta ) (\Delta +2 n+2) (2 n+4 m+5) (-\Delta +2 n+4 m+2)}
\\
& 
-\frac{(\Delta +2) (\Delta +4) m (-\Delta +2 n-2) (\Delta +2 n+4) (2 n+2 m+1) (-\Delta +2 n+4 m) (\Delta +2 n+4 m+6)^2}{1024 (\Delta +1) (\Delta +3)^2
   (\Delta +5) (2 n-1) (2 n+3) (2 n-\Delta ) (\Delta +2 n+2) (2 n+4 m+1) (\Delta +2 n+4 m+4)}
\\
c_{\Delta+6,[2m+2,2n-4,2m+2]} & = -\frac{(\Delta +2) (\Delta +4) (2 n-3) (m+1)^5 (\Delta +2 n+4) (2 n+2 m+1)^2 (-\Delta +2 n+4 m) (\Delta +2 n+4 m+6)}{32 (\Delta +3)^2 (2 n+1) (2 m+1) (4
   m+3)^2 (4 m+5)^2 (\Delta +2 n+2) (n+m) (n+m+1) (-\Delta +2 n+4 m+2) (\Delta +2 n+4 m+4)}
\\
c_{\Delta+6,[2m+2,2n,2m+2]} & = -\frac{(\Delta +2) (\Delta +4) (m+1)^5 (-\Delta +2 n-2) (\Delta +2 n+4) (-\Delta +2 n+4 m)}{2048 (\Delta +3)^2 (2 m+1) (4 m+3)^2 (4 m+5)^2 (2 n-\Delta )
   (\Delta +2 n+2) (-\Delta +2 n+4 m+2)}
\end{split}
\end{align}
}
{\tiny
\begin{align}
\begin{split}
c_{\Delta+8,[2m,2n,2m]} & =\frac{(\Delta +2) (\Delta +4)^2 (\Delta +6) (\Delta -2 n+2) (\Delta +2 n+4) (\Delta -2 n-4 m) (\Delta +2 n+4 m+6)}{65536 (\Delta +3)^2 (\Delta +5)^2 (\Delta
   -2 n) (\Delta +2 n+2) (\Delta -2 n-4 m-2) (\Delta +2 n+4 m+4)}
\end{split}
\end{align}
}